\documentclass[12pt,a4paper]{article}

\usepackage[english]{babel}
\usepackage{amsmath}
\allowdisplaybreaks[3]
\usepackage{amsfonts}
\usepackage{amssymb}
\usepackage{graphicx}
\usepackage{multirow}
\usepackage[left=2cm,right=2cm,top=2cm,bottom=2cm]{geometry}
\usepackage[normalem]{ulem}

\newcommand{\csw}{c_\mathrm{SW}}

\newcommand{\Tr}{\mathrm{Tr}}

\title{One-loop $\csw$ for Wilson and Brillouin fermions with stout smearing or Wilson flow}

\author{Maximilian Ammer$\,^{a}$, Stephan D\"urr$\,^{a,b}$}

\begin{document}
\maketitle
\begin{center}
${}^a${\sl Department of Physics, University of Wuppertal, 42119 Wuppertal, Germany}\\
${}^b${\sl J\"ulich Supercomputing Centre, Forschungszentrum J\"ulich, 52425 J\"ulich, Germany}
\end{center}
\vspace{10pt}
\begin{abstract}
We present results for the one-loop value of the improvement coefficient $c_\mathrm{SW}$ for Wilson and
Brillouin fermions subject to stout smearing or Wilson flow, in combination with Wilson or Symanzik glue.
To this end we use a recently developed method that allows one to modify an existing perturbative calculation, like the one for $c_\mathrm{SW}^{(1)}$,
to include stout smearing or Wilson flow at arbitrary stout parameters ($\varrho$, $n_\mathrm{stout}$) or flow times $t/a^2$, respectively.
Our results indicate that already a small amount of smoothing makes the perturbative series well behaved,
suggesting that a non-perturbatively determined $c_\mathrm{SW}$ might be close to its one-loop value for couplings $g_0^2\simeq 1$.
\end{abstract}

%%%%%%%%%%%%%%%%%%%%%%%%%%%%%
\section{Introduction}
%%%%%%%%%%%%%%%%%%%%%%%%%%%%%

Stout smearing and the gradient flow (in the form of Wilson or Symanzik flow) are two intimately related methods to smoothen a given gauge background in lattice gauge theory.
Such smoothed gauge links may be used both in gluonic observables (e.g.\ interpolating fields for glueball physics) and in the covariant derivative of a lattice fermion operator.

In this paper we are interested in $\csw^{(1)}$, which is a specific improvement coefficient (see below) for Wilson \cite{Wilson:1974sk} or Brillouin \cite{Durr:2010ch} fermions.
This coefficient is an essential ingredient in the Symanzik improvement program \cite{Symanzik:1983dc,Symanzik:1983gh} which effectively adds higher-dimensional counterterms to the fermion action.
Without these counterterms the leading cut-off effects in on-shell quantities would scale $\propto a$ for either fermion action, where $a$ is the lattice spacing.
The goal of the improvement program is to reduce them to $\propto \alpha a$ or better, where $\alpha=g^2/(4\pi)$ is the strong coupling constant at a scale $\mu \sim a^{-1}$.
This parametric reduction of the leading cut-off artefacts allows one to reach better precision in continuum extrapolated quantities.

Wilson and Brillouin fermions have $\csw^{(0)}=r$ at tree-level, with $r$ being the lifting parameter, regardless of the amount and/or type of smoothing (see Ref.~\cite{Ammer:2023otl} and references therein).
The one-loop value for non-smoothed Wilson fermions has been calculated in various setups, with increasing numerical precision over time \cite{Sheikholeslami:1985ij,Wohlert:1987rf,Aoki:1998qd,Horsley:2008ap}.
Here we follow the ``conventional'' perturbative approach of Ref.~\cite{Aoki:2003sj}, with some details specified in Ref.~\cite{Ammer:2023otl} (which also contains the calculation for Brillouin fermions).
The task is thus to apply the ``upgrade recipe'' of Ref.~\cite{Ammer:2024hqr} to the calculation of Refs.~\cite{Aoki:2003sj,Ammer:2023otl} to determine $\csw^{(1)}$ for Wilson or Brillouin fermions with stout smearing or gradient flow on Wilson or Symanzik glue.

Iterated stout smearing with parameters ($\varrho$, $n=n_\mathrm{stout}$) is defined through \cite{Morningstar:2003gk}
\begin{align}
U^{(n+1)}_\mu(x)&=\exp\big\{i\varrho Q^{(n)}_\mu(x)\big\}U^{(n)}_\mu(x)\label{eq:U(n+1)}\\
U^{(0)}_\mu(x)&=U_\mu(x)\,.
\label{eq:U(0)=U}
\end{align}
It generates the Wilson flow, which is defined by the differential equation \cite{Narayanan:2006rf,Luscher:2010iy,Luscher:2011bx}
\begin{align}
\partial_t U_\mu(x,t)&=iQ_\mu(x,t) U_\mu(x,t)
\label{eq:wf_de}\\
U_\mu(x,0)&=U_\mu(x)
\end{align}
at flow times $t>0$ in the limit $\varrho\to 0$, $n_\mathrm{stout}\to\infty$ with $n_\mathrm{stout}\cdot\varrho=\mathrm{const}=t/a^2$.
The Hermitian operators $Q_\mu$ in (\ref{eq:U(n+1)}) and (\ref{eq:wf_de}) are identical, except that one of them is constructed from smeared and the other one from flowed gauge backgrounds.
They ensure that the new link variables are again elements of $SU(N_c)$, with
\begin{align}
Q_\mu(x)&=\frac{1}{2i}\left(W_\mu(x)-\frac{1}{N_c}\Tr\left[W_\mu(x)\right]\right)\\
W_\mu(x)&=S_\mu(x)\tilde{U}^{\dagger}_\mu(x)-\tilde{U}_\mu(x)S^{\dagger}_\mu(x)\\
S_\mu(x)&=\sum\limits_{\nu\neq\mu}\left(\tilde{U}_\nu(x)\tilde{U}_\mu(x+\hat{\nu})\tilde{U}^{\dagger}_\nu(x+\hat{\mu})+\tilde{U}^{\dagger}_\nu(x-\hat{\nu})\tilde{U}_\mu(x-\hat{\nu})\tilde{U}_\nu(x+\hat{\mu}-\hat{\nu})\right),
\label{eq:S_mu(x)}
\end{align}
where $\tilde{U}_\mu(x)=U_\mu^{(n)}(x)$ for stout smearing and $\tilde{U}_\mu(x)=U_\mu(x,t)$ for the gradient flow setup.
In other words the $n$-dependence of $Q_\mu^{(n)}(x)$ in (\ref{eq:U(n+1)}) and the $t$-dependence of $Q_\mu(x,t)$ in (\ref{eq:wf_de}) ensure that the infinitesimal change to the gauge field is computed from the latest version of the smoothed or flowed background.
Mathematically speaking the stout smearing procedure (\ref{eq:U(n+1)}) represents a first oder integration scheme to solve the flow equation (\ref{eq:wf_de}).
A non-Abelian version of the midpoint rule would define a second order scheme, and higher-order integration schemes have been proposed in Refs.~\cite{Luscher:2010iy,Ce:2015qha}.

The framework of Ref.~\cite{Ammer:2024hqr} allows for a unified treatment of both stout smearing and gradient flow in lattice perturbation theory.
To this end one defines new gluon fields $\tilde{A}_\mu(g_0,x)$ through
\begin{align}
\tilde{U}_\mu(x)=e^{ig_0T^a\tilde{A}_\mu(g_0,x)}
\end{align}
where $\tilde{U}_\mu(x)$ is defined after (\ref{eq:S_mu(x)}), and $\tilde{A}_\mu(g_0,x)=\tilde{A}_\mu^{(n)}(g_0,x)$ in the case of stout smearing or $\tilde{A}_\mu(g_0,x)=\tilde{A}_\mu(g_0,x,t)$ with the gradient flow.
Throughout $\tilde{A}_\mu=\tilde{A}_\mu^a T^a$ where the adjoint color index $a=1,...,N_c^2-1$ is summed over, and $T^a$ is the Lie algebra generator (e.g.\ $T^a=\frac{1}{2}\lambda^a$ for $N_c=3$).
This unified treatment is based on a $SU(N_c)$ expansion of the link variables which coincides with the perturbative expansion in the bare coupling $g_0^2=2N_c/\beta$ only for $n_\mathrm{stout}=0$ or $t=0$, respectively, so
\begin{align}
U^{(0)}_\mu(x)=U_\mu(x,0)=U_\mu(x)=e^{ig_0T^aA^a_\mu(x)}.
\end{align}
In consequence the new gluon fields $\tilde{A}_\mu$ need to be expanded in $g_0^2$, and the result
\begin{align}
\tilde{A}_\mu^{a}(g_0)&=\sum\limits_\nu \mathcal{F}_{\mu\nu}A^{a}_\nu
-\frac{g_0}{2}f^{abc}\sum\limits_{\nu\rho}\mathcal{F}_{\mu\nu\rho}A^a_\nu A^b_\rho
\nonumber\\&
-\frac{g_0^2}{6}
\bigg[\frac{1}{N_c}\delta^{ae}\delta^{bc}
+\frac{1}{2}\big(d^{bcd}d^{eda}-f^{bcd}f^{eda}\big)\bigg]
\sum\limits_{\nu\rho\sigma}
\mathcal{F}_{\mu\nu\rho\sigma}
A^e_\nu A^b_\rho A^c_\sigma
+\mathcal{O}(g_0^3)
\end{align}
was given in Ref.~\cite{Ammer:2024hqr}.
It depends on the functions $\mathcal{F}_{\mu\nu}$, $\mathcal{F}_{\mu\nu\rho}$ and $\mathcal{F}_{\mu\nu\rho\sigma}$ which depend on either $(\varrho,n)$ or $t$.
The $\mathcal{F}$ are thus coefficients multiplying the original gauge field $A_\mu$, while $f$ and $d$ denote the structure constants of $SU(N_c)$.
In the case of stout smearing $\mathcal{F}_{\mu\nu}$ is also called $\tilde{g}_{\mu\nu}^{(n)}$, in the case of the Wilson flow it is $B_{\mu\nu}(t)$ (and analogous for $\mathcal{F}_{\mu\nu\rho}$ and $\mathcal{F}_{\mu\nu\rho\sigma}$).
Below we will discuss how these functions are to be included in the Feynman rules of a given lattice fermion operator such that they reflect the effect of stout smearing or gradient flow in a perturbative calculation.

Let us illustrate why a solid perturbative knowledge of $\csw^{(0)},\csw^{(1)}$ will remain useful even if $\csw$ might be determined non-perturbatively, at some point in the future, for stout-smeared or gradient-flowed Wilson or Brillouin fermions.
The Sheikholeslami-Wohlert coefficient $\csw$ of the clover improvement term
\begin{align}
\csw \frac{a^5}{2}\sum\limits_x\sum\limits_{\nu>\mu}\bar{\psi}(x) \sigma_{\mu\nu} F_{\mu\nu}(x)\psi(x)
\end{align}
is often expressed in terms of a rational function, for example through
\begin{align}
\csw(g_0) = \frac{c_0(1+c_1 g_0^2 + c_2 g_0^4+c_3 g_0^6)}{1 + c_4 g_0^2}
\label{eq:csw_rational}
\end{align}
which is then used to fit some data points in the intermediate coupling regime ($g_0^2\simeq1$) obtained from non-perturbative lattice simulations.
On the other hand lattice perturbation theory in the weak coupling regime yields the first few terms in the expansion
\begin{align}
\csw = \csw^{(0)} + \csw^{(1)} g_0^2 +\mathcal{O}(g_0^4)
\end{align}
where we include only the tree-level and one-loop values which are currently known for Wilson and Brillouin fermions.
In order to reconcile the two ansaetze, we expand Eq.~(\ref{eq:csw_rational})
\begin{align}
\csw(g_0) =c_0 +c_0 (c_1-c_4)g_0^2 +\mathcal{O}(g_0^4)
\end{align}
and find $\csw^{(0)}=c_0$ as well as $\csw^{(1)}=c_0(c_1-c_4)$.
This implies that the coefficient $c_0$ can be eliminated from the rational ansatz (\ref{eq:csw_rational}) in favor of $\csw^{(0)}$, and likewise $c_4$ is traded for $c_1-\csw^{(1)}/\csw^{(0)}$.
This reduces the number of degrees of freedom in the fit by two units, which typically comes with reduced uncertainties of the remaining coefficients and thus of the overall fit curve.

The remainder of this article is organized as follows.
The Feynman rules for either Wilson or Brillouin fermions coupled to either Wilson or Symanzik glue are sketched in Section~\ref{sec:feynman_rules}, with all details left to Ref.~\cite{Ammer:2024hqr}.
The key elements how $\csw^{(1)}$ may be calculated for either fermion action are summarized in Section~\ref{sec:csw}, with all details left to Ref.~\cite{Ammer:2023otl}.
Section~\ref{sec:csw_stout} presents our results for $\csw^{(1)}$ for stout smeared Wilson and Brillouin fermions on Wilson and Symanzik glue, with tables and plots organized such as to facilitate a comparison between the four options implied.
Section~\ref{sec:csw_wf} contains analogous results with the gradient flow (instead of stout smearing) being the smoothing recipe.
A few concluding remarks are arranged in Section~\ref{sec:conclusion}, while additional steps of the calculation and/or results are presented in various appendices.

%%%%%%%%%%%%%%%%%%%%%%%%%%%%%
\section{Feynman rules for fermions with smoothed links\label{sec:feynman_rules}}
%%%%%%%%%%%%%%%%%%%%%%%%%%%%%

This section contains a brief exposition of the Feynman rules for a lattice fermion coupled to a smoothed gauge background.
The smoothing is encapsulated in the functions $\tilde{g}_{\mu\nu}$, $\tilde{g}_{\mu\nu\rho}$, $\tilde{g}_{\mu\nu\rho\sigma}$ in case of stout smearing,
or $B_{\mu\nu}$, $B_{\mu\nu\rho}$, $B_{\mu\nu\rho\sigma}$ in case of the gradient flow.
For all details the reader is referred to Ref.~\cite{Ammer:2024hqr} and references therein.

\begin{figure}[!tb]
\centering
\includegraphics[scale=1.0]{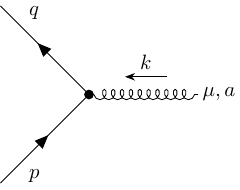}  \hspace{1.0cm}
\includegraphics[scale=1.0]{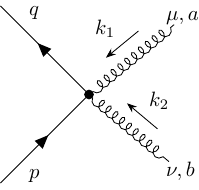} \hspace{1.5cm}
\includegraphics[scale=1.0]{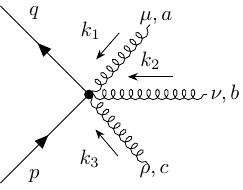}
\caption{Momentum assignments for the vertices with one, two and three gluons.}
\label{fig:vertices}
\end{figure}

Given the Feynman rules up to the order $g_0^3$ for a lattice fermion action in the form
\begin{align}
V^a_{1\mu}(p,q)&=T^aV_{1\mu}(p,q)
\label{eq:15}
\\
V^{ab}_{2\mu\nu}(p,q,k_1,k_2)&=T^aT^bV_{2\mu\nu}(p,q,k_1,k_2)
\\
V^{abc}_{3\mu\nu\rho}(p,q,k_1,k_2,k_3)&=T^aT^bT^cV_{3\mu\nu\rho}(p,q,k_1,k_2,k_3)
\label{eq:17}
\end{align}
with the momentum assignments as depicted in Figure~\ref{fig:vertices}, the vertices for the variety with stout smearing ($n$ steps with parameter $\varrho$) are obtained through
\begin{align}
V^{(n)a}_{1\mu}(p,q)&=T^a\sum\limits_\nu \tilde{g}^n_{\mu\nu}(\varrho,q-p)V_{1\nu}(p,q)
\label{eq:V1_stout}
\\%%%%%%%%%%%%%
V^{(n)ab}_{2\mu\nu}(p,q,k_1,k_2)&=
T^aT^b\sum\limits_{\rho\sigma}
\tilde{g}^n_{\mu\rho}(\varrho,k_1) \tilde{g}^n_{\nu\sigma}(\varrho,k_2) V_{2\rho\sigma}(p,q,k_1,k_2)
\nonumber\\&
+\frac{g_0}{2i}[T^a,T^b]\sum\limits_\rho \tilde{g}^{(n)}_{\rho\mu\nu}(k_1,k_2)V_{1\rho}(p,q)
\label{eq:V2_stout}
\\%%%%%%%%%%%%
V^{(n)abc}_{3\mu\nu\rho}(p,q,k_1,k_2,k_3)&=T^aT^bT^c\sum\limits_{\alpha\beta\gamma}
\tilde{g}^n_{\mu\alpha}(\varrho,k_1) \tilde{g}^n_{\nu\beta}(\varrho,k_2)  \tilde{g}^n_{\rho\gamma}(\varrho,k_3)V_{3\alpha\beta\gamma}(p,q,k_1,k_2,k_3)
\nonumber\\
-\frac{g_0}{2i}\{T^a,[T^b,T^c]\}&\sum\limits_{\alpha\beta} \tilde{g}^n_{\mu\alpha}(\varrho,k_1)\tilde{g}^{(n)}_{\beta\nu\rho}(\varrho,k_2,k_3)
V_{2\alpha\beta}(p,q,k_1,k_2+k_3)
\nonumber\\
-\frac{g_0^2}{6}
\bigg[T^aT^bT^c+T^c&T^bT^a-\frac{1}{N_c}\Tr\big(T^aT^bT^c+T^cT^bT^a\big)\bigg] \sum\limits_\alpha \tilde{g}^{(n)}_{\alpha\mu\nu\rho}(\varrho,k_1,k_2,k_3)V_{1\alpha}(p,q)
\label{eq:V3_stout}
\end{align}
where the functions $\tilde{g}_{\mu\nu}$, $\tilde{g}_{\mu\nu\rho}^{(n)}$, $\tilde{g}_{\mu\nu\rho\sigma}^{(n)}$ are given in Appendix~\ref{app:form_functions}.
Similarly, in case of Wilson flow (with $t$ denoting the flow time in lattice units) the vertices take the form
\begin{align}
V^{a}_{1\mu}(p,q,t)&=T^a\sum\limits_\nu B_{\mu\nu}(q-p,t) V_{1\nu}(p,q)
\label{eq:V1_flow}
\\%%%%%%%%%%%%%
V^{ab}_{2\mu\nu}(p,q,k_1,k_2,t)&=
T^aT^b\sum\limits_{\rho\sigma}
B_{\mu\rho}(k_1,t)B_{\nu\sigma}(k_2,t) V_{2\rho\sigma}(p,q,k_1,k_2)
\nonumber\\&
+\frac{g_0}{2i}[T^a,T^b]\sum\limits_\rho B_{\rho\mu\nu}(k_1,k_2,t)V_{1\rho}(p,q)
\label{eq:V2_flow}
\\%%%%%%%%%%%%
V^{abc}_{3\mu\nu\rho}(p,q,k_1,k_2,k_3,t)&=T^aT^bT^c\sum\limits_{\alpha\beta\gamma}
B_{\mu\alpha}(k_1,t) B_{\nu\beta}(k_2,t)  B_{\rho\gamma}(k_3,t)V_{3\alpha\beta\gamma}(p,q,k_1,k_2,k_3)
\nonumber\\
-\frac{g_0}{2i}\{T^a,[T^b,T^c]\}&\sum\limits_{\alpha\beta} B_{\mu\alpha}(k_1,t)B_{\beta\nu\rho}(k_2,k_3,t)
V_{2\alpha\beta}(p,q,k_1,k_2+k_3)
\nonumber\\
-\frac{g_0^2}{6}
\bigg[T^aT^bT^c+T^c&T^bT^a-\frac{1}{N_c}\Tr\big(T^aT^bT^c+T^cT^bT^a\big)\bigg] \sum\limits_\alpha B_{\alpha\mu\nu\rho}(k_1,k_2,k_3,t)V_{1\alpha}(p,q)
\label{eq:V3_flow}
\end{align}
where the functions $B_{\mu\nu}(k_1,t)$, $B_{\mu\nu\rho}(k_1,k_2,t)$, $B_{\mu\nu\rho\sigma}(k_1,k_2,k_3,t)$ are given in Appendix~\ref{app:form_functions}.

This defines the Feynman rules needed for the computation of the 1-loop improvement coefficient $\csw^{(1)}$ for both Wilson and Brillouin fermions.
Let us add that the unsmoothed Feynman rules (\ref{eq:15}-\ref{eq:17}) have a part which is specific to either the Wilson or the Brillouin action, and a shared clover contribution
\begin{align}
V_{1\mu}(p,q)&=V_{1(\mathrm{W/B})\mu}(p,q)+V_{1(\mathrm{c})\mu}(p,q)
\\
V_{2\mu\nu}(p,q,k_1,k_2)&=V_{2(\mathrm{W/B})\mu\nu}(p,q)+V_{2(\mathrm{c})\mu\nu}(p,q,k_1,k_2)
\\
V_{3\mu\nu\rho}(p,q,k_1,k_2,k_3)&=V_{3(\mathrm{W/B})\mu\nu\rho}(p,q)+V_{3(\mathrm{c})\mu\nu\rho}(p,q,k_1,k_2,k_3)
\;.
\end{align}
The unsmoothed Wilson and clover vertices are fairly short and listed in Appendix~\ref{app:feynman_rules}, while the unsmoothed Brillouin vertices are much longer and given in Ref.~\cite{Ammer:2023otl}.
Putting everything together this is a practical recipe how the Feynman rules for an unsmoothed (``thin-link'') fermion action may be upgraded to include the effect of some link smoothing \cite{Ammer:2024hqr}.

%%%%%%%%%%%%%%%%%%%%%%%%%%%%%
\section{Strategy for the determination of $\csw^{(1)}$\label{sec:csw}}
%%%%%%%%%%%%%%%%%%%%%%%%%%%%%

\begin{figure}[!tb]
\centering
\begin{tabular}{ccc}
 (a) & (b) & (c) \\
\includegraphics[scale=1.0]{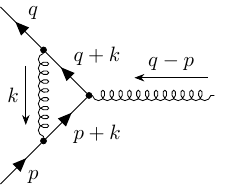} &
\includegraphics[scale=1.0]{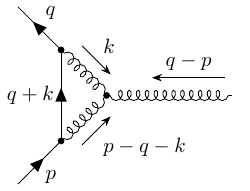} &
\includegraphics[scale=1.0]{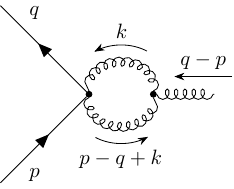} \\
 (d) & (e) & (f) \\
\includegraphics[scale=1.0]{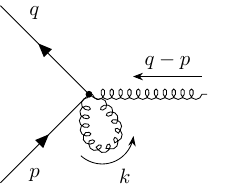} &
\includegraphics[scale=1.0]{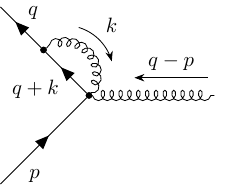} &
\includegraphics[scale=1.0]{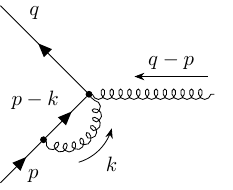}
\end{tabular}
\caption{The six one-loop diagrams contributing to the vertex function.}
\label{fig:diagrams}
\end{figure}

The Sheikholeslami-Wohlert coefficient has a perturbative expansion
\begin{align}
\csw=\csw^{(0)}+g_0^2\csw^{(1)}+\mathcal{O}(g_0^4)
\label{expansion_csw}
\end{align}
in powers of the bare coupling $g_0^2=2N_c/\beta$.
It can be calculated order by order from the perturbative expansion of the quark-quark-gluon-vertex function
\begin{align}
\Lambda^a_\mu(p,q)=\sum\limits_{L=0}^\infty g_0^{2L+1} \Lambda^{a(L)}_\mu(p,q)
\;.
\label{def_Lambda}
\end{align}
The tree-level value is $\csw^{(0)}=r$ for both Wilson and Brillouin fermion actions (see Ref.~\cite{Ammer:2023otl}).

Our method for calculating $\csw^{(1)}$ is detailed in Refs.~\cite{Aoki:2003sj,Ammer:2023otl}.
In short we use the formula
\begin{align}
g_0^3T^aG_1=-\frac{1}{8}\text{Tr}
\bigg[\bigg(\frac{\partial}{\partial p_\mu}+\frac{\partial}{\partial q_\mu}\bigg)\Lambda^{a(1)}_\mu - \bigg(\frac{\partial}{\partial p_\nu}-\frac{\partial}{\partial q_\nu}\bigg)\Lambda^{a(1)}_\mu\gamma_\nu\gamma_\mu \bigg]^{\mu\neq\nu}_{p,q\rightarrow 0}
\label{eq:G1}
\end{align}
and finish the calculation with $\csw^{(1)}=2G_1$.

In the following we present our results for $\csw^{(1)}$ with stout smearing or Wilson flow and four combinations of fermion (Wilson or Brillouin) and gluon (plaquette or L\"uscher-Weisz) action.

Even though most of the individual one-loop diagrams are divergent, the sum is always finite (see Refs.~\cite{Aoki:2003sj,Ammer:2023otl} for details).
Moreover, the final result can be split into two parts (proportional to $N_c$ and $1/N_c$, or equivalently proportional to $N_c$ and $C_F$).
Hence we give results for $\csw^{(1)}$ with $N_c=3$, but also the coefficients of the parts $\propto N_c$ and $\propto 1/N_c$, to allow the reader to reconstruct the values for arbitrary $N_c$.

There is an interesting feature of the Wilson and Brillouin fermion actions which facilitates the computation of $\csw^{(1)}$.
The (finite) tadpole diagram (d) in Figure~\ref{fig:diagrams} is the only place where the $\bar{q}qggg$-vertex is needed.
This vertex is of order $g_0^3$ and, in turn, the only place where the third order of the perturbative expansion of stout smearing or Wilson flow appears.
Due to a cancellation between the Wilson/Brillouin part and the clover part in the vertex, the expression to which the third order form factor couples, vanishes.
Thus the stout smearing or Wilson flow of order $g_0^3$, i.e.\ the form factors $\tilde{g}^{(n)}_{\mu\nu\rho\sigma}(\varrho,k_1,k_2,k_3)$ from Equations~(\ref{eq:V3_stout}) and (\ref{eq:g_tilde_NNLO})
or $B_{\mu\nu\rho\sigma}(k_1,k_2,k_3,t)$ from Equations~(\ref{eq:V3_flow}) and (\ref{eq:B_NNLO}) is not needed to compute $\csw^{(1)}$ for the two fermion actions under consideration (see Appendix~\ref{app:proof} for a proof).
This is true as long as the same smearing/flowing is applied to the covariant derivative, the Laplace operator and the clover term of the action.
For the determination of $\csw^{(1)}$ with SLiNC fermions in Ref.~\cite{Horsley:2008ap}, for example, this is not the case.

%%%%%%%%%%%%%%%%%%%%%%%%%%%%%
\section{Results for $\mathbf{c_{SW}^{(1)}}$ with stout smearing\label{sec:csw_stout}}
%%%%%%%%%%%%%%%%%%%%%%%%%%%%%

We are now in a position to give results for $\csw^{(1)}$ with up to four steps of stout smearing for the Wilson fermion action on plaquette (plaq) or L\"uscher-Weisz (sym) gluon background.
Also for the Brillouin fermion action with plaquette gluons results for $\csw^{(1)}$ with up to four steps of stout smearing will be given.
The combination of Brillouin fermions and Symanzik improved gluons tends to produce large analytic expressions, which become very time and memory intensive to integrate.
For this reason we give only results with up to three stout steps for this particular combination of actions.

\graphicspath{{./figures/}}
\begin{figure}[!htb]
\centering
\includegraphics[scale=0.48]{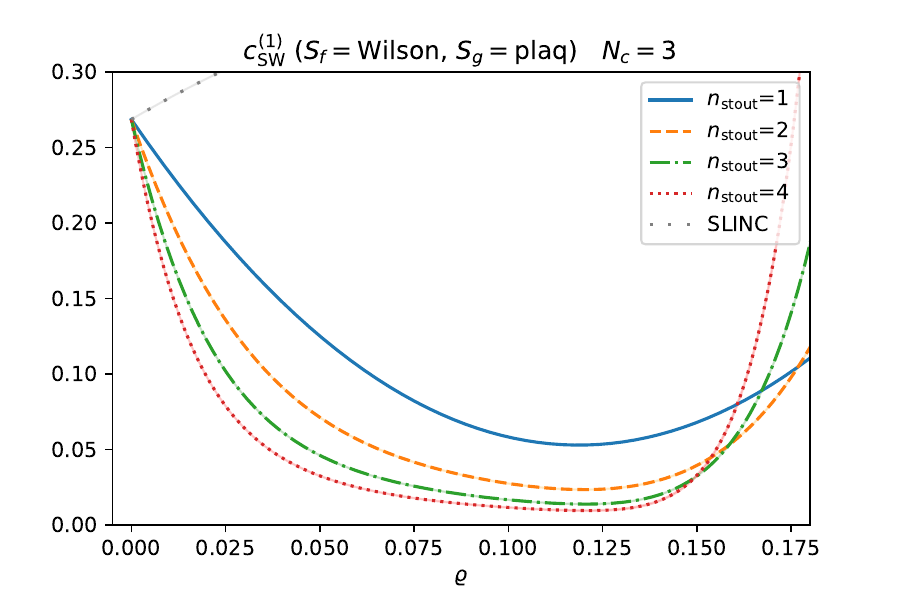}
\includegraphics[scale=0.48]{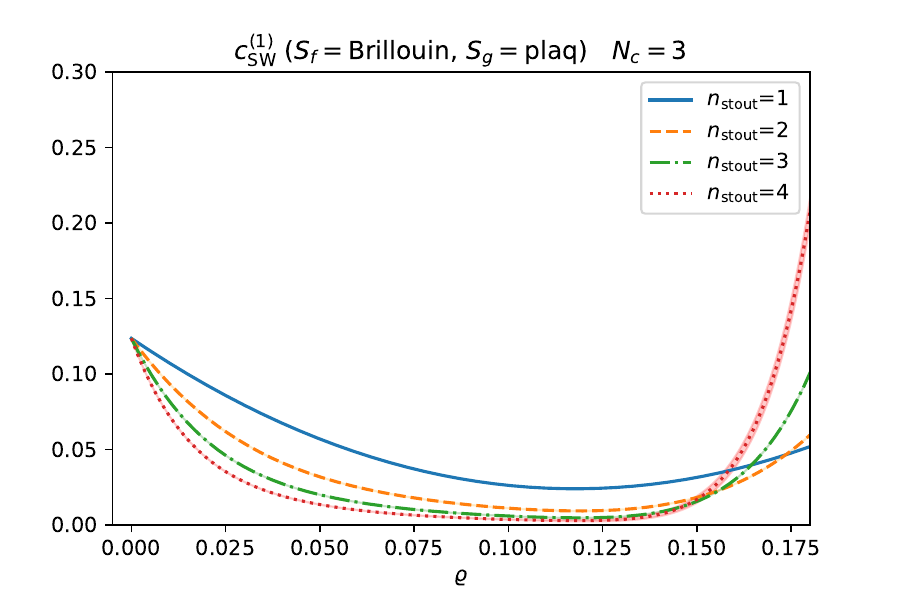}\\
\includegraphics[scale=0.48]{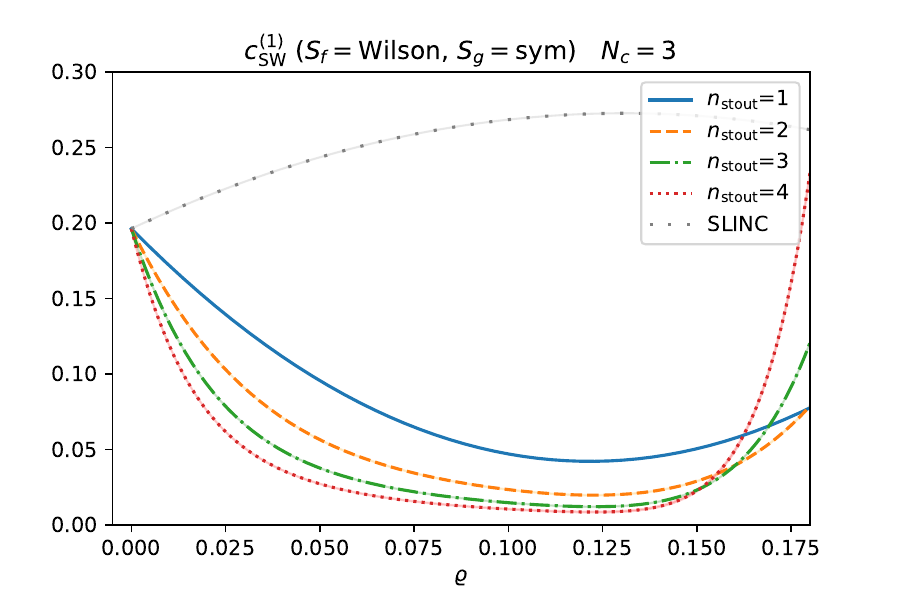}
\includegraphics[scale=0.48]{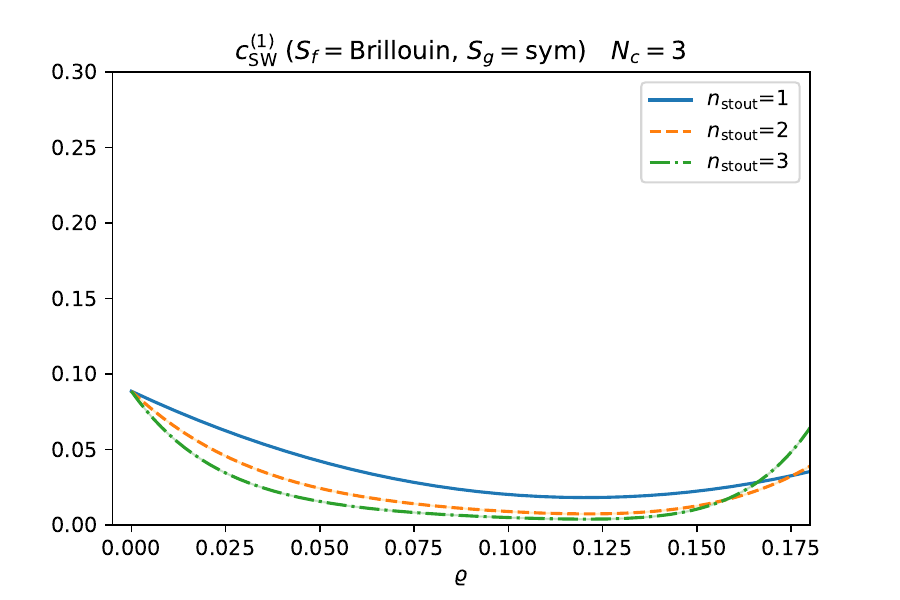}
\caption{The one-loop improvement coefficient $\csw^{(1)}$ with $N_c=3$ and up to four steps of stout smearing as a function of the smearing parameter $\varrho$.
The four panels reflect the four combinations of the Wilson or Brillouin fermion (left vs.\ right) with plaquette or L\"uscher-Weisz glue (top vs.\ bottom).
\label{fig:csw_stout}}
\end{figure}

In Figure~\ref{fig:csw_stout} the coefficient $\csw^{(1)}$ is plotted against the smearing parameter $\varrho$ for $N_c=3$.
In each panel the three or four curves shown start from a joint value which agrees with the unsmeared $\csw^{(1)}$ for the respective action combination as presented in Ref.~\cite{Ammer:2023otl},
and the slope at $\varrho=0$ is proportional to $n_\mathrm{stout}$.

\begin{table}
\centering
\begin{tabular}{|c|c|c|c|c|c|}
\hline
& & $n=1$ & $n=2$ & $n=3$ & $n=4$ \\
 \hline
 \multirow{2}{*}{Wil./plaq} &
 $\varrho_\mathrm{min}$ &
0.1186789(3) &
0.12009(3) &
0.1205(3) &
0.1207(3)\\
\cline{2-6}
& ${\csw^{(1)}}_\mathrm{min}$  &
0.052915(2)&
 0.0234(1)&
0.0138(1)&
0.0095(2)\\
\hline
\multirow{2}{*}{Wil./sym} &
 $\varrho_\mathrm{min}$ &
 0.12166(2)&
0.12268(4)&
0.1228(1)&
0.123(1)\\
\cline{2-6}
&  ${\csw^{(1)}}_\mathrm{min}$  &
0.0422(1)&
 0.0197(1)&
0.0121(1)&
 0.0086(3)\\
\hline
\multirow{2}{*}{Bri./plaq} &
 $\varrho_\mathrm{min}$&
0.117679(1) &
0.1188(1)&
0.1186(1)&
0.118(3)\\
\cline{2-6}
 &${\csw^{(1)}}_\mathrm{min}$  &
0.02399(2)&
0.0093(1)&
0.0047(1)&
0.003(1)\\
\hline
\multirow{2}{*}{Bri./sym} &
 $\varrho_\mathrm{min}$  &
0.12046(2)&
0.121(1)& & \\
\cline{2-6}
& ${\csw^{(1)}}_\mathrm{min}$  &
0.0182(1)&
0.0073(1)&&\\
\hline
\end{tabular}
\caption{Position and values of the minima of the curves displayed in Figure~\ref{fig:csw_stout}.
\label{tab:csw_stout_minima}}
\end{table}

Each curve in Figure~\ref{fig:csw_stout} reaches a minimum near $\varrho\simeq0.12$ (see Table~\ref{tab:csw_stout_minima} for details) before rising again for values larger than $0.125$.
If $n_\mathrm{stout}$ is increased, the curves approach zero faster and start to form a plateau closer to the $x$-axis.
As we have observed in Ref.~\cite{Ammer:2023otl} for the unsmeared values of $\csw^{(1)}$, switching to the Brillouin action approximately halves the value.
This continues to be true for $\varrho>0$ and the improved gluon action reduces the value further by another 30\% or so.

\begin{table}[!htb]
\centering
\begin{tabular}{|c|c|c|c|}
\hline
& $C_F\cdot\varrho^0$ &  $C_F\cdot\varrho^1$ & $C_F\cdot\varrho^2$\\
\hline
This work (plaq) & 0.167635  & 1.079148 & -3.68668  \\
\hline
Ref.~\cite{Horsley:2008ap} (plaq) & 0.167635 & 1.079148 & -3.697285 \\
\hline
& $N_c\cdot\varrho^0$ &  $N_c\cdot\varrho^1$ & $N_c\cdot\varrho^2$\\
\hline
This work (plaq) & 0.015025 &  0.009618  & -0.284785 \\
\hline
Ref.~\cite{Horsley:2008ap} (plaq) & 0.015025 & 0.009617 & -0.284786 \\
\hline
\hline
& $C_F\cdot\varrho^0$ &  $C_F\cdot\varrho^1$ & $C_F\cdot\varrho^2$\\
\hline
This work (sym) &  0.116185  &  0.857670 & -2.85147 \\
\hline
Ref.~\cite{Horsley:2008ap} (sym) & 0.116185 & 0.828129 & -2.455080 \\
\hline
& $N_c\cdot\varrho^0$ &  $N_c\cdot\varrho^1$ & $N_c\cdot\varrho^2$\\
\hline
This work (sym) & 0.013777 &  0.008520   & -0.2228016 \\
\hline
Ref.~\cite{Horsley:2008ap} (sym) & 0.013777 & 0.015905 & -0.321899 \\
\hline
\end{tabular}
\caption{Comparison of our result for SLiNC fermions (one step of stout smearing in the Wilson action but not in the clover term) to those of Horsley et al.~\cite{Horsley:2008ap}.
Coefficients of $C_F=(N_c^2-1)/(2N_c)$ and $N_c$ in combination with three powers of the smearing parameter $\varrho$ (called $\omega$ in \cite{Horsley:2008ap}) are shown.
Uncertainties are of the order of the last digit shown.
\label{tab:csw_slinc}}
\end{table}

In the panels for the Wilson action we have also included SLiNC results, where only the kinetic part of the action undergoes one step of stout smearing but not the link variables in the clover term.
Our numbers in this case show some discrepancies to those given by \emph{Horsley et al}.\ in Ref.~\cite{Horsley:2008ap}, which we were unable to resolve.
In Table~\ref{tab:csw_slinc} we show our results next to those of Ref.~\cite{Horsley:2008ap} in the representation used there, i.e.\ coefficients of $C_F$ and $N_c$.
In the plaquette case there is a difference in the coefficient of $C_F\varrho^2$, while in the Symanzik case there are differences for both coefficients of $\varrho$ and $\varrho^2$.
The absolute differences in the coefficients of $C_F$ are four%
\footnote{Explicitly $\frac{0.8576704-0.828129}{0.015905-0.00852}\approx 4.0002$ and $\frac{2.85147-2.455080}{0.321899-0.22280159}\approx 4.000004$.}
times the absolute differences in the coefficients of $N_c$.
This suggests that they might stem from the same place, which is most likely the (d) or (e) and (f) diagrams, as their color factors are the only ones that include both $C_F$ and $N_c$.
We were however unable to reproduce the numbers of Ref.~\cite{Horsley:2008ap}, even when using (as far as possible) the exact expressions for the Feynman rules given in the appendix of Ref.~\cite{Horsley:2008ap} and the methods described there.
Hence, we are unable to settle this issue definitively, but have performed many error searches and cross-checks in our Mathematica code to be reasonably confident in our methods and results.

\begin{figure}[!htb]
\centering
\includegraphics[scale=0.48]{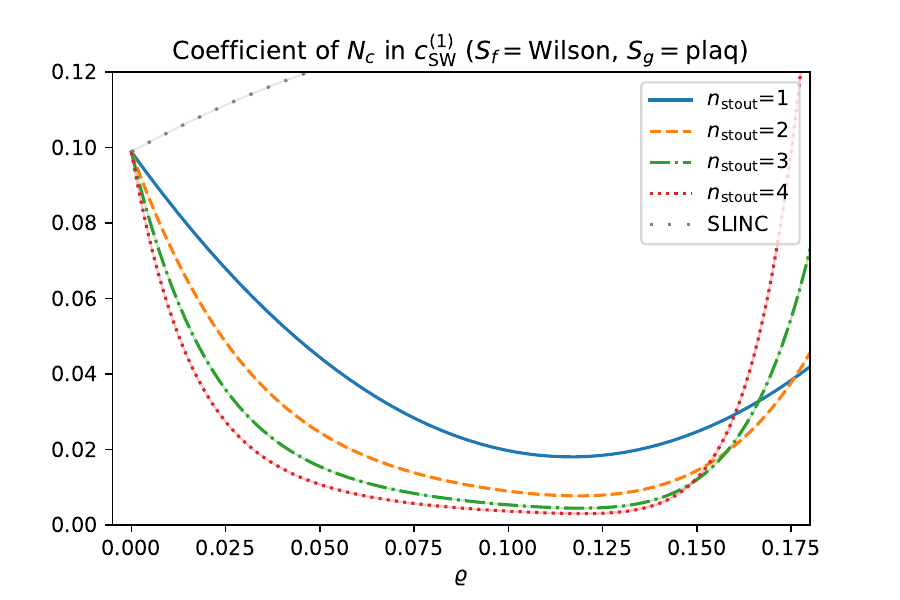}
\includegraphics[scale=0.48]{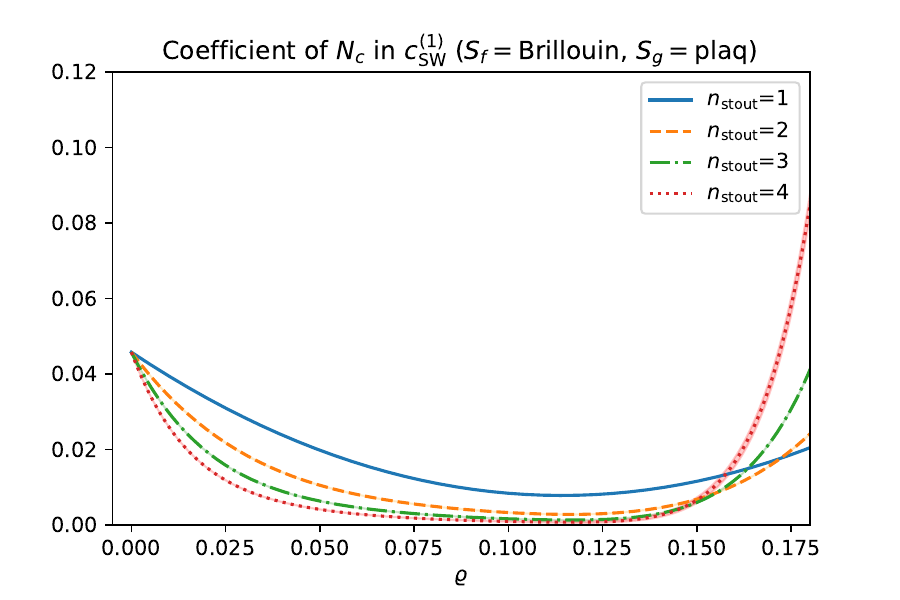}\\
\includegraphics[scale=0.48]{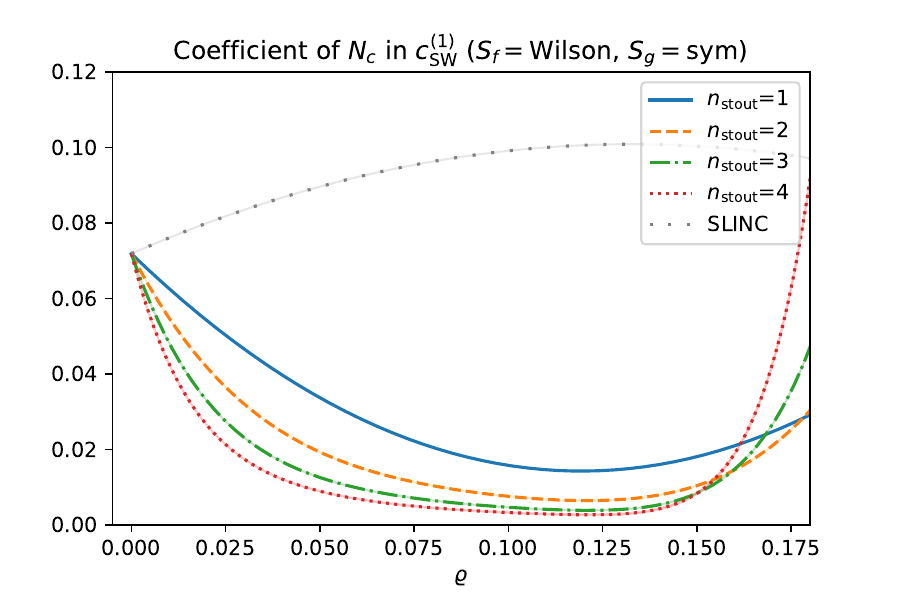}
\includegraphics[scale=0.48]{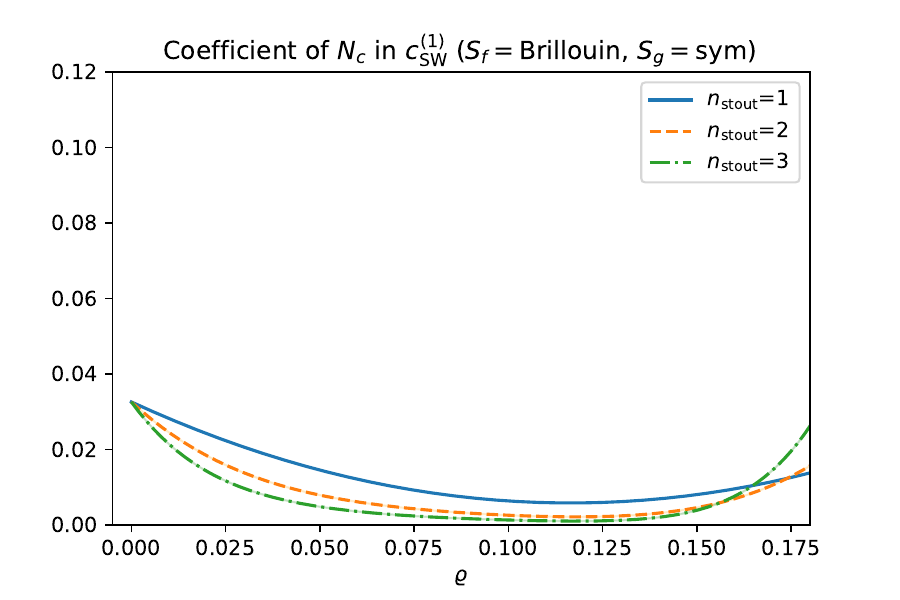}
\caption{Same as in Figure~\ref{fig:csw_stout} but for the $\csw^{(1)}$ contribution linear in $N_c$.
\label{fig:csw_nc_stout}}
\end{figure}

\begin{figure}[!htb]
\centering
\includegraphics[scale=0.48]{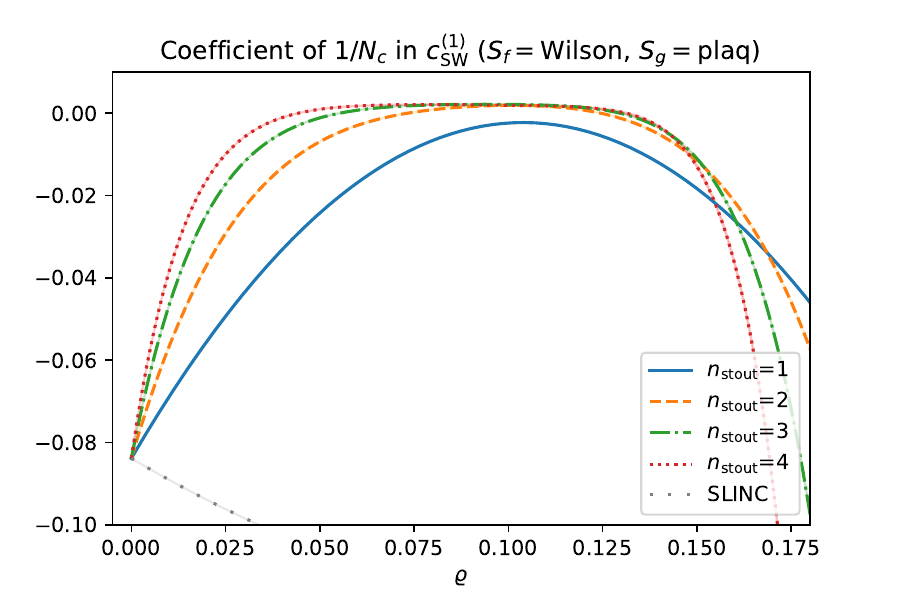}
\includegraphics[scale=0.48]{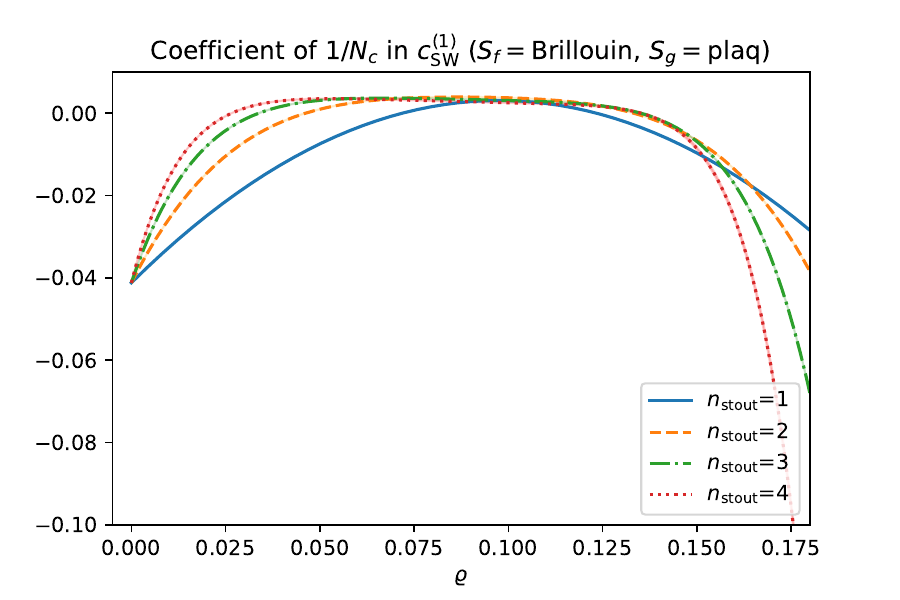}\\
\includegraphics[scale=0.48]{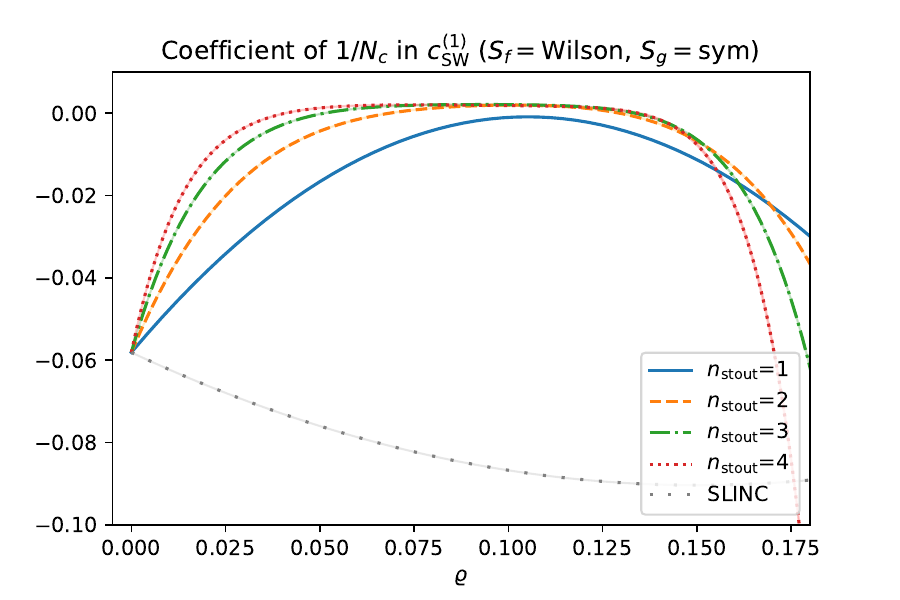}
\includegraphics[scale=0.48]{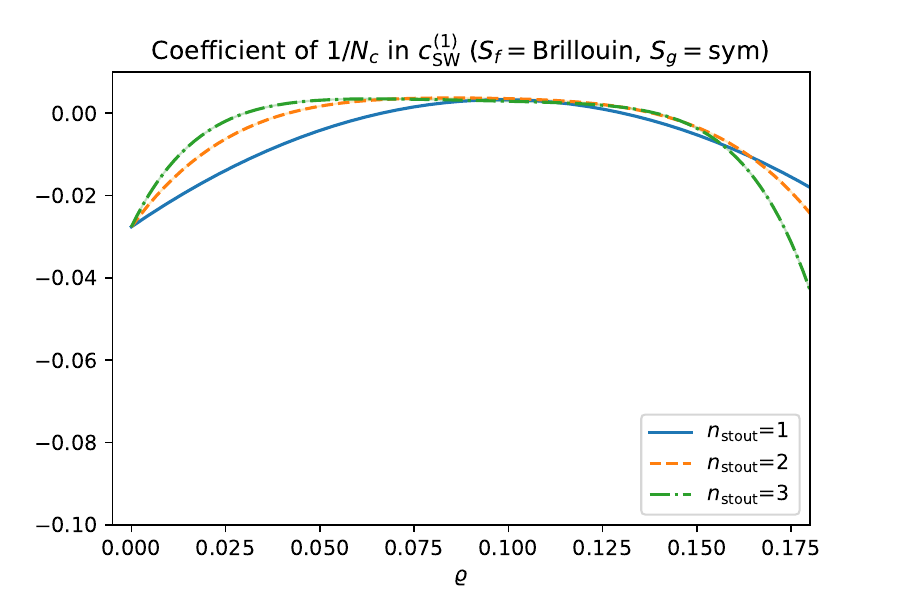}
\caption{Same as in Figure~\ref{fig:csw_stout} but for the $\csw^{(1)}$ contribution linear in $1/N_c$.
\label{fig:csw_ncinv_stout}}
\end{figure}

It is interesting to dissect the physical result, as shown in Figure~\ref{fig:csw_stout}, into the portions $\propto N_c$ and $\propto 1/N_c$.
The result is shown in Figure~\ref{fig:csw_nc_stout} and Figure~\ref{fig:csw_ncinv_stout}, respectively.
In every panel each curve has its minimum near $\varrho\simeq0.12$; this indicates that the near-vanishing of $\csw^{(1)}$ is a generic feature which holds for all $N_c$.
In other words, one-loop perturbation theory suggests that for both Wilson and Brillouin fermions coupled to stout-smeared plaquette or Symanzik glue, $\csw$ is near its tree-level value.

\begin{table}[!htb]
\centering
\begin{tabular}{|c|c|c|c|c|c|}
\hline
 & $\varrho$ & $n=1$ & $n=2$ & $n=3$ & $n=4$ \\
\hline
\multirow{6}{*}{$\csw^{(1)}$ (Wil./plaq))}
 & 0.05 & 0.1251411(9) & 0.07125(5) & 0.04611(4) & 0.03251(5)\\
 & 0.09 & 0.065509(2) & 0.03191(7) & 0.01953(5) & 0.0136(1)\\
 & 0.11 & 0.054068(2) & 0.02455(8) & 0.01470(6) & 0.0102(2)\\
 & 0.12 & 0.052941(2) & 0.02343(9) & 0.01382(7) & 0.0095(2)\\
 & 0.125 & 0.053526(2) & 0.02374(9) & 0.01404(7) & 0.0097(2)\\
 & 0.13 & 0.054877(2) & 0.02478(9) & 0.01495(7) & 0.0105(3)\\
\hline
\multirow{6}{*}{$\csw^{(1)}$ (Wil./sym))}
 & 0.05 & 0.09562(4) & 0.05649(5) & 0.03768(4) & 0.02724(8)\\
 & 0.09 & 0.05259(5) & 0.02693(7) & 0.01706(6) & 0.0121(2)\\
 & 0.11 & 0.04357(5) & 0.02098(9) & 0.01307(7) & 0.0093(2)\\
 & 0.12 & 0.04219(6) & 0.01979(9) & 0.01216(8) & 0.0086(3)\\
 & 0.125 & 0.04227(6) & 0.01977(9) & 0.01214(8) & 0.0086(3)\\
 & 0.13 & 0.04288(6) & 0.0202(1) & 0.01257(9) & 0.0090(3)\\
\hline
\multirow{6}{*}{$\csw^{(1)}$ (Bri./plaq))}
 & 0.05 & 0.056945(2) & 0.03189(4) & 0.02003(4) & 0.01353(6)\\
 & 0.09 & 0.029502(2) & 0.01327(5) & 0.00728(5) & 0.0045(2)\\
 & 0.11 & 0.024414(2) & 0.00974(6) & 0.00503(6) & 0.0031(4)\\
 & 0.12 & 0.024028(2) & 0.00931(6) & 0.00473(7) & 0.0029(5)\\
 & 0.125 & 0.024375(2) & 0.00956(7) & 0.00495(7) & 0.0031(7)\\
 & 0.13 & 0.025082(2) & 0.01019(7) & 0.00555(8) & 0.0037(9)\\
\hline
\multirow{6}{*}{$\csw^{(1)}$ (Bri./sym))}
 & 0.05 & 0.04226(3) & 0.02429(4) & 0.01558(4) & \\
 & 0.09 & 0.02266(4) & 0.01056(5) & 0.00596(6) & \\
 & 0.11 & 0.01869(4) & 0.00781(6) & 0.00418(7) & \\
 & 0.12 & 0.01816(4) & 0.00734(6) & 0.00386(8) & \\
 & 0.125 & 0.01826(4) & 0.00741(6) & 0.00395(8) & \\
 & 0.13 & 0.01860(5) & 0.00773(6) & 0.00427(8) & \\
\hline
\end{tabular}
\caption{Results for $\csw^{(1)}$ at $N_c=3$ with up to four stout steps and several values $\varrho$.
\label{tab:csw_stout_values}}
\end{table}

In Table~\ref{tab:csw_stout_values} the values of $\csw^{(1)}$ are listed for a number of $n_\mathrm{stout}$ and a selection of $\varrho$ values.
For a user interested in the value of $\csw^{(1)}$ for an arbitrary value of $\varrho$ at $r=1$ and $N_c=3$, we believe a spline interpolation of this information will do fine.

\begin{figure}[!htb]
\centering
\includegraphics[scale=0.48]{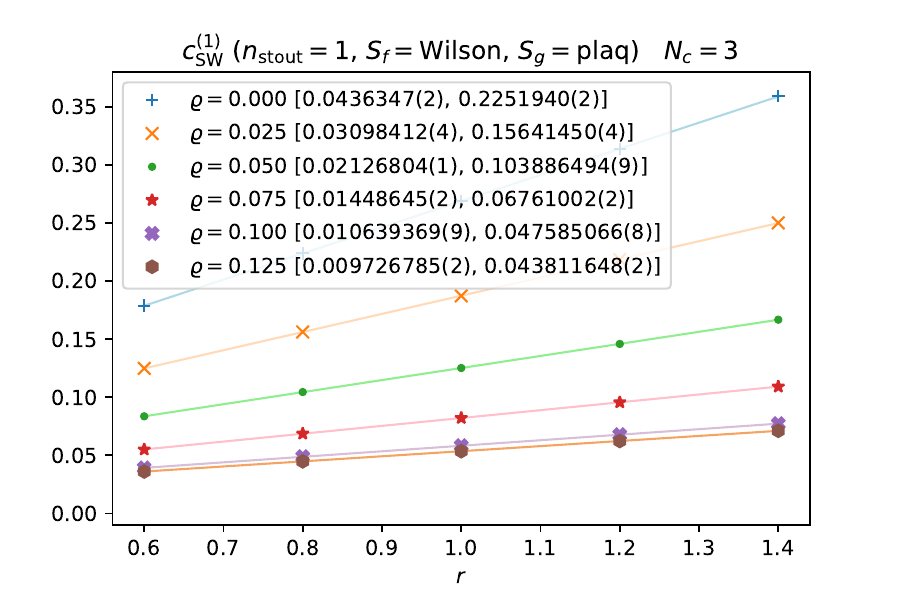}
\includegraphics[scale=0.48]{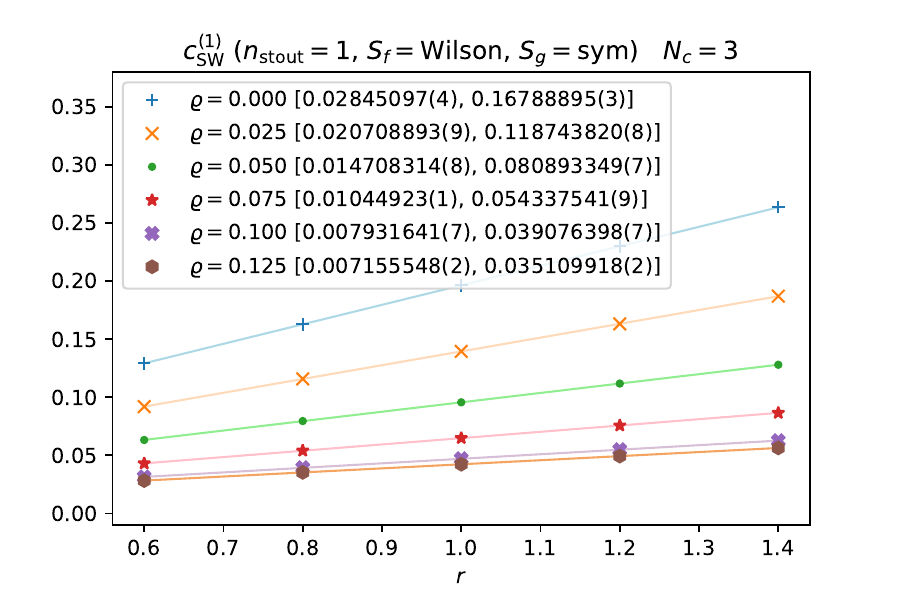}\\
\includegraphics[scale=0.48]{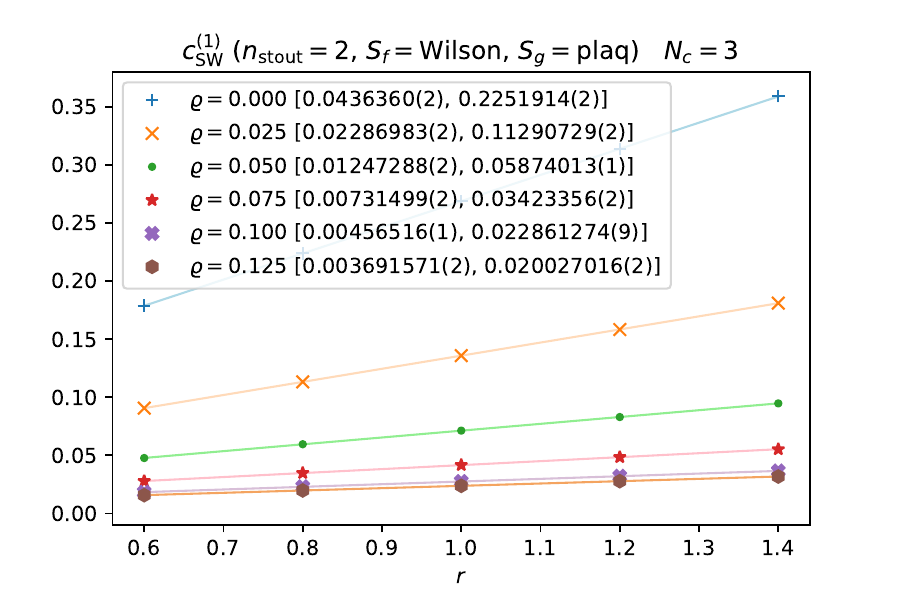}
\includegraphics[scale=0.48]{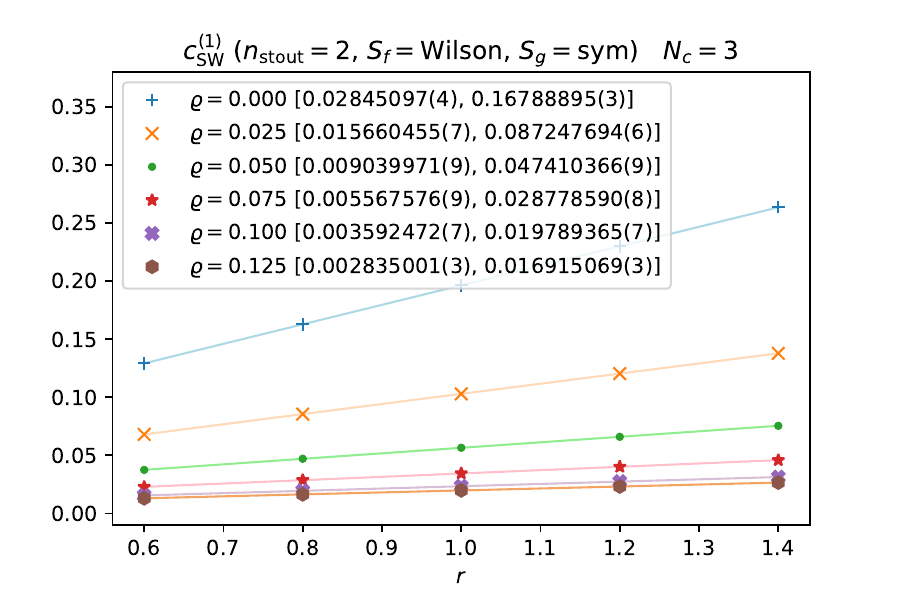}\\
\includegraphics[scale=0.48]{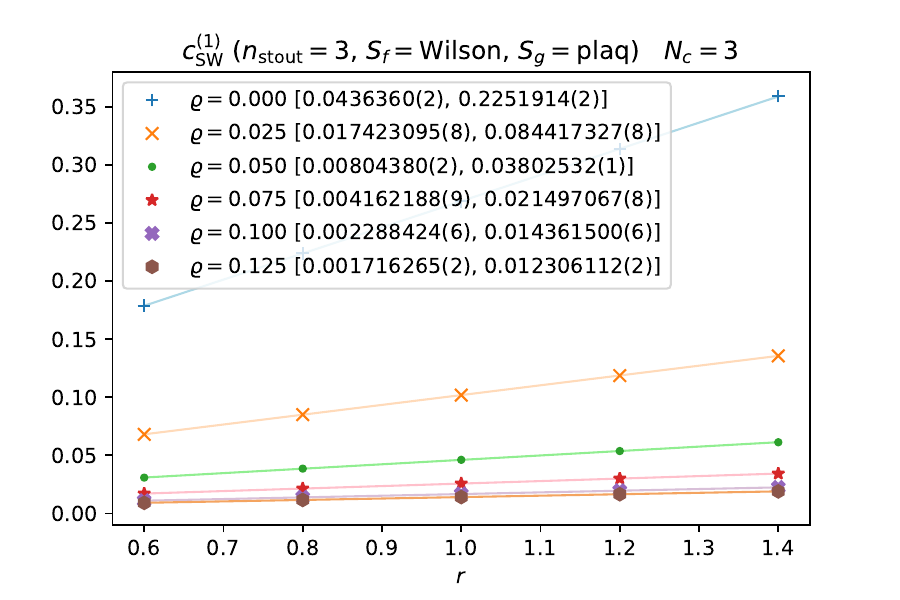}
\includegraphics[scale=0.48]{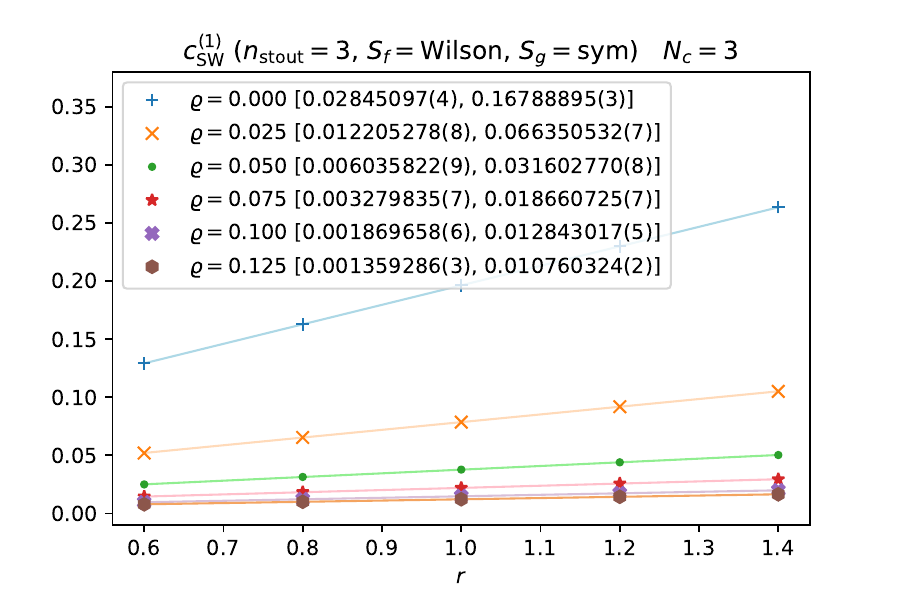}\\
\includegraphics[scale=0.48]{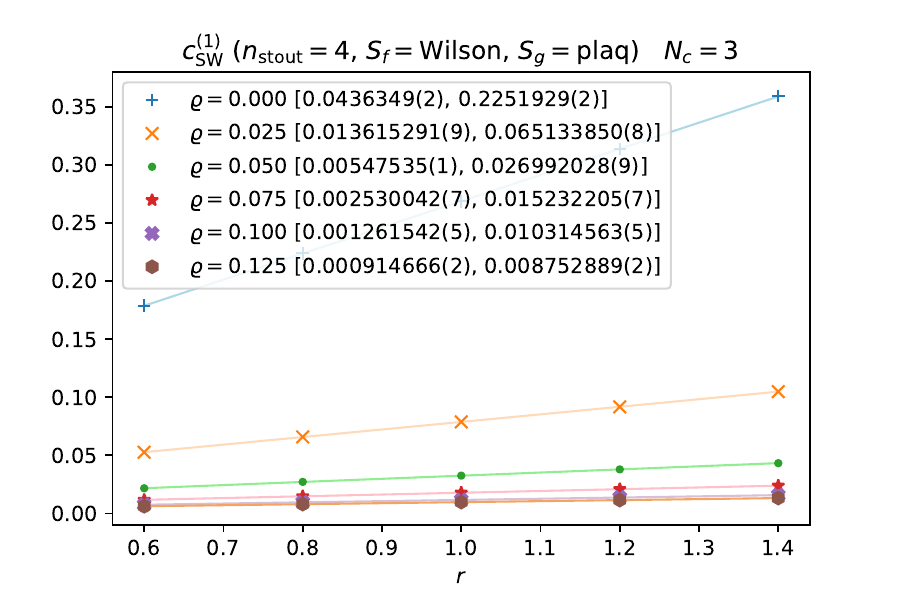}
\includegraphics[scale=0.48]{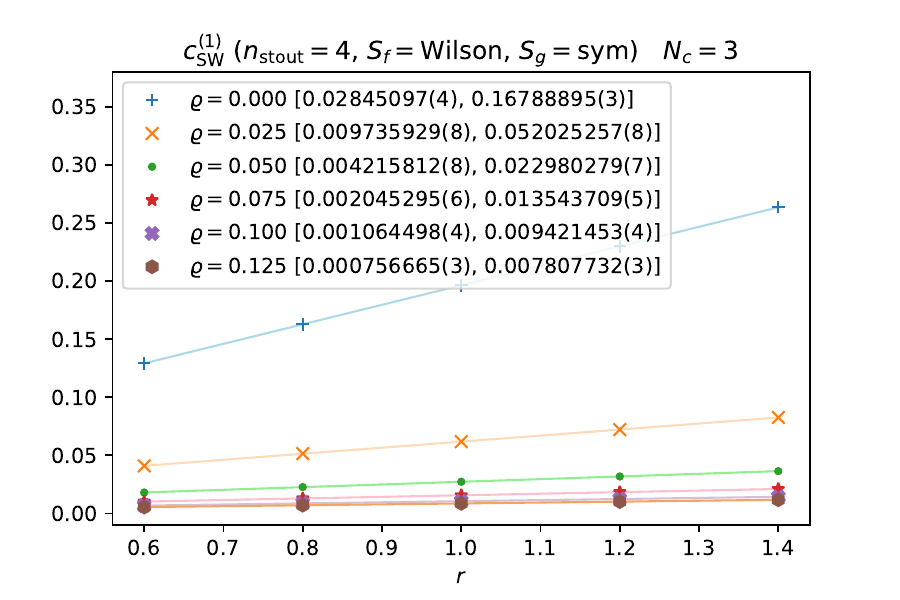}
\caption{$\csw^{(1)}$ as a function of the Wilson parameter $r$ for Wilson fermions with $N_c=3$.
The number of stout smearing steps increases from top to bottom, the left column shows results with plaquette gauge action, the right one with L\"uscher-Weisz action.
Linear least square fits of the form $c_0+c_1 \cdot r$ are also shown and the coefficients $[c_0,c_1]$ are given in brackets.
\label{fig:csw_of_r_wil_stout}}
\end{figure}

\begin{figure}[!htb]
\centering
\includegraphics[scale=0.48]{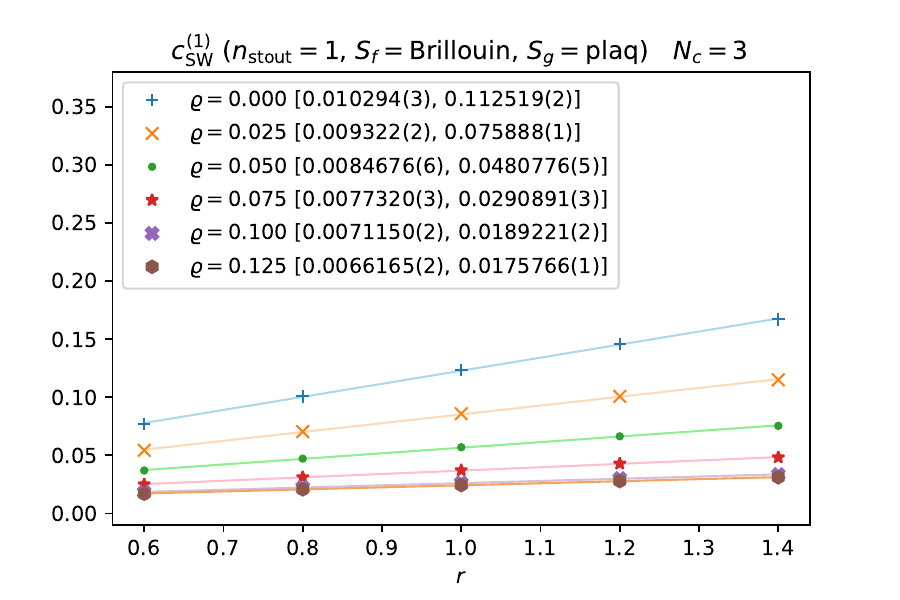}
\includegraphics[scale=0.48]{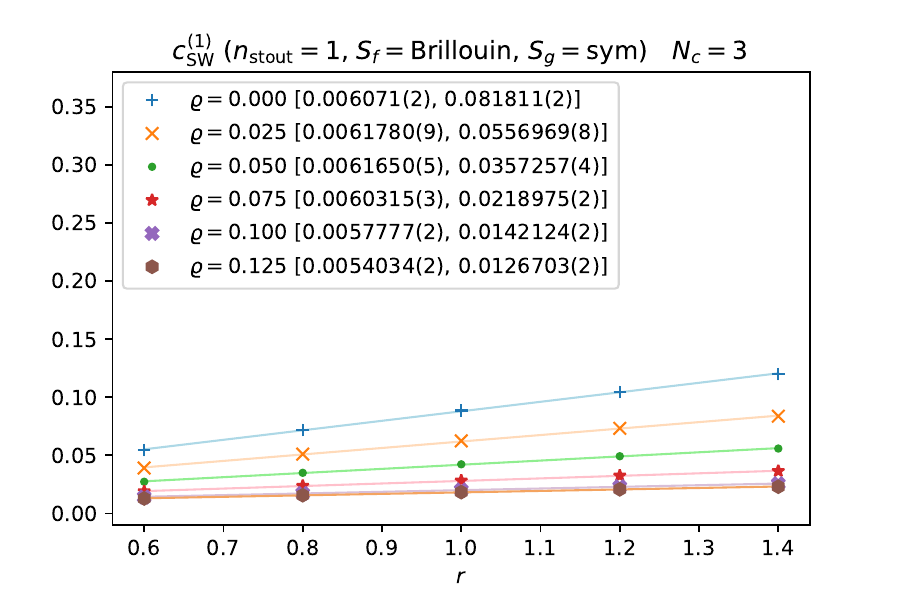}\\
\includegraphics[scale=0.48]{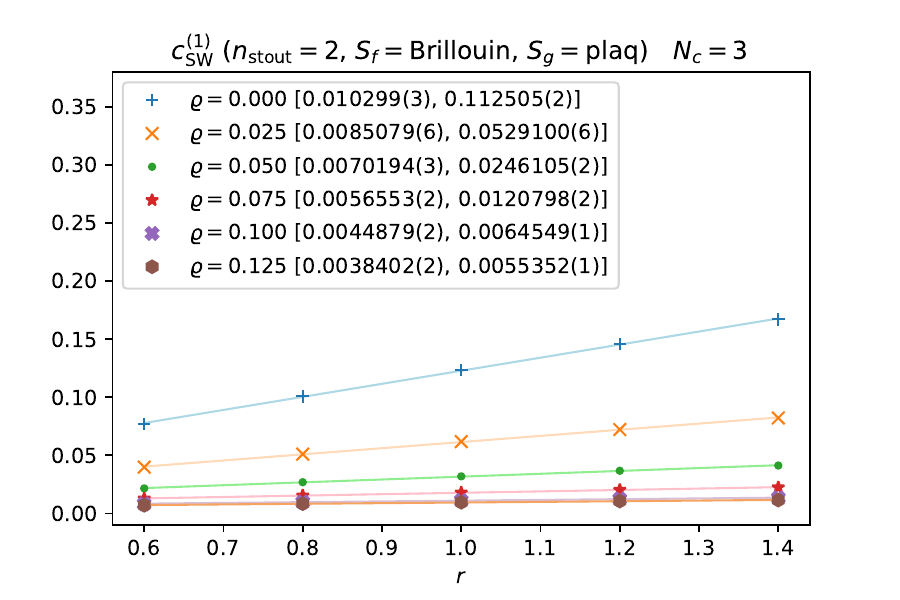}
\includegraphics[scale=0.48]{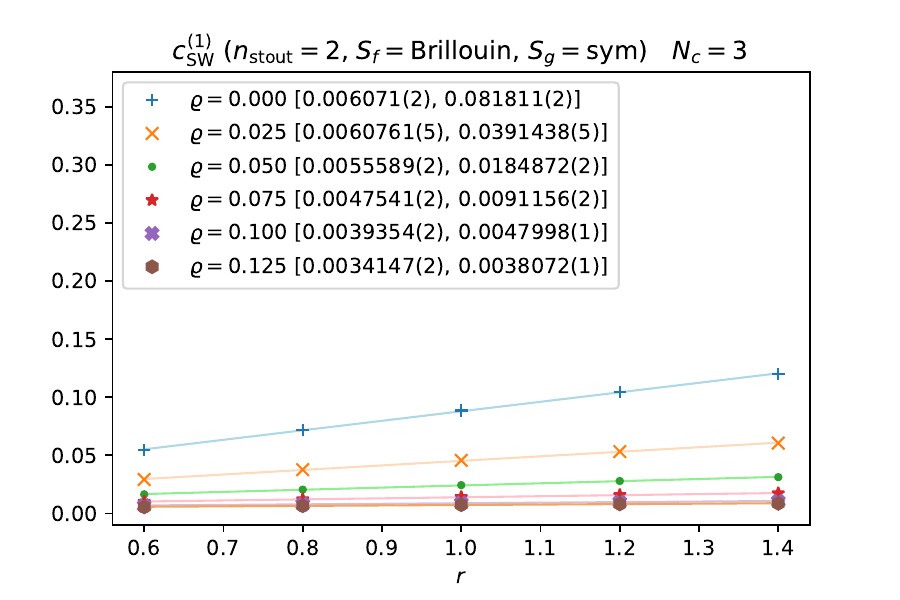}\\
\includegraphics[scale=0.48]{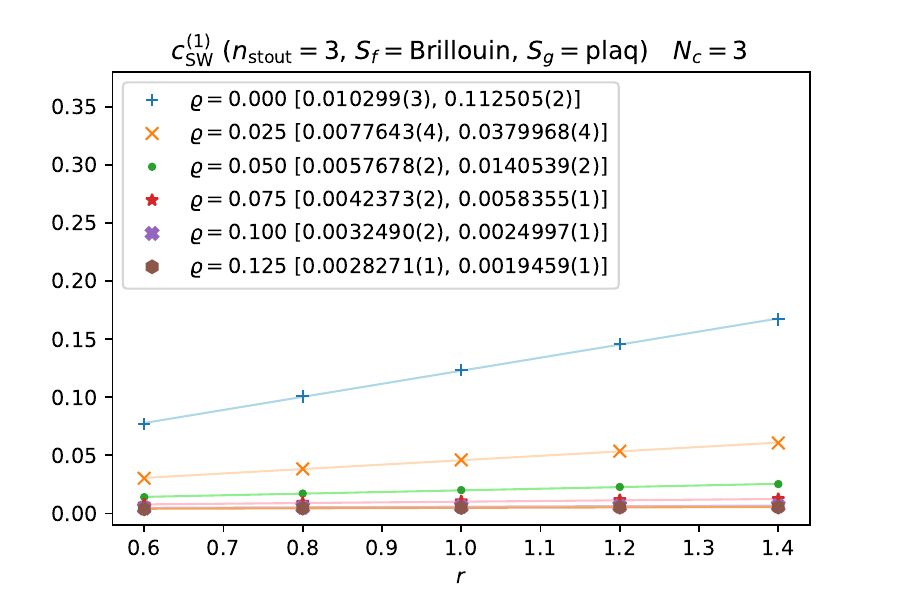}
\includegraphics[scale=0.48]{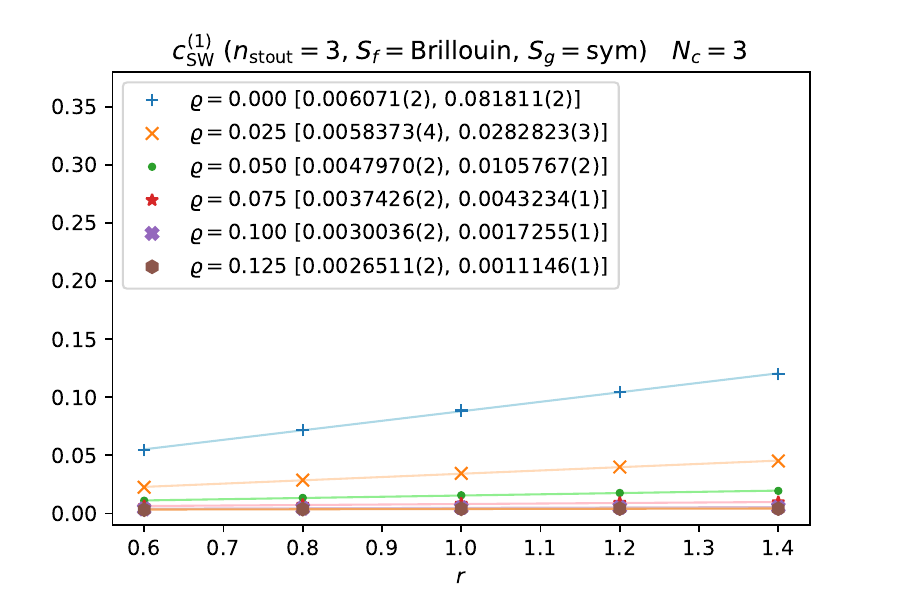}\\
\includegraphics[scale=0.48]{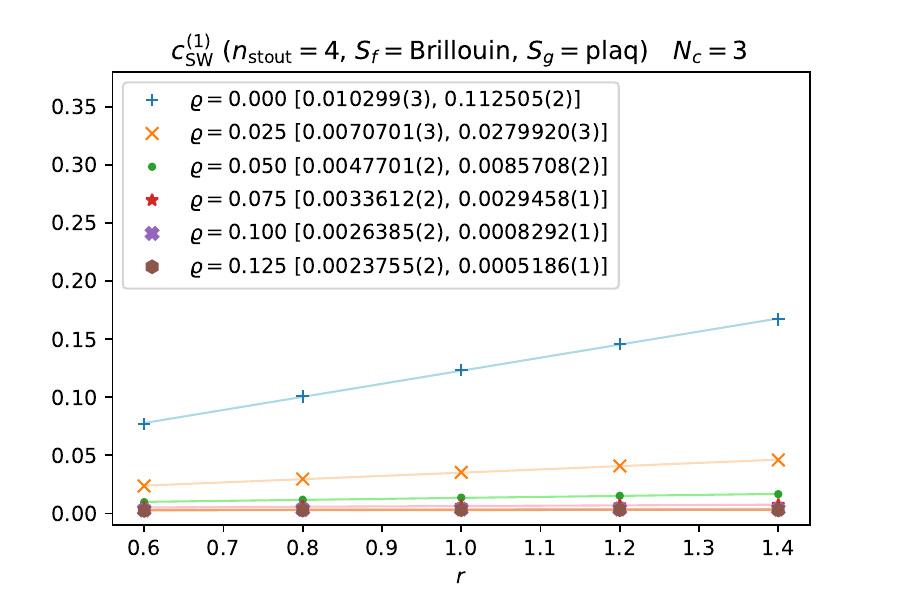}
\caption{The same as Figure~\ref{fig:csw_of_r_wil_stout} but for Brillouin fermions.
\label{fig:csw_of_r_bri_stout}}
\end{figure}

Figures~\ref{fig:csw_of_r_wil_stout} and \ref{fig:csw_of_r_bri_stout} show the behavior of $\csw^{(1)}$ as a function of $r$.
Although $\csw^{(1)}$ is not a linear function of $r$, it shows an almost linear behavior over the range $r\in[0.6,1.4]$ shown there.
Therefore, a reader interested in $\csw^{(1)}$ for $r\neq1$ should be able to infer its value from the linear fit specified in the respective panel.
Again, as for the fermion self energy, we see the lines approaching a constant for increasing $\varrho$, which happens faster and faster, the more stout steps are performed.
Choosing either the Brillouin fermion action or the L\"uscher-Weisz gauge action (or the combination of these modifications) decreases both the overall values and the slopes.

%%%%%%%%%%%%%%%%%%%%%%%%%%%%%
\section{Results for $\mathbf{c_{SW}^{(1)}}$ with Wilson Flow\label{sec:csw_wf}}
%%%%%%%%%%%%%%%%%%%%%%%%%%%%%

We begin the discussion of the Wilson flow results, like we did for the self energy in Ref.~\cite{Ammer:2024hqr}, by illustrating the approximation of the gradient flow by smearing.
The bottom line of this discussion was that $n_\mathrm{stout}$ steps of stout smearing with parameter $\varrho$ have almost the same effect as the gradient flow at flow-time $t$,
provided the two are matched through
\begin{align}
t/a^2=n_\mathrm{stout}\cdot\varrho
\;.
\label{flow_stout_correspondence}
\end{align}
In other words, the flow time (in lattice units) equals, in good approximation, the \emph{cumulative sum} of the stout parameters applied.

\begin{figure}[!htb]
\centering
\includegraphics[scale=0.48]{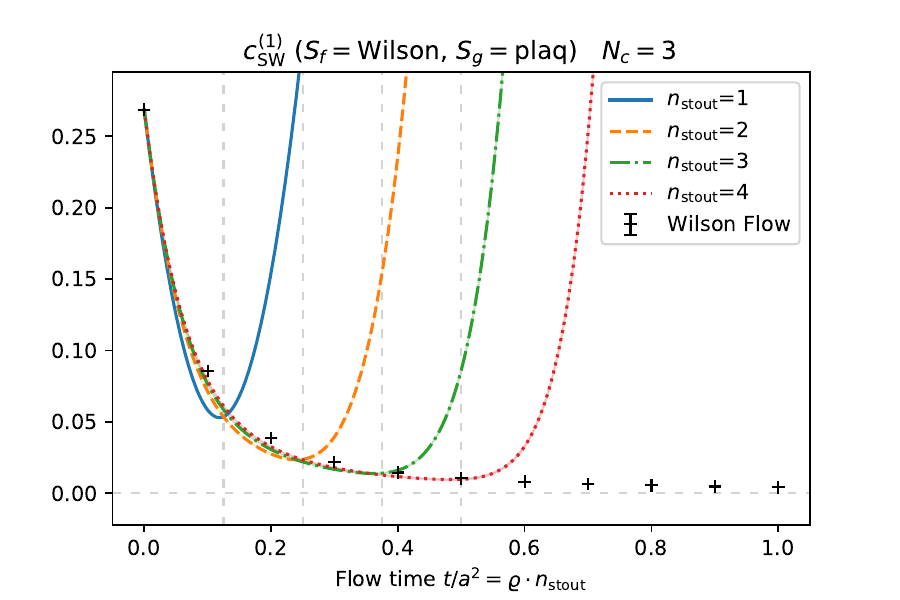}
\includegraphics[scale=0.48]{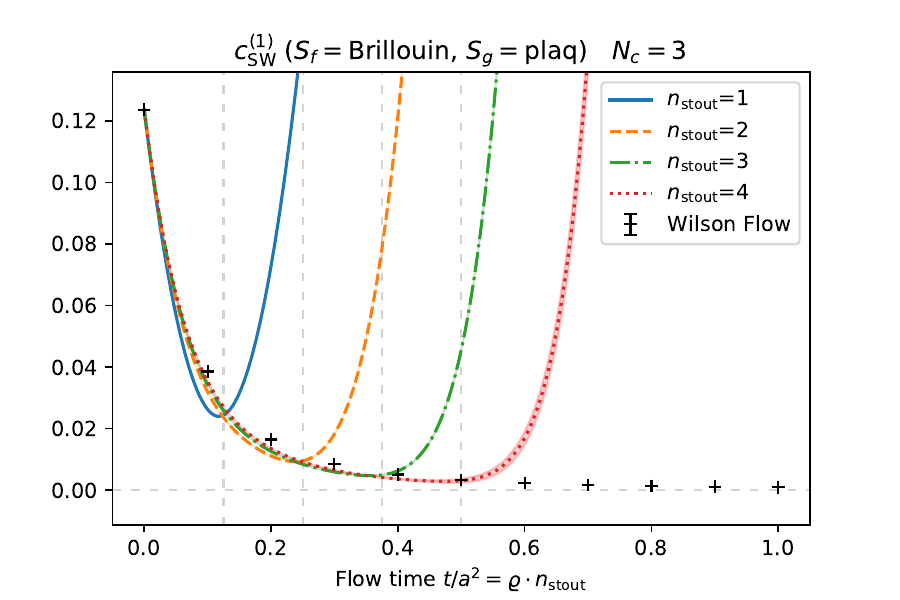}\\
\includegraphics[scale=0.48]{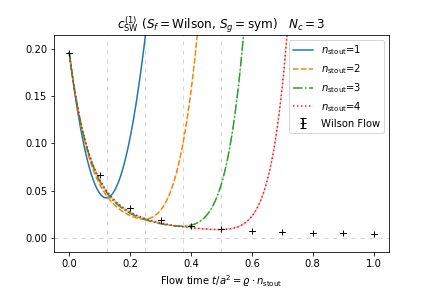}
\includegraphics[scale=0.48]{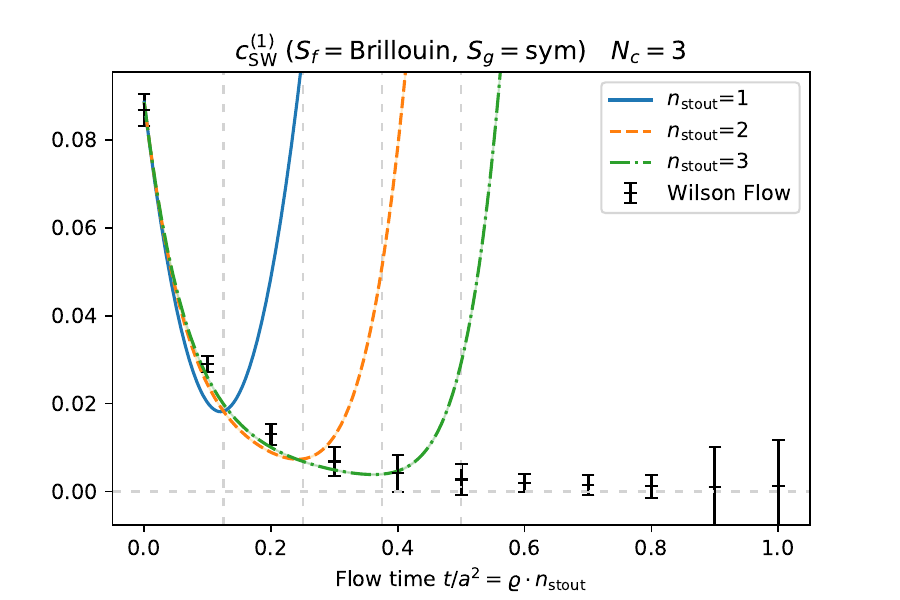}
\caption{$\csw^{(1)}$ as a function of the product $n_\mathrm{stout}\cdot\varrho$, which has the meaning of $t/a^2$, for up to four stout steps.
In addition, results for the direct Wilson flow calculation are shown for selected values of $t$.
The four panels reflect the four combinations of Wilson or Brillouin fermion (left vs.\ right) with plaquette or L\"uscher-Weisz glue (top vs.\ bottom).
\label{fig:csw_flow}}
\end{figure}

In Figure~\ref{fig:csw_flow} we plot the stout curves against the cumulative stout parameter $n_\mathrm{stout}\cdot\varrho$, and we include the new Wilson flow results versus $t/a^2$.
The stout data follow the flow data increasingly well as the number of stout steps is increased, up to $t\approx n_\mathrm{stout}\cdot0.125$ (indicated by vertical dotted lines).
Overall, we see the announced behavior; the flow result at any $t/a^2$ is well approximated by some iterated stout smoothing, provided Equation~(\ref{flow_stout_correspondence}) holds true.
The only extra condition is that none of the stout steps involved shall exceed $\varrho_\mathrm{max}=0.125$.
This finding nicely agrees with the perturbative upper bound of Ref.~\cite{Capitani:2006ni} for stout smearing.

\begin{figure}[!htb]
\centering
\includegraphics[scale=0.48]{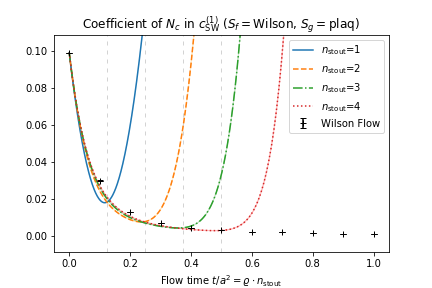}
\includegraphics[scale=0.48]{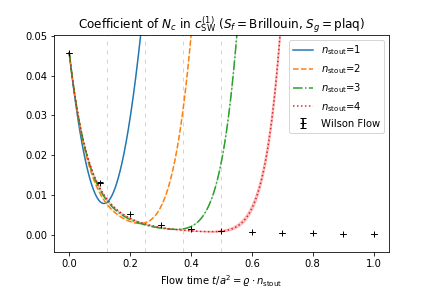}\\
\includegraphics[scale=0.48]{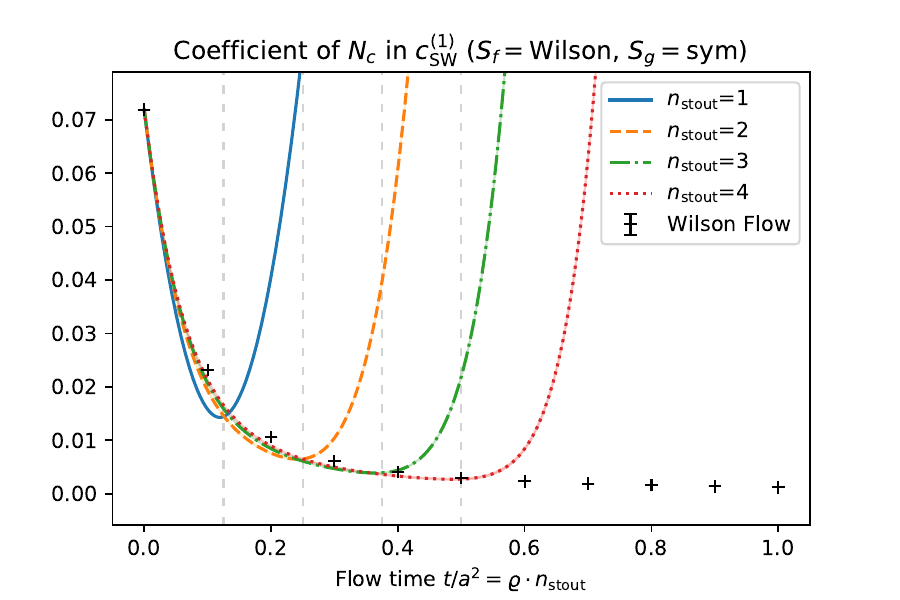}
\includegraphics[scale=0.48]{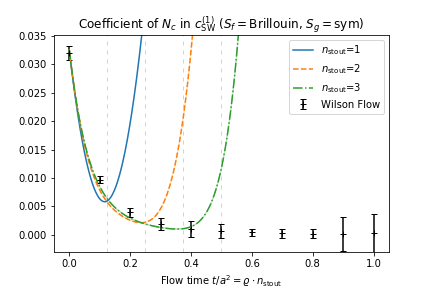}
\caption{Same as in Figure~\ref{fig:csw_flow}, but for the coefficient in front of $N_c$.
\label{fig:csw_nc_flow}}
\end{figure}

\begin{figure}[!htb]
\centering
\includegraphics[scale=0.48]{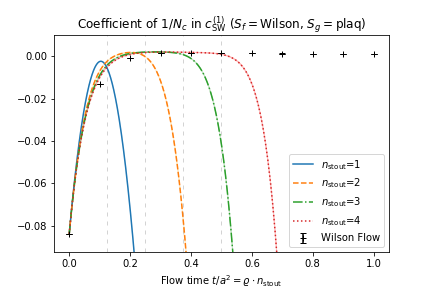}
\includegraphics[scale=0.48]{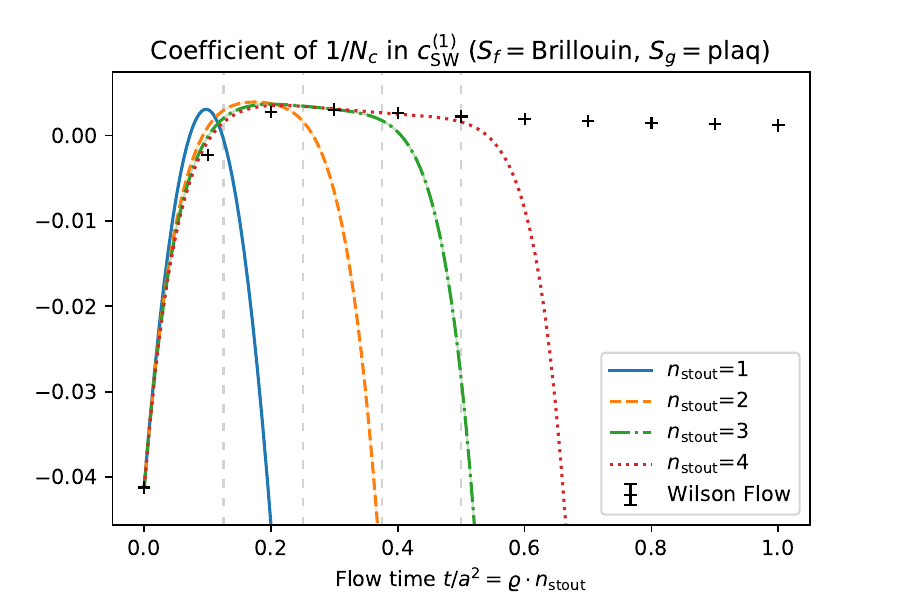}\\
\includegraphics[scale=0.48]{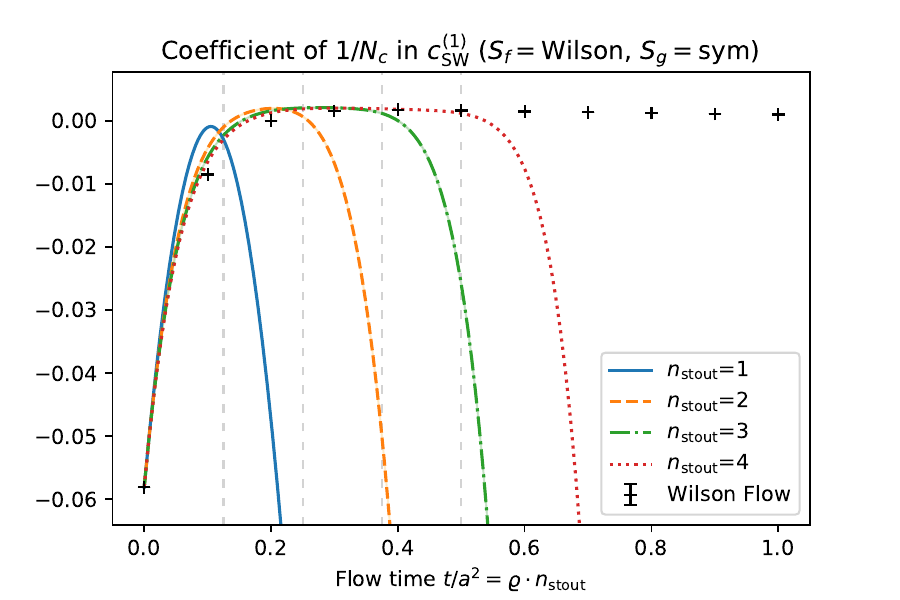}
\includegraphics[scale=0.48]{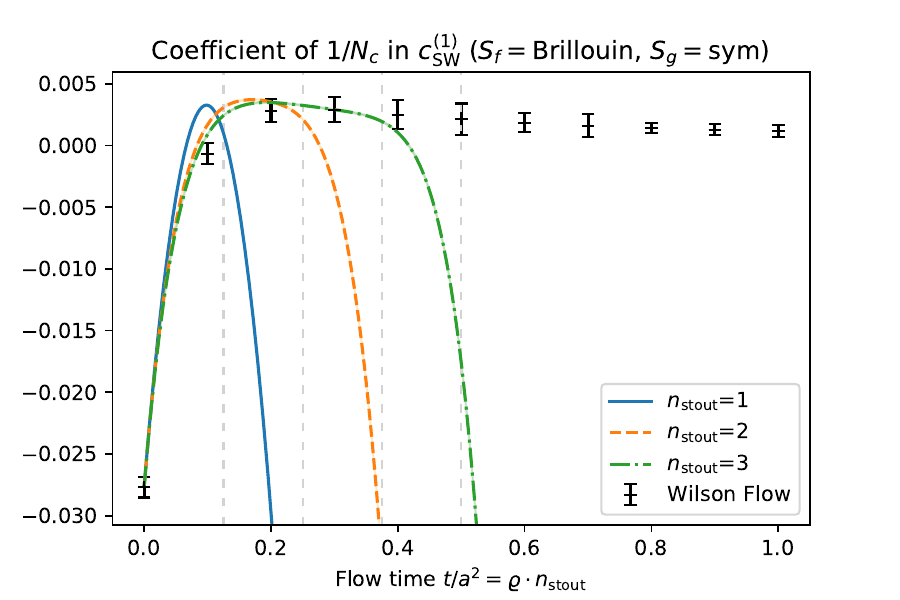}
\caption{Same as in Figure~\ref{fig:csw_flow}, but for the coefficient in front of $1/N_c$.
\label{fig:csw_ncinv_flow}}
\end{figure}

As was done in the previous section, Figures~\ref{fig:csw_nc_flow} and \ref{fig:csw_ncinv_flow} split the physical result into its contributions $\propto N_c$ and $\propto 1/N_c$, respectively.
The reader interested in the value of $\csw^{(1)}$ with gradient flow for an arbitrary value of $N_c$ should be able to infer its value from the latter two plots.
Qualitatively, whatever has been observed in Figure~\ref{fig:csw_flow} holds again, and this means that the features observed in Figure~\ref{fig:csw_flow} hold true for arbitrary $N_c$.

Overall, we are pleased to find the equivalence relation (\ref{flow_stout_correspondence}) between the Wilson flow and the stout smearing results so well obeyed,
since it provides a non-trivial check on our calculations.

The information shown in Figure~\ref{fig:csw_flow} might be split into its contributions from diagrams (a) through (f), respectively.
We remind the reader that out of the individual diagrams only (d) is finite, while the divergences of the remaining diagrams cancel, so that the overall sum is finite.
Accordingly, plots similar to Figure~\ref{fig:csw_flow} can be drawn for (the finite part of) each of the six diagrams contributing to $\csw^{(1)}$.
The interested reader finds this information in Figures~\ref{app_fig:csw_a_flow}-\ref{app_fig:csw_e_flow} in Appendix~\ref{app:csw_wf}.

\begin{figure}[!htb]
\centering
\includegraphics[scale=0.46]{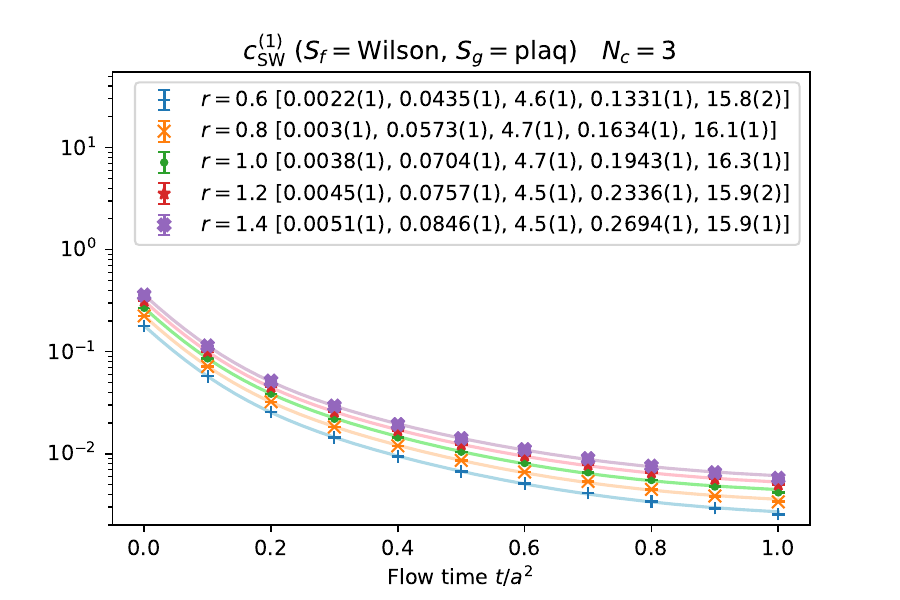}
\includegraphics[scale=0.46]{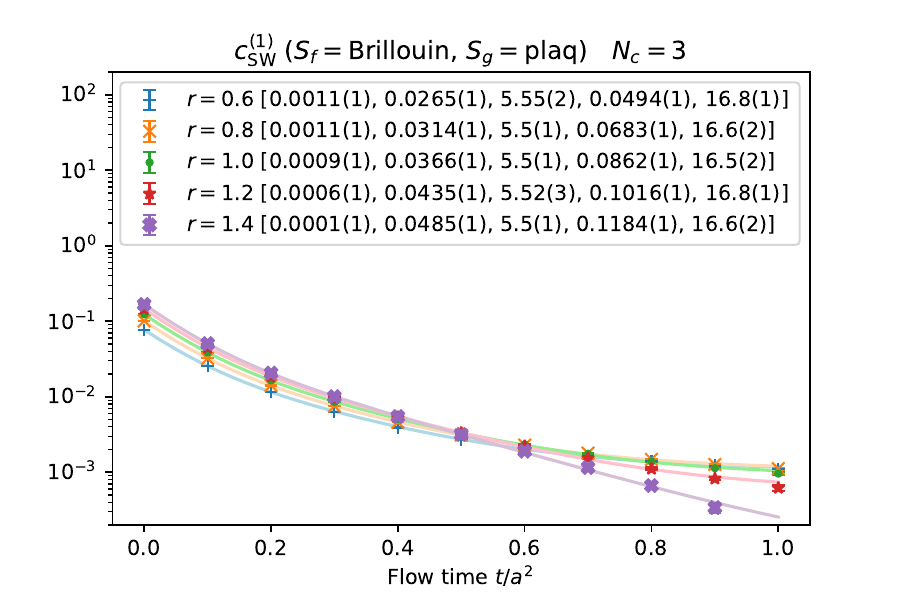}\\
\includegraphics[scale=0.46]{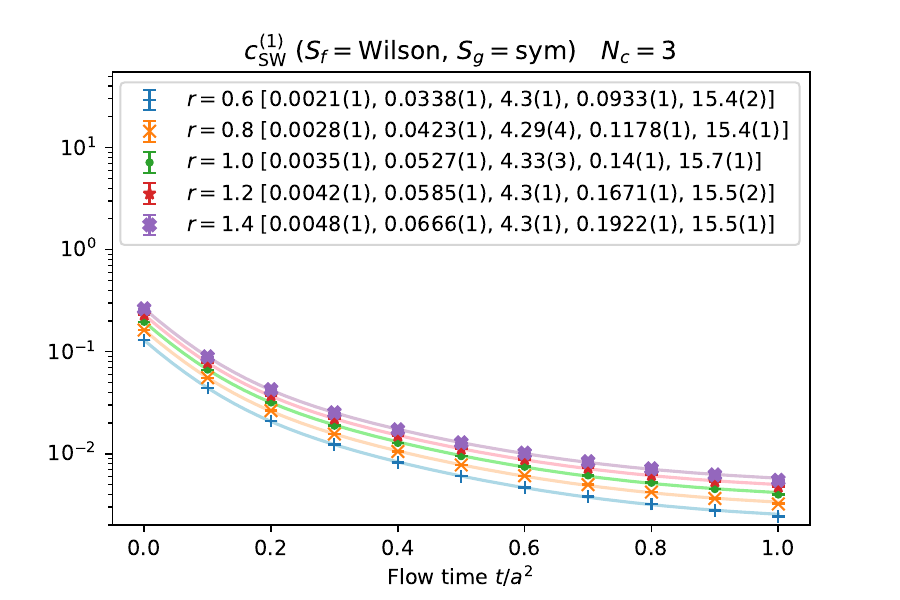}
\includegraphics[scale=0.46]{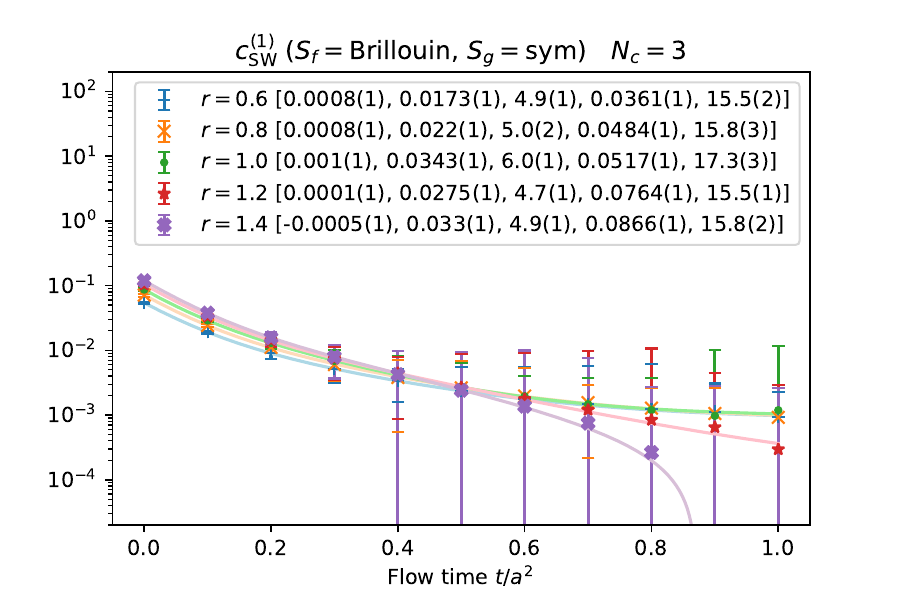}
\caption{Logarithmic plot of $\csw^{(1)}$ with Wilson flow and $N_c=3$ as a function of the flow time $t$ for five different values of $r$.
Two-exponential fits of the form $c_0+c_1e^{-c_2t}+c_3e^{-c_4t}$ are included and values of the the coefficients $[c_0,c_1,c_2,c_3,c_4]$ are given in brackets.
\label{fig:csw_flow_fits}}
\end{figure}

It turns out that the gradient flow results for $\csw^{(1)}$ can be reasonably well described by a fit function with two exponentials.
All Wilson flow results with $r\in\{0.6,0.8,1.0,1.2,1.4\}$ are shown in Figure~\ref{fig:csw_flow_fits}, with a logarithmic scale on the $y$-axis.
The coefficients of the two-exponential fits are given in the legends of the plots.
The attentive reader might be surprised to see that the coefficient $c_0$ is small but non-zero.
We take it for granted that $c_0\to0$ if the number of exponentials increases, so the constant term is a proxy for additional (small) exponential terms.
Similar plots for general $N_c$ are arranged in Appendix~\ref{app:csw_wf}.

\begin{figure}[!htb]
\centering
\includegraphics[scale=0.46]{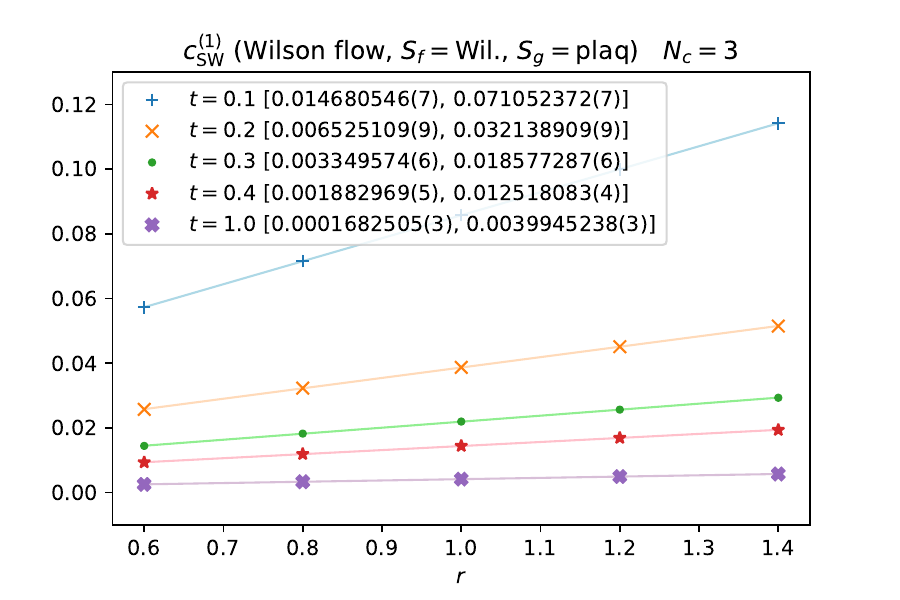}
\includegraphics[scale=0.46]{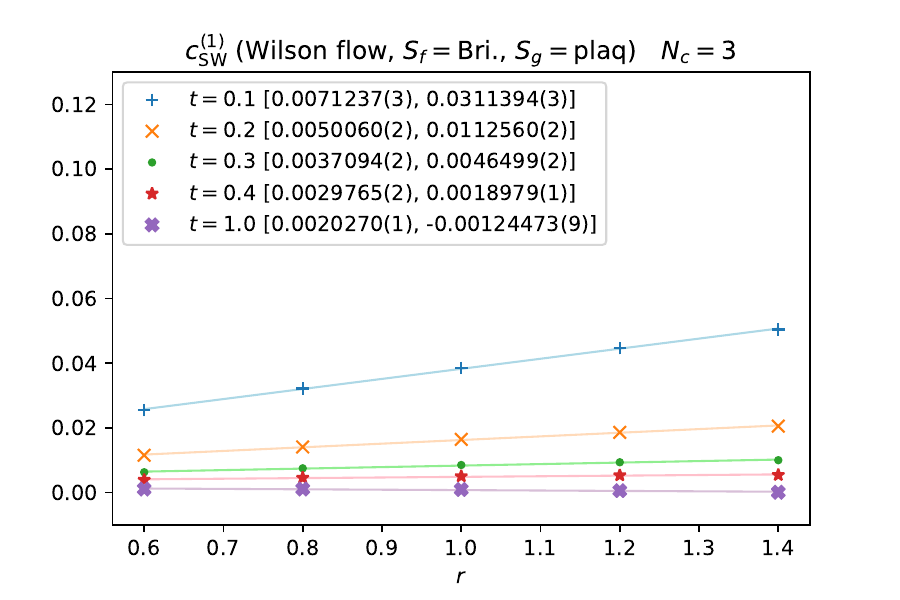}\\
\includegraphics[scale=0.46]{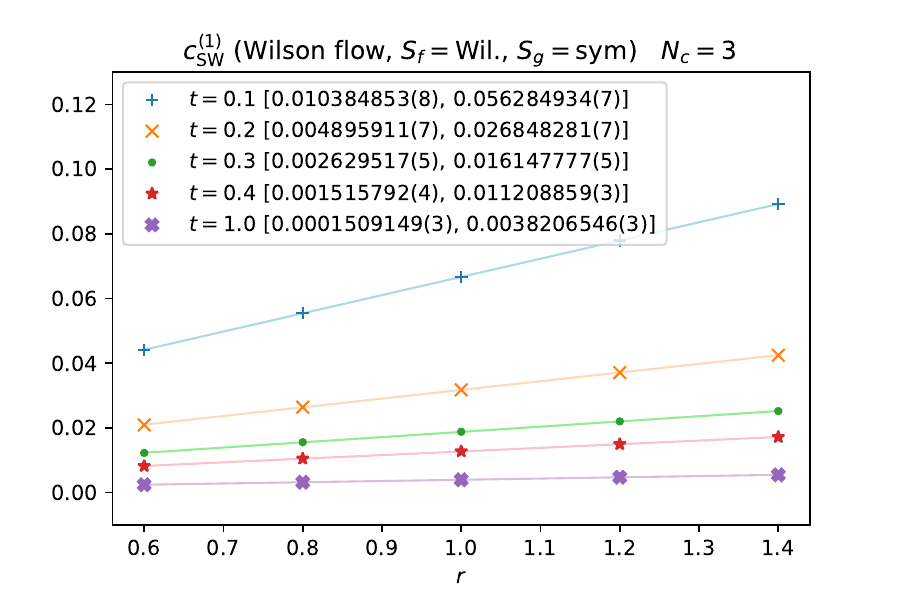}
\includegraphics[scale=0.46]{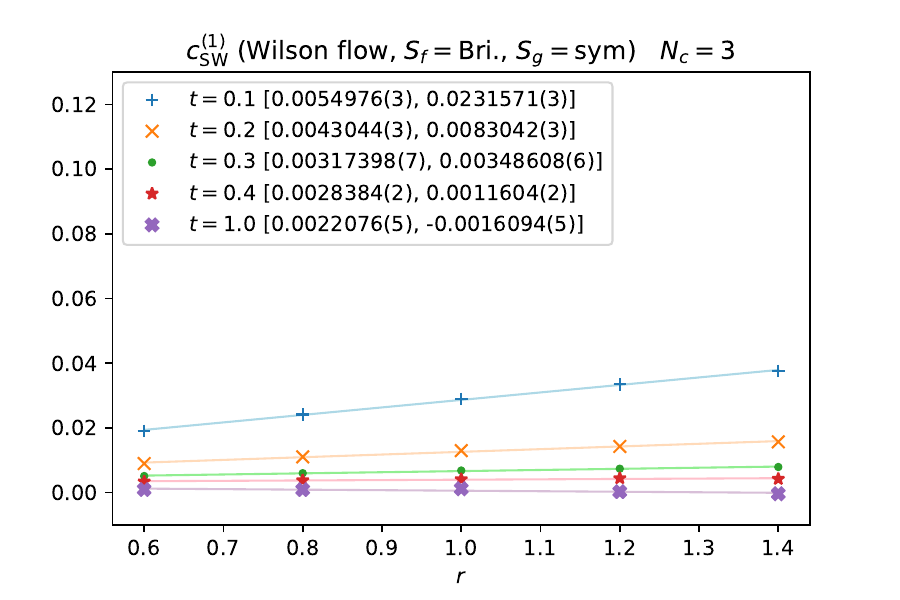}
\caption{$\csw^{(1)}$ for $N_c=3$ as a function of $r$ for a few values of the flow time $t/a^2$.
Linear least-squares fits of the form $c_0+c_1\cdot r$ are included and the values of the coefficients $[c_0,c_1]$ are given in brackets.
\label{fig:csw_of_r_flow}}
\end{figure}

In Figure~\ref{fig:csw_of_r_flow} we show how $\csw^{(1)}$ with Wilson flow depends on $r$.
Again we see an almost linear behavior over the range shown, and the respective fit parameters are included in the plots.
Note that we do not show the data for $t=0$, as it would create a large gap to the results at $t\geq 0.1$.
The interested reader finds this information in Figures~\ref{fig:csw_of_r_wil_stout} and \ref{fig:csw_of_r_bri_stout} under the label $\varrho=0$ or directly in Ref.~\cite{Ammer:2023otl}.

%%%%%%%%%%%%%%%%%%%%%%%%%%%%%
\section{Conclusion\label{sec:conclusion}}
%%%%%%%%%%%%%%%%%%%%%%%%%%%%%

The purpose of this paper has been to extend the calculation of the one-loop improvement coefficient $\csw^{(1)}$ for Wilson and Brillouin fermions on plaquette or Symanzik glue \cite{Aoki:2003sj,Ammer:2023otl} to the situation
where the quark-(multi)gluon-antiquark coupling is smoothed through either stout smearing \cite{Morningstar:2003gk} or gradient flow with respect to the Wilson gauge action \cite{Narayanan:2006rf,Luscher:2010iy,Luscher:2011bx}.

Technically, we applied the ``upgrade recipe'' of Ref.~\cite{Ammer:2024hqr} to our calculation of $\csw^{(1)}$ in Ref.~\cite{Ammer:2023otl},
where the upgrade refers to the perturbative implementation of stout smearing and/or Wilson flow according to the recipe Ref.~\cite{Ammer:2024hqr}.
A felicitous cancellation between contributions from the Wilson/Brillouin action and contributions from the clover term eliminated the order $g_0^3$ form factor $\tilde{g}_{\mu\nu\rho\sigma}$ or $B_{\mu\nu\rho\sigma}$ and made the calculation of diagram (d) a bit easier.

We presented the calculated one-loop values of $\csw^{(1)}$ with up to four stout steps and various flow times $t\in [0,1]$ in a large number of plots and tables.
We emphasize that all of our results refer to the situation where the covariant derivative, the Laplacian and the clover term in the fermion action are treated with the same smoothing recipe.
Our results suggest that even a mild amount of smearing (say three stout steps with $\varrho=0.11$, tantamount to $t/a^2=0.33$) makes the one-loop contribution significantly smaller than the tree-level result.
This is in line with similar observations in the self energy calculations presented in Ref.~\cite{Ammer:2024hqr} and lets us hope that the perturbative series at any $t/a^2>0$ is well behaved.

In our view Figure~\ref{fig:csw_flow} summarizes the main results of this paper.
For each one of the four action combinations considered, the value $\csw^{(1)}$ approaches zero if the flow time $t/a^2$ is chosen large enough.
The correspondence (\ref{flow_stout_correspondence}) is found to be well-observed, provided none of the stout steps exceeds $\varrho_\mathrm{max} \approx 0.125$.
Compared to Wilson fermions, the use of Brillouin fermions lowers the value of $\csw^{(1)}$ by about 50\%, while the use of improved gluons lowers it by another 25\% (roughly).
For any of these actions with $n_\mathrm{stout}\cdot\varrho \geq 0.5$ or $t/a^2 \geq 0.5$, the value of $\csw$ is close to the tree-level value $\csw^{(0)}=r$ over a wide range of $g_0^2$.
This is especially true for the combination of Brillouin fermions and L\"uscher-Weisz gluons, which seem to have a very well behaved perturbative expansion (as was already pointed out in Ref~\cite{Ammer:2024hqr}).

We believe that even with a more modest smoothing recipe (e.g.\ three stout steps at $\varrho=0.11$) one can simulate any of the four action combinations with $\csw$ dialed to the one-loop value calculated
in this paper, and one will obtain data with a decent continuum behavior, that is with manageable leading cut-off effects $\propto \alpha^2 a$ and well behaved higher-order terms.

%%%%%%%%%%%%%%%%%%%%%%%%%%%%%
\appendix\section{Feynman rules\label{app:feynman_rules}}
%%%%%%%%%%%%%%%%%%%%%%%%%%%%%

The vertices of the Wilson action are
\begin{align}
V^a_{1W\mu}(p,q)&=-g_0T^a\Big(i\gamma_\mu c(p_\mu+q_\mu)+r s(p_\mu+q_\mu)\Big)
\\
V^{ab}_{2W\mu\nu}(p,q)&=\frac{g_0^2}{2}T^aT^b\delta_{\mu\nu}\Big(i\gamma_\mu s(p_\mu+q_\mu)-r c(p_\mu+q_\mu)\Big)
\\
V^{abc}_{3W\mu\nu\rho}(p,q)&=\frac{g_0^3}{6}T^aT^bT^c\delta_{\mu\nu}\delta_{\mu\rho}\Big(i\gamma_\mu c(p_\mu+q_\mu)+r s(p_\mu+q_\mu)\Big)
\end{align}
The vertices of the clover term are
\begin{align}
&V_{1c\,\mu}(p,q)=i\frac{g_0}{2}\csw^{(0)}\sum\limits_{\nu}\sigma_{\mu\nu}c(p_\mu-q_\mu)\bar{s}(p_\nu-q_\nu)
\\
&V_{2c\,\mu\nu}(p,q,k_1,k_2)=
i g_0^2 \csw^{(0)}\Bigg(\frac{1}{4}\delta_{\mu\nu}
\sum\limits_\rho \sigma_{\mu\rho}s(q_\mu-p_\mu)\big(\bar{s}(k_{2\rho})-\bar{s}(k_{1\rho})\big)
\nonumber\\&\quad
+\sigma_{\mu\nu}\bigg(c(k_{1\nu})c(k_{2\mu})c(q_\mu-p_\mu)c(p_\nu-p_\nu)-\frac{1}{2}c(k_{1\mu})c(k_{2\nu})\bigg)
\Bigg)
\\
&V_{3c\,\mu\nu\rho}(p,q,k_1,k_2,k_2)=
ig_0^3\csw^{(0)} T^aT^bT^c\Bigg(\delta_{\mu\nu}\delta_{\mu\rho}\sum\limits_\alpha \sigma_{\mu\alpha}
\bigg[-\frac{1}{12}c(q_\mu-p_\mu)\bar{s}(q_\alpha-p_\alpha)
\nonumber\\&\quad
-\frac{1}{2}c(q_\mu-p_\mu)c(q_\alpha-p_\alpha)c(k_{3\alpha}-k_{1\alpha})s(k_{2\alpha})\bigg]
\nonumber\\&\quad
+\delta_{\mu\nu}\sigma_{\mu\rho}\bigg[-\frac{1}{2}c(q_\mu-p_\mu)c(q_\rho-p_\rho)c(k_{1\rho}+k_{2\rho})s(k_{3\mu})
+\frac{1}{4}s(k_{1\mu}+k_{2\mu})c(2k_{2\rho}+k_{3\rho})\bigg]
\nonumber\\&\quad
+\delta_{\nu\rho}\sigma_{\mu\nu}\bigg[-\frac{1}{2}c(q_\mu-p_\mu)c(q_\nu-p_\nu)c(k_{2\mu}+k_{3\mu})s(k_{1\nu})
+\frac{1}{4}s(k_{2\nu}+k_{3\nu})c(k_{1\mu}+2k_{2\mu})\bigg]
\nonumber\\&\quad
+\delta_{\mu\rho}\sigma_{\mu\nu}
\bigg[\frac{1}{2}c(q_\nu-p_\nu)c(k_{3\nu}-k_{1\nu})s(k_{1\mu}+2k_{2\mu}+k_{3\mu})
\bigg]\Bigg)
\;.
\end{align}

%%%%%%%%%%%%%%%%%%%%%%%%%%%%%
\section{Form Functions\label{app:form_functions}}
%%%%%%%%%%%%%%%%%%%%%%%%%%%%%

The form factors of stout smearing used in Equations~(\ref{eq:V1_stout}-\ref{eq:V3_stout}) are
\begin{align}
\tilde{g}^{n}_{\mu\nu}(\varrho,k)&=
\bigg(\delta_{\mu\nu}+\varrho g_{\mu\nu}(k)\bigg)^n
=\big(1-\varrho\hat{k}^2\big)^n\delta_{\mu\nu}-\Big(\big(1-\varrho\hat{k}^2\big)^n-1\Big)\frac{\hat{k}_\mu\hat{k}_\nu}{\hat{k}^2}
\\
\tilde{g}^{(n)}_{\mu\nu\rho}(\varrho,k_1,k_2)&=\sum\limits_{\alpha\beta\gamma}\varrho g_{\alpha\beta\gamma}(k_1,k_2)\sum\limits_{m=0}^{n-1}\tilde{g}_{\mu\alpha}^{n-m-1}(\varrho,k_1+k_2)\tilde{g}^m_{\beta\nu}(\varrho,k_1)\tilde{g}^m_{\gamma\rho}(\varrho,k_2)
\\
\tilde{g}^{(n)}_{\mu\nu\rho\sigma}(\varrho,k_1,k_2,k_3)&=
6\varrho\sum\limits_{\alpha\beta\gamma}g_{\alpha\beta\gamma}(k_1,k_2+k_3)\sum\limits_{m=1}^{n-1}\tilde{g}^{n-m-1}_{\mu\alpha}(\varrho,k_1+k_2+k_3) \tilde{g}^m_{\beta\nu}(\varrho,k_1)\tilde{g}^{(m)}_{\gamma\rho\sigma}(k_2,k_3)
\nonumber\\
+\sum\limits_{\alpha\beta\gamma\delta} &\tilde{G}_{\alpha\beta\gamma\delta}(\varrho,k_1,k_2,k_3)
\sum\limits_{m=0}^{n-1} \tilde{g}^{n-m-1} _{\mu\alpha}(\varrho,k_1+k_2+k_3)
\tilde{g}^m_{\beta\nu}(\varrho,k_1)\tilde{g}^m_{\gamma\rho}(\varrho,k_2)\tilde{g}^m_{\delta\sigma}(\varrho,k_3)
\;.
\label{eq:g_tilde_NNLO}
\end{align}
For the Wilson flow the functions that are needed in Equations (\ref{eq:V1_flow}-\ref{eq:V3_flow}) are
\begin{align}
B_{\mu\nu}(k,t)&=e^{-\hat{k}^2t}\delta_{\mu\nu}-\Big(e^{-\hat{k}^2t}-1\Big)\frac{\hat{k}_\mu\hat{k}_\nu}{\hat{k}^2}
\\
B_{\mu\nu\rho}(k_1,k_2,t)&=\sum\limits_{\alpha\beta\gamma}g_{\alpha\beta\gamma}(k_1,k_2)\int\limits_0^t B_{\mu\alpha}(k_1+k_2,t-t')B_{\beta\nu}(k_1,t^\prime)B_{\gamma\rho}(k_2,t^\prime)dt^\prime
\\
B_{\mu\nu\rho\sigma}(k_1,k_2,k_3,t)&=
 \sum\limits_{\alpha\beta\gamma}g_{\alpha\beta\gamma}(k_1,k_2+k_3)
 \int\limits_0^t B_{\mu\alpha}(k_1+k_2+k_3,t-t^\prime)
 B_{\beta\nu}(k_1,t^\prime)B_{\gamma\rho\sigma}(k_2,k_3,t^\prime)
\,dt^\prime
\nonumber\\
+\sum\limits_{\alpha\beta\gamma\delta}&G_{\alpha\beta\gamma\delta}(k_1,k_2,k_3)
 \int\limits_0^t B_{\mu\alpha}(k_1+k_2+k_3,t-t^\prime)
 B_{\beta\nu}(k_1,t^\prime)B_{\gamma\rho}(k_2,t^\prime)B_{\delta\sigma}(k_3,t^\prime)
\,dt^\prime
\;.
\label{eq:B_NNLO}
\end{align}
The quantities that appear in the $\tilde{g}$ and the $B$ functions, which do not depend on the stout parameter or the flow time, are defined as
\begin{align}
g_{\mu\nu}(k)&=-\delta_{\mu\nu}\hat{k}^2+\hat{k}_\mu\hat{k}_\nu
=4\big(-\delta_{\mu\nu}s^2(k)+s(k_\mu)s(k_\nu)\big)
\\
g_{\mu\nu\rho}(k_1,k_2)
&=2i\Big(
\delta_{\mu\rho}c(k_{1\mu})s( 2 k_{2\nu}+k_{1\nu})
-\delta_{\mu\nu}c(k_{2\mu})s(2 k_{1\rho}+k_{2\rho})
\nonumber\\&
+\delta_{\nu\rho}s( k_{1\mu}-k_{2\mu})c( k_{1\nu}+k_{2\nu})\Big)
\\
\tilde{G}_{\mu\nu\rho\sigma}(\varrho,k_1,k_2,k_3)&=
\frac{1}{2}\varrho^2\Big(g_{\mu\nu}(k_1)g_{\mu\rho}(k_2)\delta_{\mu\sigma}-g_{\mu\nu}(k_1)\delta_{\mu\rho}g_{\mu\sigma}(k_3)\Big)
+\varrho\; G_{\mu\nu\rho\sigma}(k_1,k_2,k_3)
\\
G_{\mu\nu\rho\sigma}(k_1,k_2,k_3)&=
\frac{1}{2} g_{\mu\nu}(k_1)\delta_{\mu\rho}\delta_{\mu\sigma}
+\delta_{\mu\nu}g_{\mu\rho}(k_2)\delta_{\mu\sigma}
+3h_{\mu\nu\rho}(k_1,k_2)\delta_{\mu\sigma}
+\frac{3}{2}\Big(g^{(i)}_{\mu\nu}(k_1,k_2)\delta_{\nu\rho}\delta_{\mu\sigma}
\nonumber\\&
+g^{(ii)}_{\mu\nu\rho}(k_1,k_2+k_3)\delta_{\rho\sigma}
+g^{(iii)}_{\mu\nu\rho\sigma}(k_1,k_2,k_3)\Big)
+\frac{1}{2}g_{\mu\nu}(k_1+k_2+k_3)\delta_{\nu\rho}\delta_{\nu\sigma}
\\
h_{\mu\nu\rho}(k_1,k_2)
&=2\Big(
-\delta_{\mu\nu}s(k_{2\mu})s(2 k_{1\rho}+k_{2\rho})
-\delta_{\mu\rho}s(k_{1\mu})s( 2 k_{2\nu}+k_{1\nu})
\nonumber\\&
+\delta_{\nu\rho}c( k_{1\mu}-k_{2\mu})c( k_{1\nu}+k_{2\nu})\Big)
\\
g^{(i)}_{\mu\nu}(k)&=4\big(\delta_{\mu\nu}4s^2(k)+c (k_\mu)c(k_\nu)\big)
\\
g^{(ii)}_{\mu\nu\rho}(k_1,k_2)&=4\Big(
\delta_{\mu\nu}c(2k_{1\rho}+k_{2\rho})c(k_{2\mu})
+\delta_{\mu\rho}s(k_{1\mu})s(k_{1\nu}+2k_{2\nu})
\nonumber\\&
+\delta_{\nu\rho}s(k_{1\mu}-k_{2\mu})s(k_{1\nu}+k_{2\nu})\Big)
\\
g^{(iii)}_{\mu\nu\rho\sigma}(k_1,k_2,k_3)&=-4\delta_{\mu\rho}\delta_{\nu\sigma}c( k_{1\nu} +2k_{2\nu} + k_{3\nu})c(k_{1\mu}-k_{3\mu})
\;.
\end{align}

%%%%%%%%%%%%%%%%%%%%%%%%%%%%%
\section{Proof\label{app:proof}}
%%%%%%%%%%%%%%%%%%%%%%%%%%%%%

Without knowing the particular form of $\tilde{g}^{(n)}_{\mu\nu\rho\sigma}$ (or $B_{\mu\nu\rho\sigma}$) we can show, that they are not needed in the computation of $\csw^{(1)}$.
The only place, where the third order could contribute is in the tadpole diagram (d).
It is given by
\begin{align}
\Lambda_\mu^{a(1)(d)}(p,q)=\sum\limits_{b,\nu\rho}V^{abb}_{3\mu\nu\rho}(p,q,q-p,k,-k)G_{\nu\rho}(k)
+2\ \mathrm{perms}
\;.
\end{align}
The term in question (ignoring pre-factors and permutations) is
\begin{align}
\sum\limits_{\nu\rho\alpha} \tilde{g}^{(n)}_{\alpha\mu\nu\rho}(\varrho,q-p,k,-k)V_{1\alpha}(p,q)G_{\nu\rho}(k)
\;.
\end{align}
For $V_{1\mu}(p,q)$ we insert the sum of Wilson and clover Feynman rules.
They are the only parts that have a Dirac structure, thus they determine the structure after taking the trace in (\ref{eq:G1})
\begin{align}
\Tr\big[V_{1W\alpha}(p,q)+V_{1c\alpha}(p,q)]&=-4g_0r s(p_\alpha+q_\alpha)
\\
\Tr\big[\big(V_{1W\alpha}(p,q)+V_{1c\alpha}(p,q)\big)\gamma_\nu\gamma_\mu]&=-2g_0\csw^{(0)}\big(
\delta_{\mu\alpha}c(p_\mu-q_\mu)\bar{s}(p_\nu-q_\nu)
\nonumber\\&
-\delta_{\nu\alpha}c(p_\nu-q_\nu)\bar{s}(p_\mu-q_\mu)\big)
\;.
\end{align}
All of these terms involve sine functions.
Thus terms involving the derivative of $\tilde{g}^{(n)}_{\alpha\mu\nu\rho}$ will vanish when we set $p,q$ to zero.
The only derivatives that survive are (with $\mu\neq\nu$)
\begin{align}
\bigg(\frac{\partial}{\partial p_\mu}+\frac{\partial}{\partial q_\mu}\bigg)\Tr\big[V_{1W\alpha}(p,q)+V_{1c\alpha}(p,q)]&=-4g_0r \delta_{\mu\alpha} c(p_\alpha+q_\alpha)
\\
\bigg(\frac{\partial}{\partial p_\nu}-\frac{\partial}{\partial q_\nu}\bigg)\Tr\big[\big(V_{1W\alpha}(p,q)+V_{1c\alpha}(p,q)\big)\gamma_\nu\gamma_\mu]&=-4g_0\csw^{(0)}
\delta_{\mu\alpha}c(q_\mu-p_\mu)\bar{c}(q_\nu-p_\nu)
\;.
\end{align}
Setting $p$ and $q$ to zero and plugging in the tree-level value $\csw^{(0)}=r$ makes both terms identical and they cancel each other in $G_1$.

The whole argument above did not depend on the particular form of the form factor $\tilde{g}$ (or $B$) nor on the number of smearing steps or whether it was the stout smearing or Wilson flow form factor.
The same happens for any Wilson-like fermion action which fulfils the condition $\lambda_1+6\lambda_2+12\lambda_3+8\lambda_4=1$.
This does not mean, that there is no contribution at all coming from diagram (d).
There are still terms coupling to the first and second order form factors in the $\bar{q}qggg$-vertex (see Equations~(\ref{eq:V3_stout}) and (\ref{eq:V3_flow})).

%%%%%%%%%%%%%%%%%%%%%%%%%%%%%
\section{Additional Data for $\mathbf{c_{SW}^{(1)}}$ with Stout Smearing\label{app:csw_stout}}
%%%%%%%%%%%%%%%%%%%%%%%%%%%%%

\begin{figure}[!htb]
\centering
\includegraphics[scale=0.48]{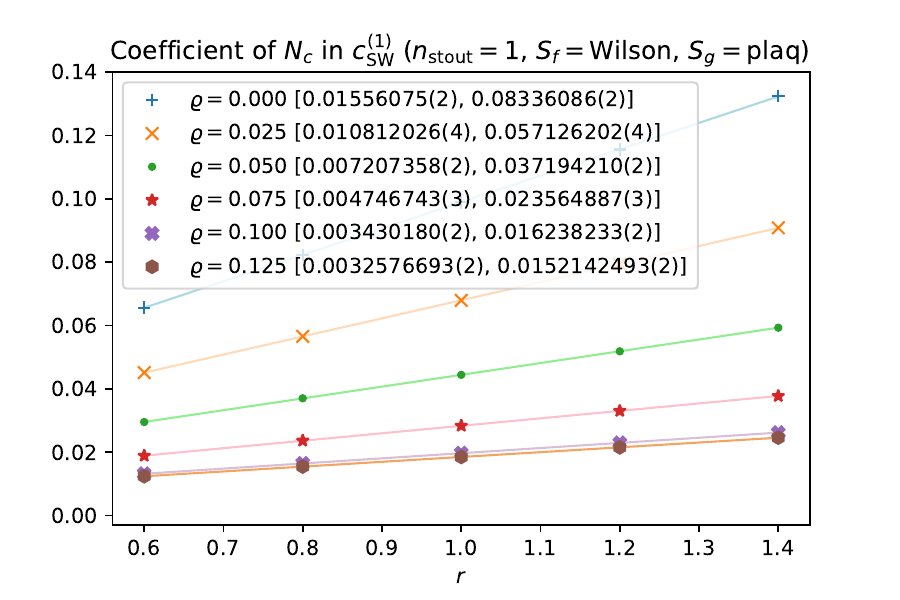}
\includegraphics[scale=0.48]{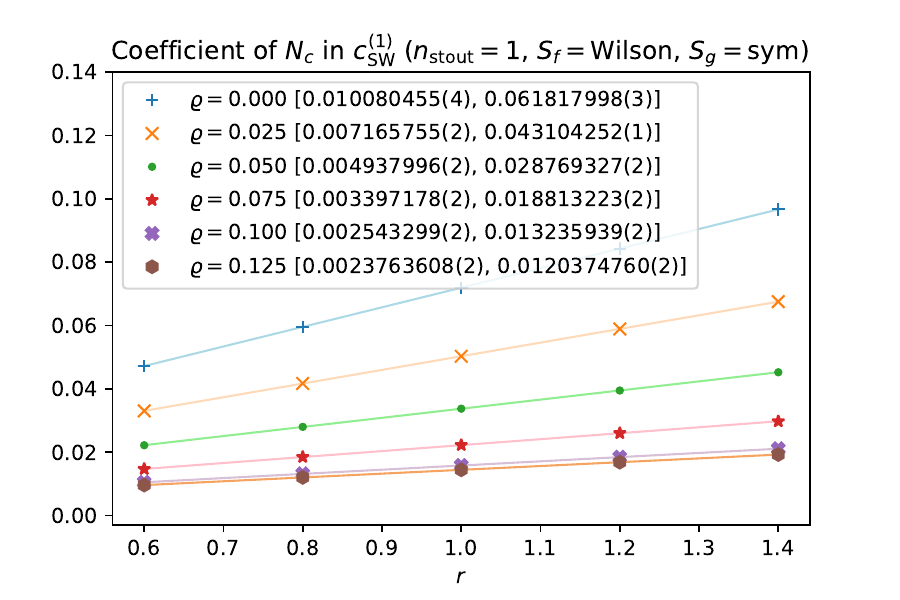}\\
\includegraphics[scale=0.48]{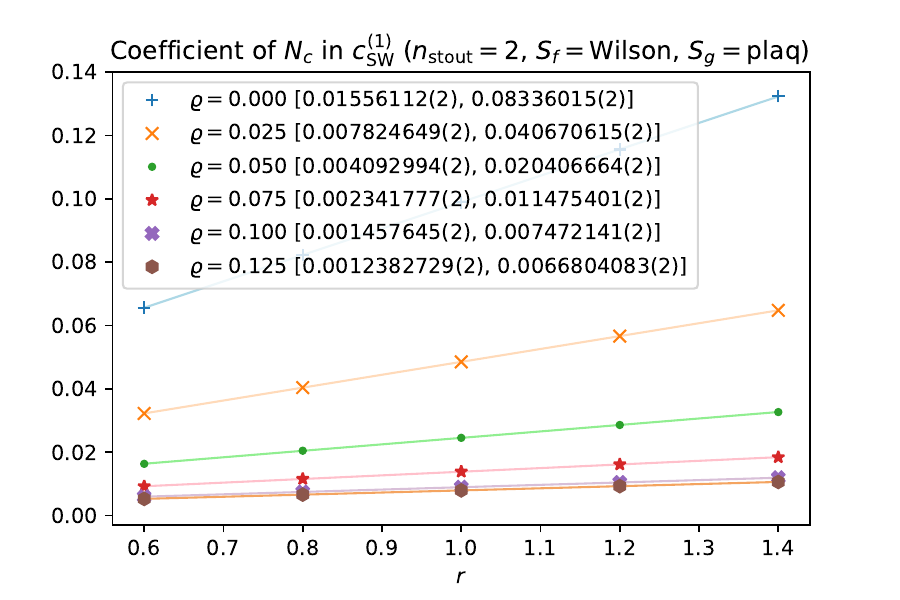}
\includegraphics[scale=0.48]{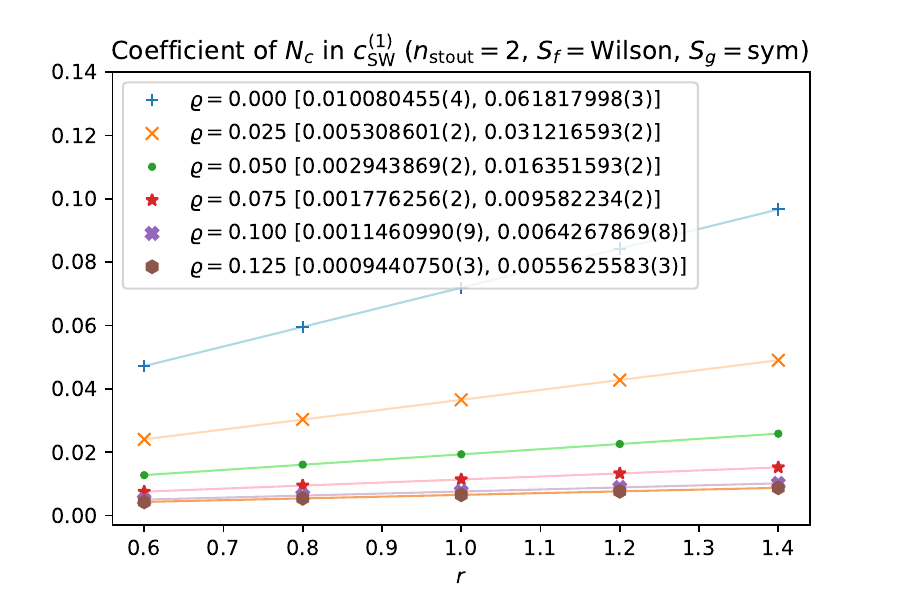}\\
\includegraphics[scale=0.48]{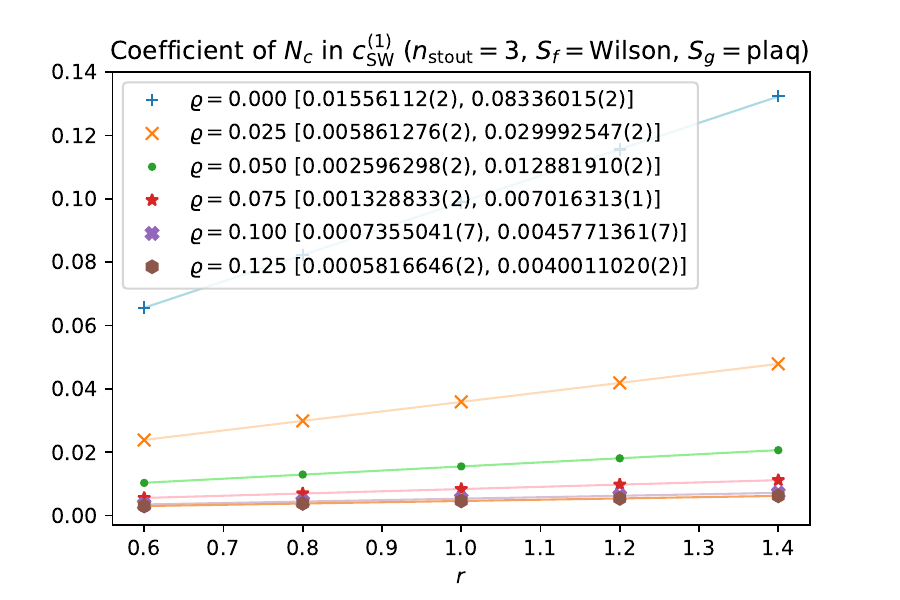}
\includegraphics[scale=0.48]{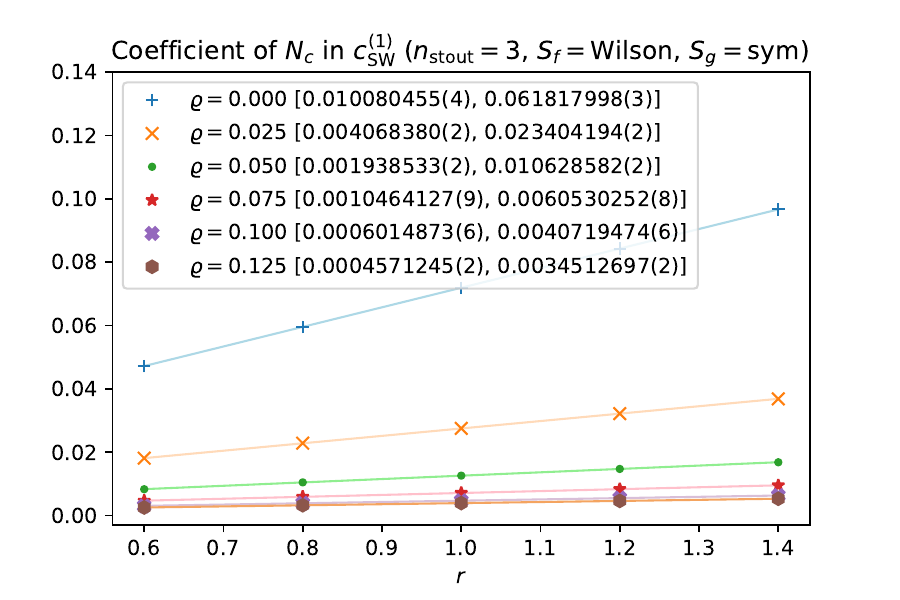}\\
\includegraphics[scale=0.48]{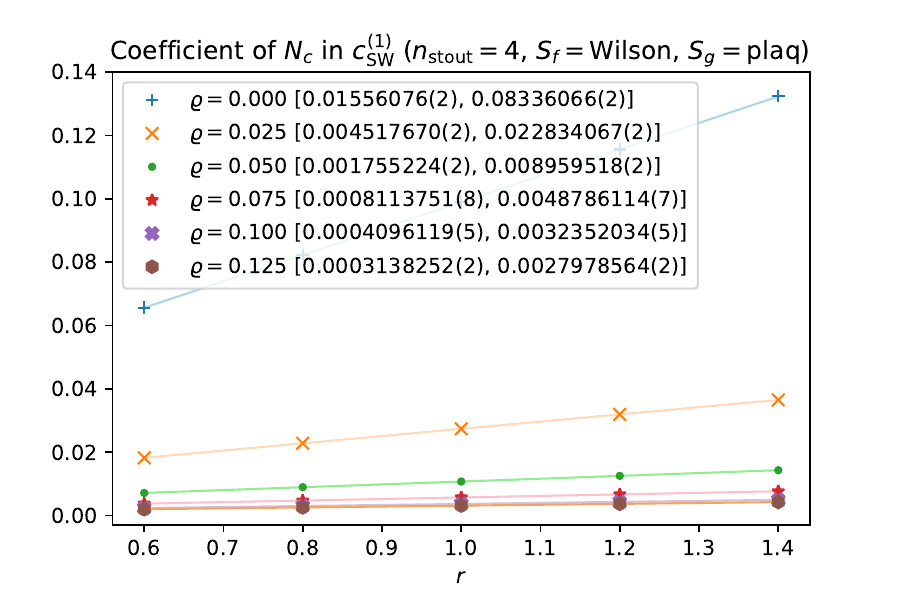}
\includegraphics[scale=0.48]{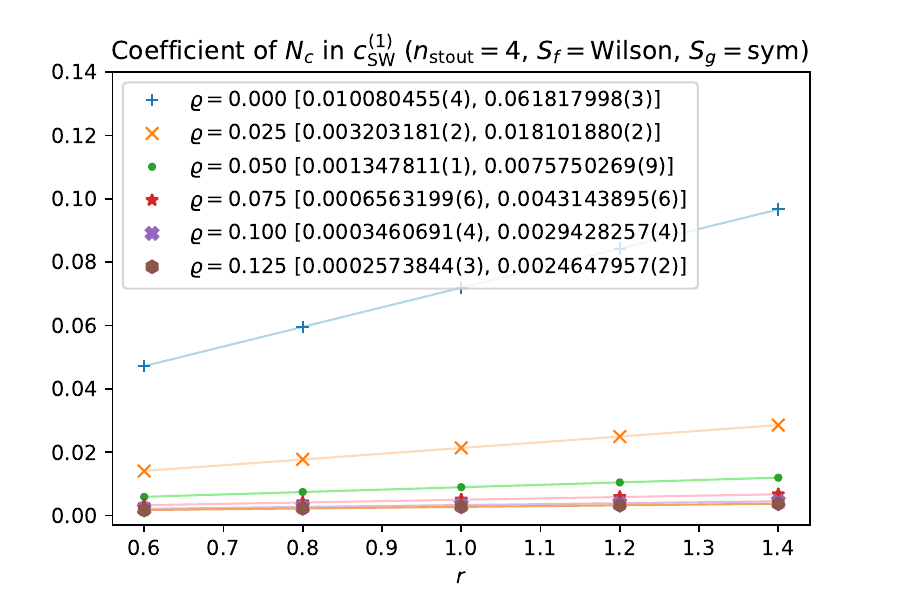}
\caption{Same as in Figure~\ref{fig:csw_of_r_wil_stout} but for the part of $\csw^{(1)}$ linear in $N_c$.
\label{app_fig:csw_nc_of_r_wil_stout}}
\end{figure}

\begin{figure}[!htb]
\centering
\includegraphics[scale=0.48]{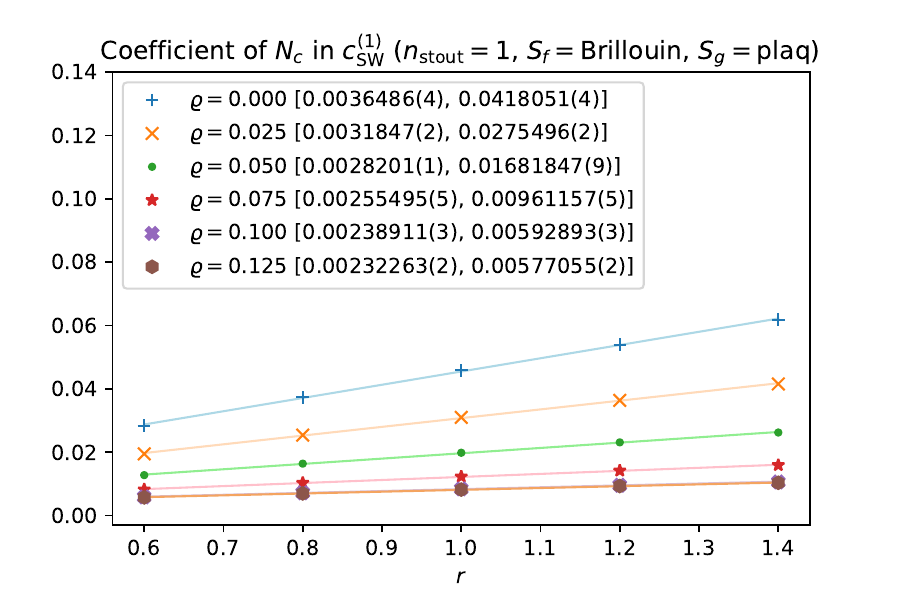}
\includegraphics[scale=0.48]{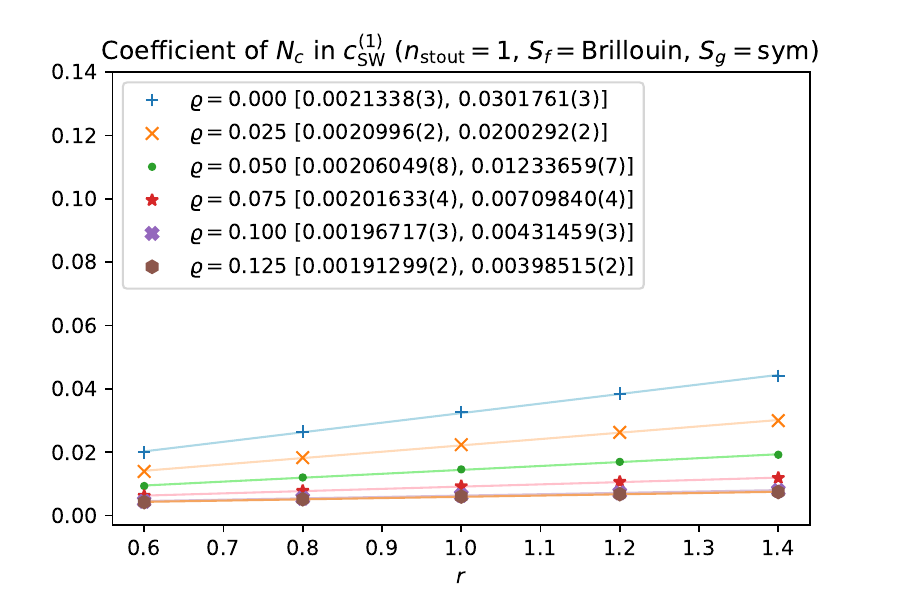}\\
\includegraphics[scale=0.48]{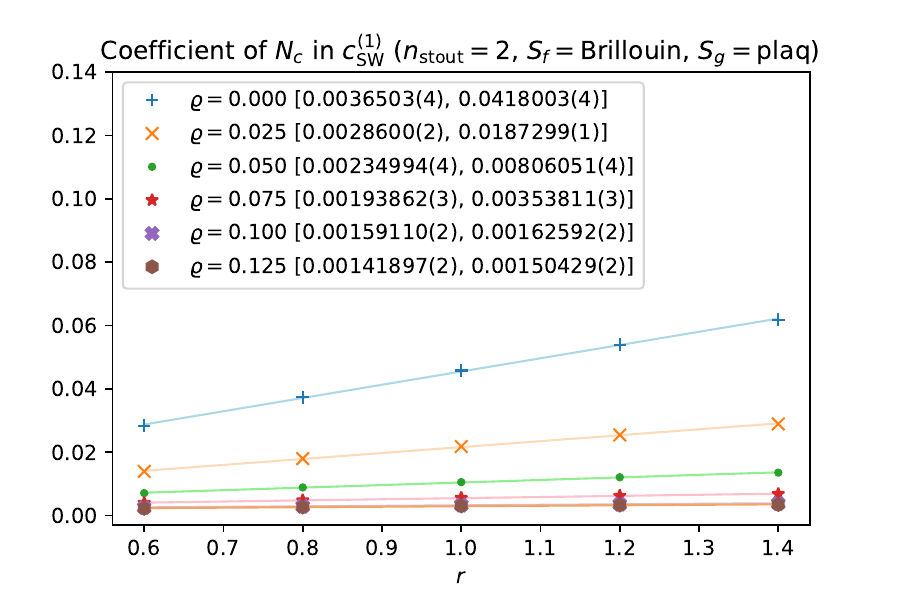}
\includegraphics[scale=0.48]{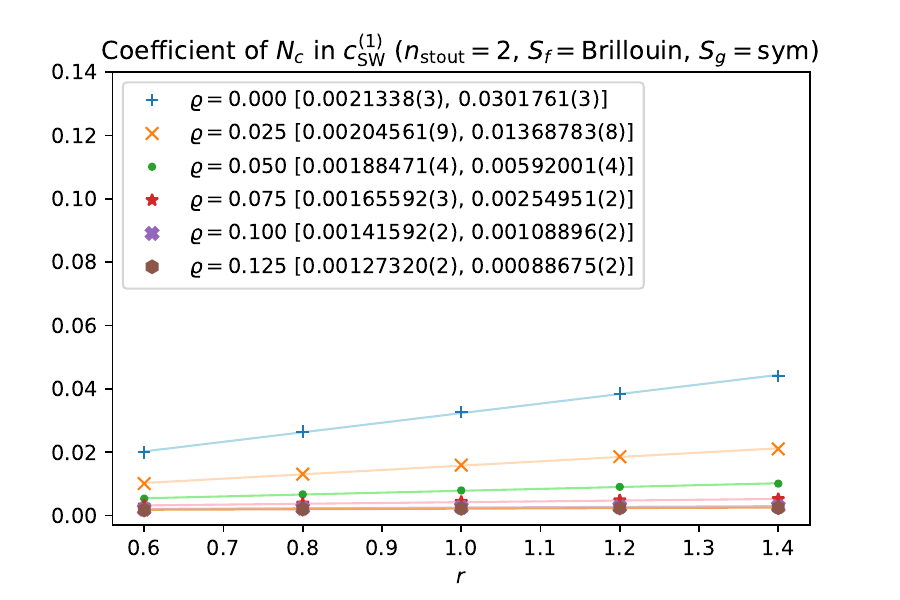}\\
\includegraphics[scale=0.48]{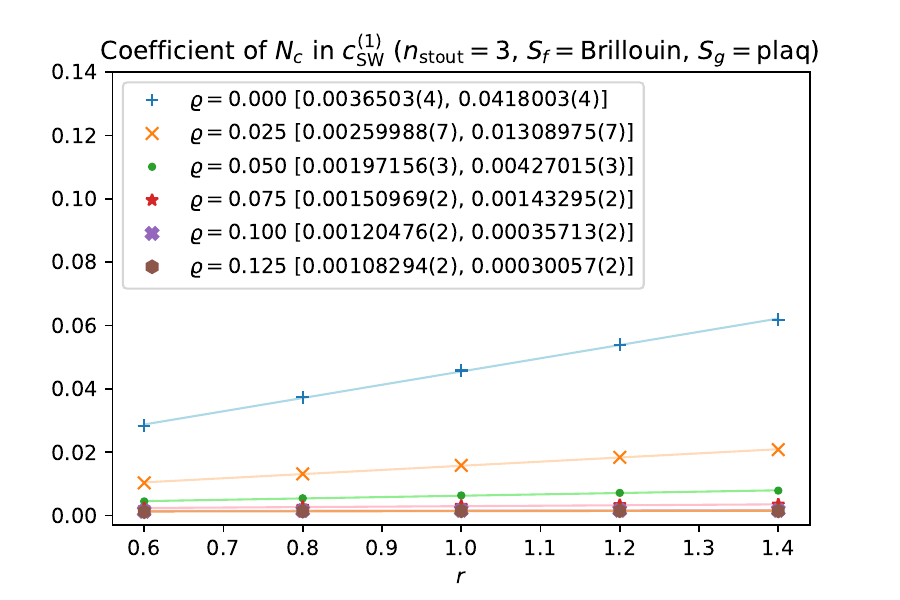}
\includegraphics[scale=0.48]{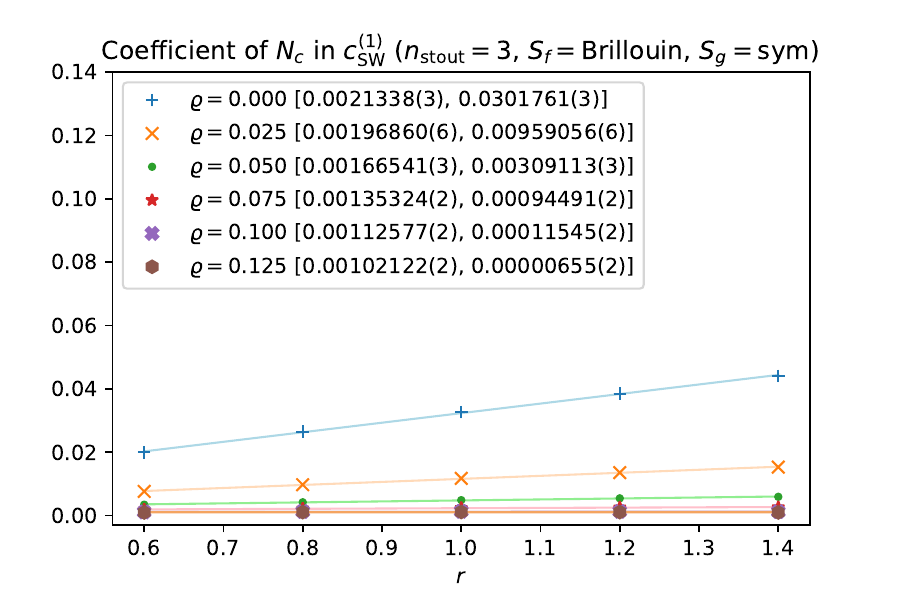}\\
\includegraphics[scale=0.48]{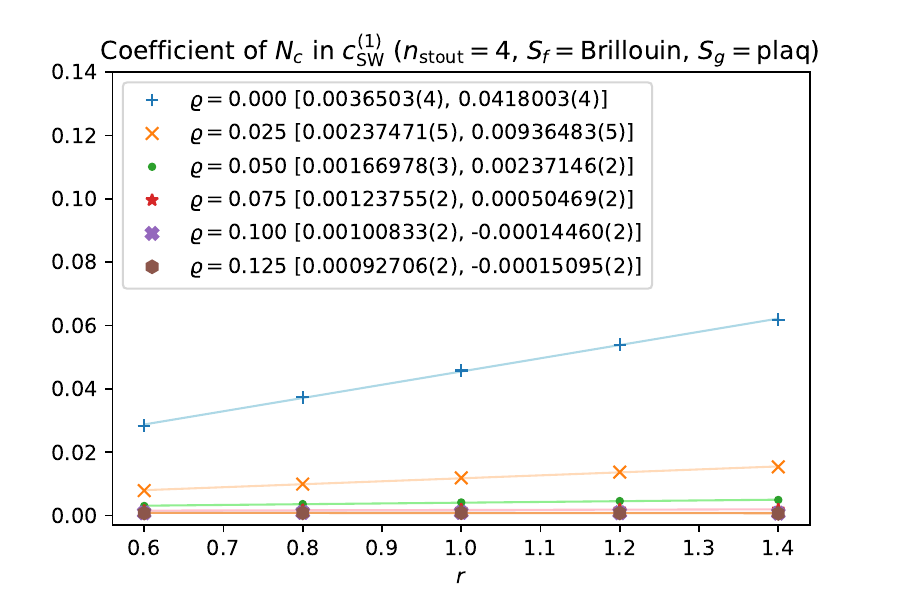}
\hspace*{7.35cm}\
\caption{Same as in Figure~\ref{fig:csw_of_r_bri_stout} but for the part of $\csw^{(1)}$ linear in $N_c$, or
the same as Figure~\ref{app_fig:csw_nc_of_r_wil_stout} but for Brillouin fermions.
\label{app_fig:csw_nc_of_r_bri_stout}}
\end{figure}

\begin{figure}[!htb]
\centering
\includegraphics[scale=0.48]{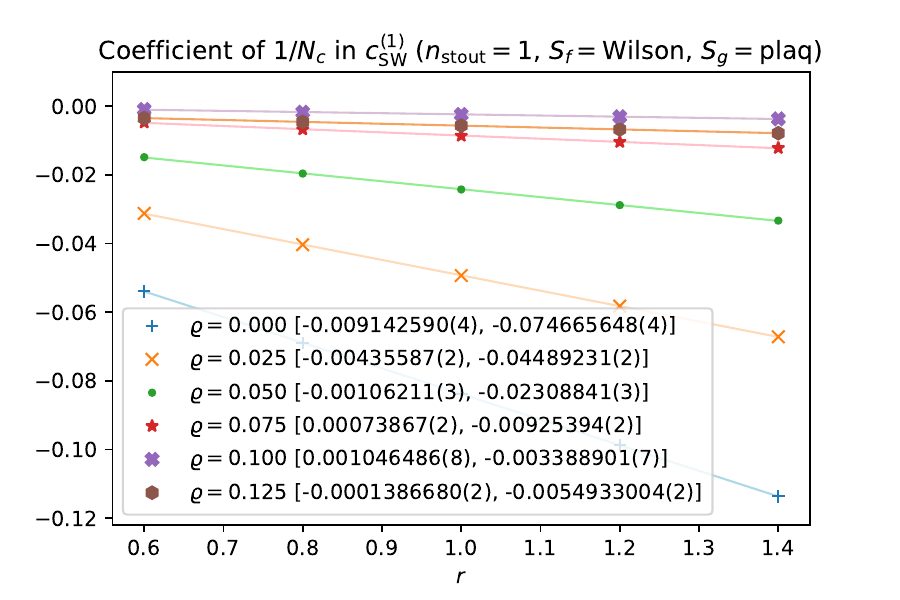}
\includegraphics[scale=0.48]{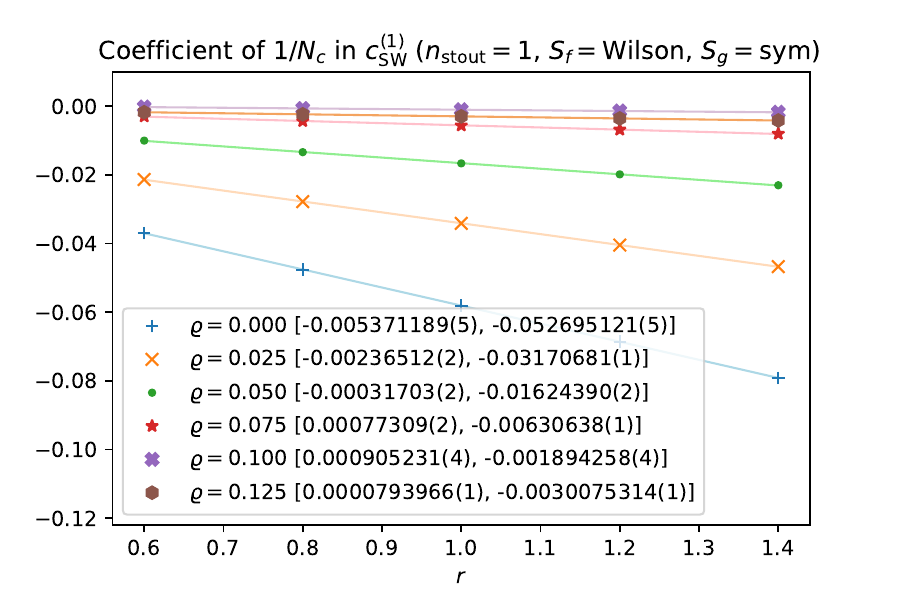}\\
\includegraphics[scale=0.48]{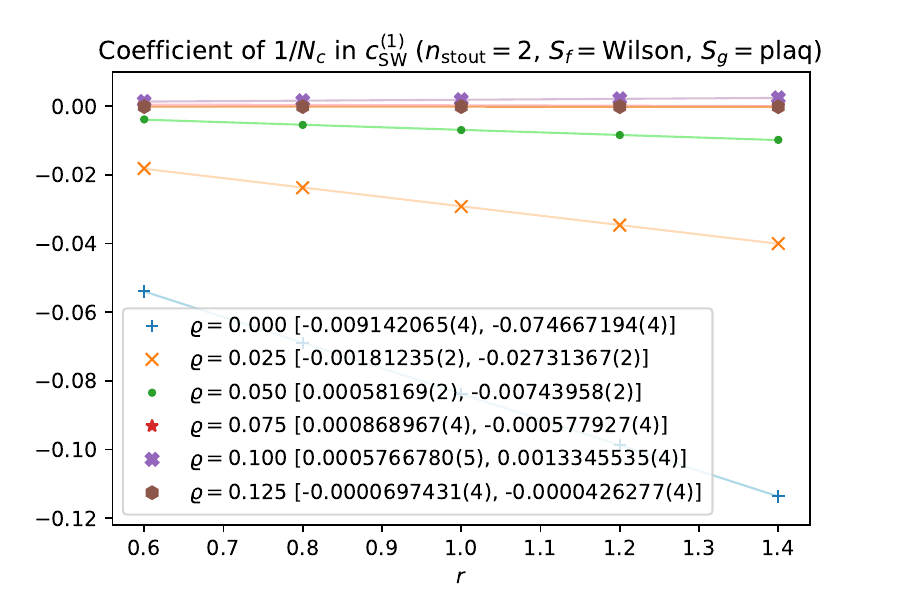}
\includegraphics[scale=0.48]{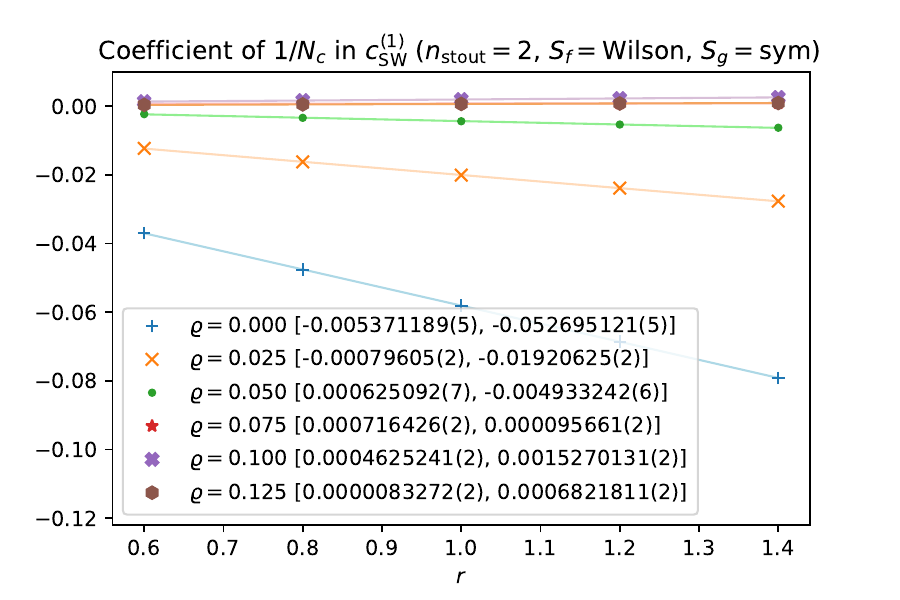}\\
\includegraphics[scale=0.48]{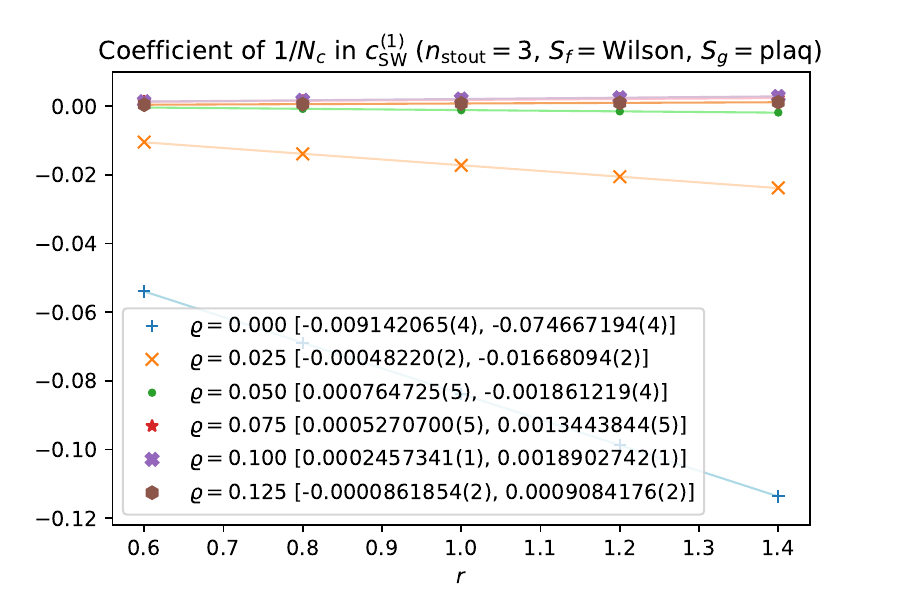}
\includegraphics[scale=0.48]{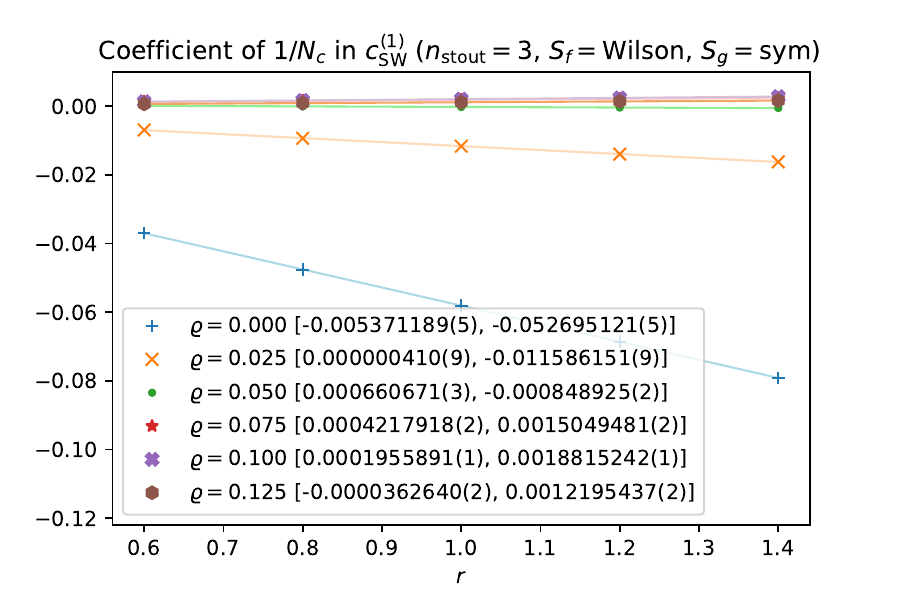}\\
\includegraphics[scale=0.48]{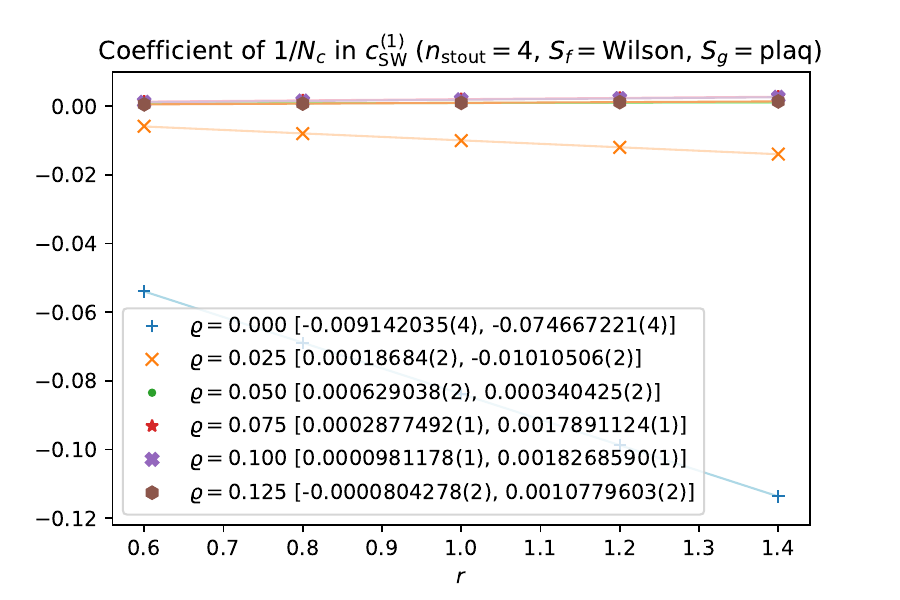}
\includegraphics[scale=0.48]{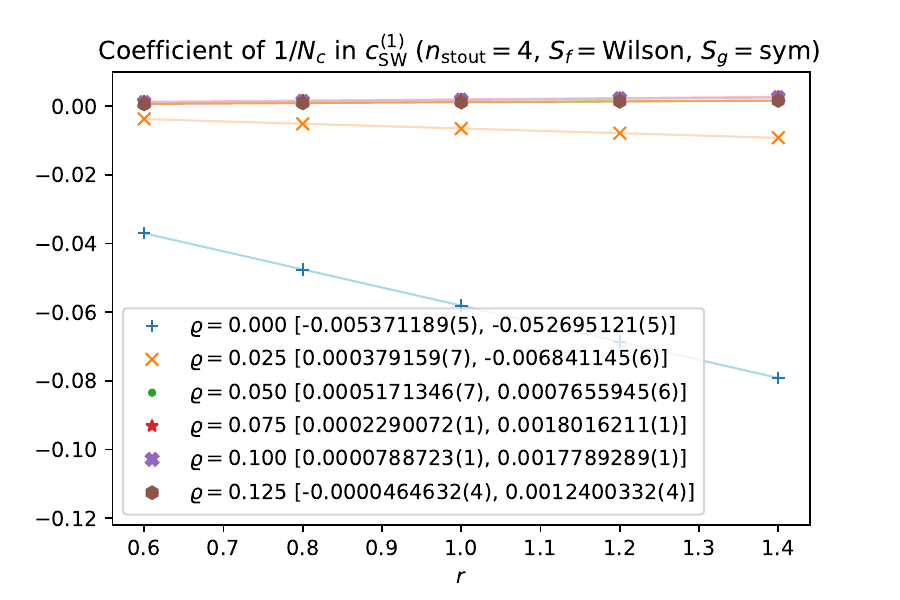}
\caption{Same as in Figure~\ref{fig:csw_of_r_wil_stout} but for the part of $\csw^{(1)}$ linear in $1/N_c$.
\label{app_fig:csw_ncinv_of_r_wil_stout}}
\end{figure}

\begin{figure}[!htb]
\centering
\includegraphics[scale=0.48]{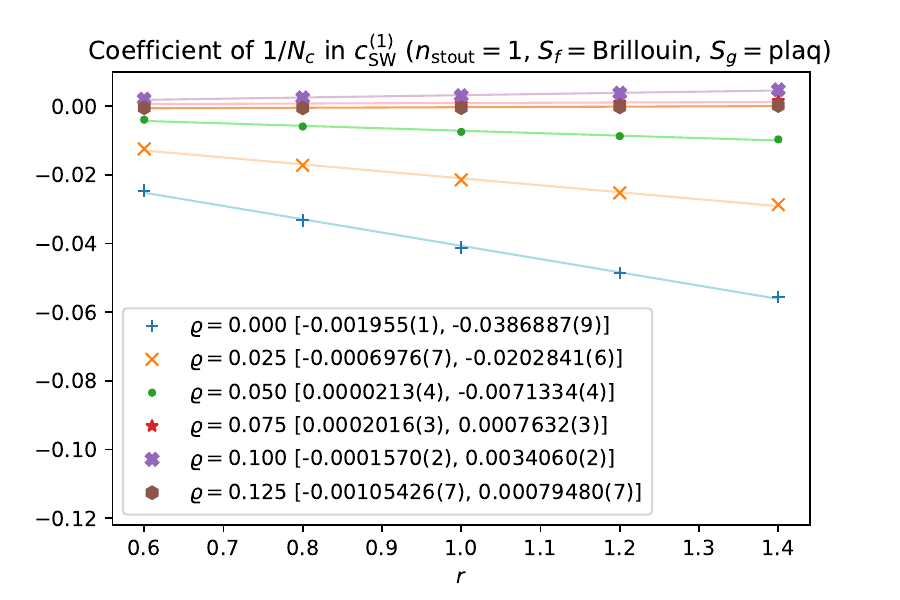}
\includegraphics[scale=0.48]{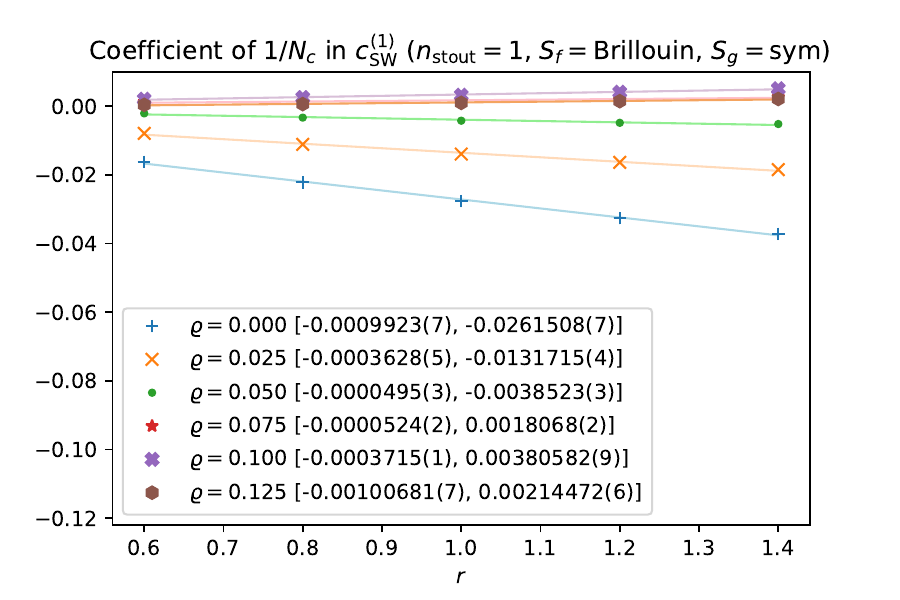}\\
\includegraphics[scale=0.48]{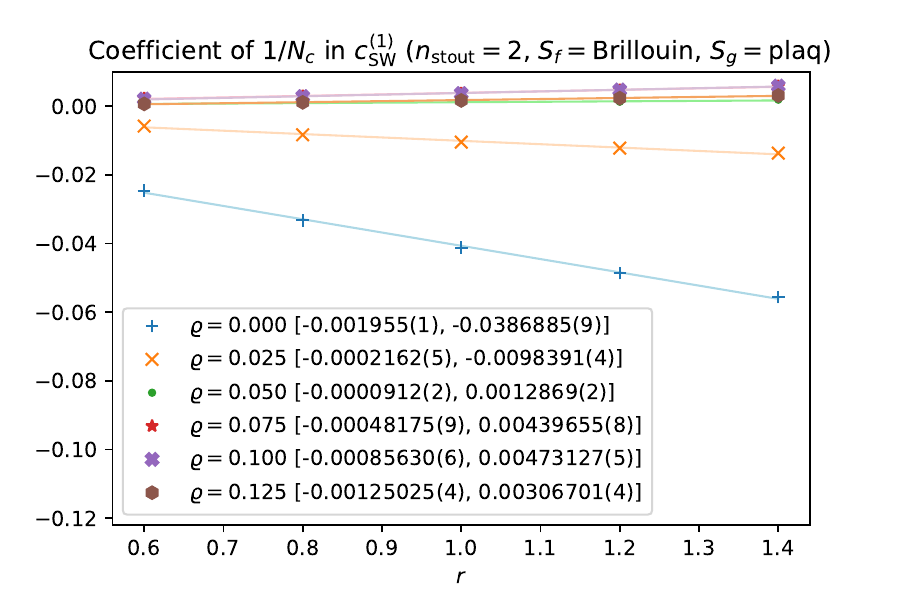}
\includegraphics[scale=0.48]{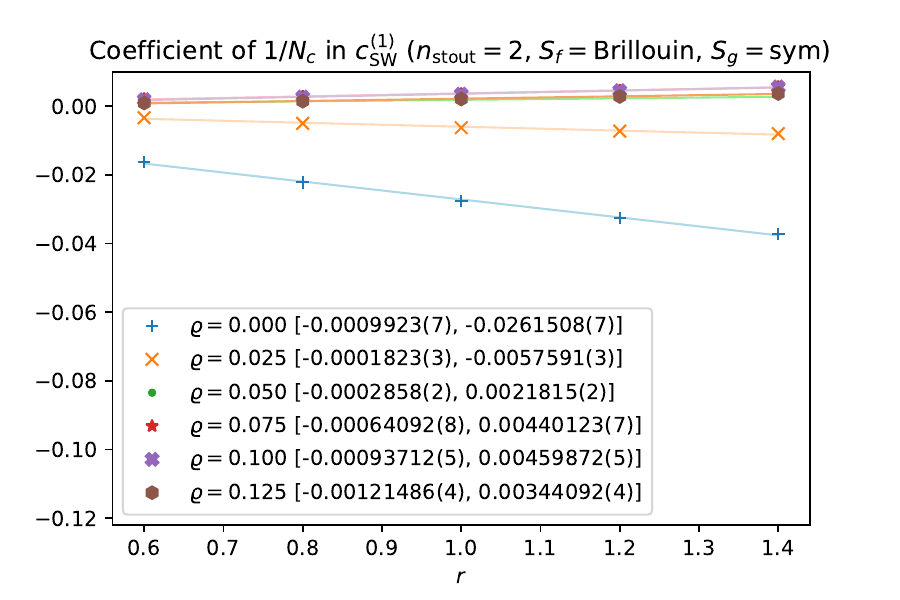}\\
\includegraphics[scale=0.48]{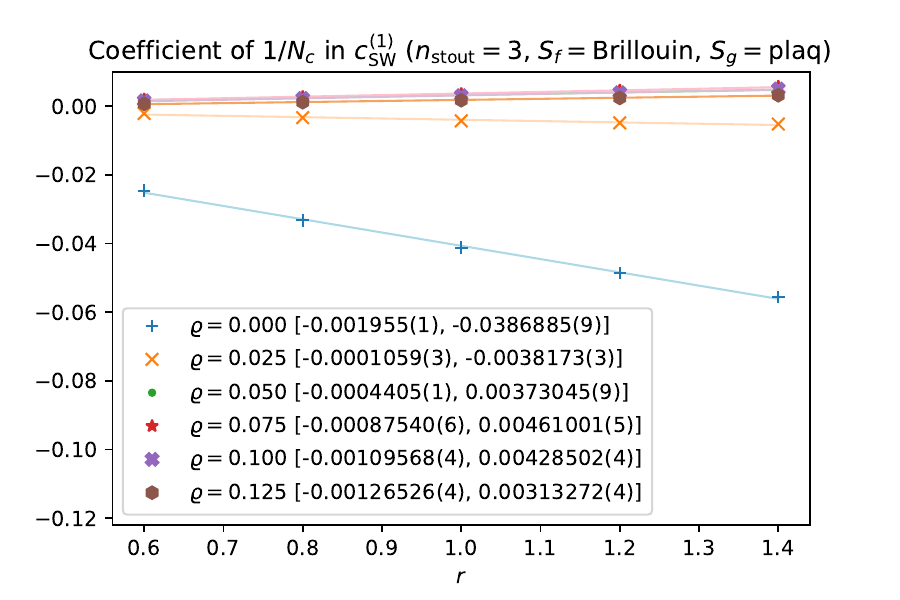}
\includegraphics[scale=0.48]{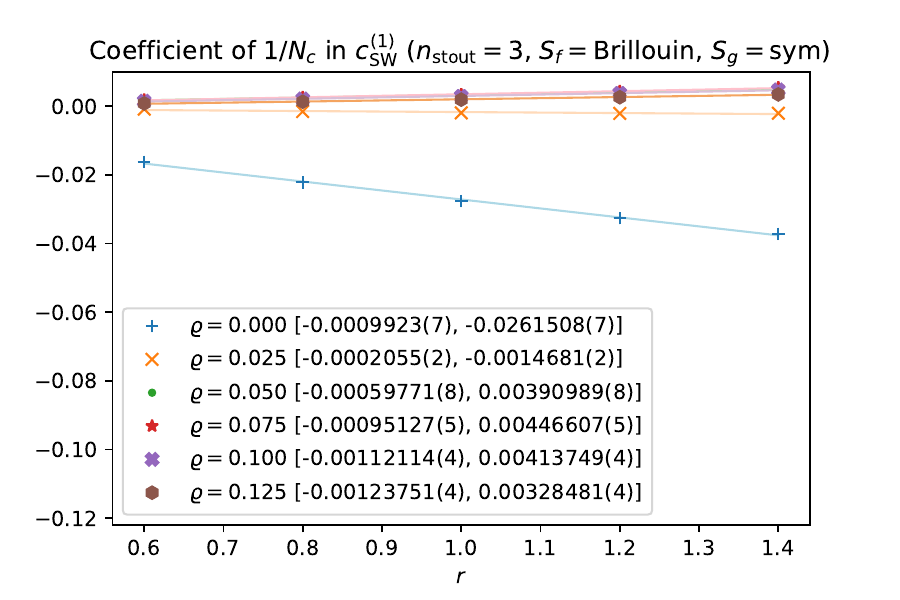}\\
\includegraphics[scale=0.48]{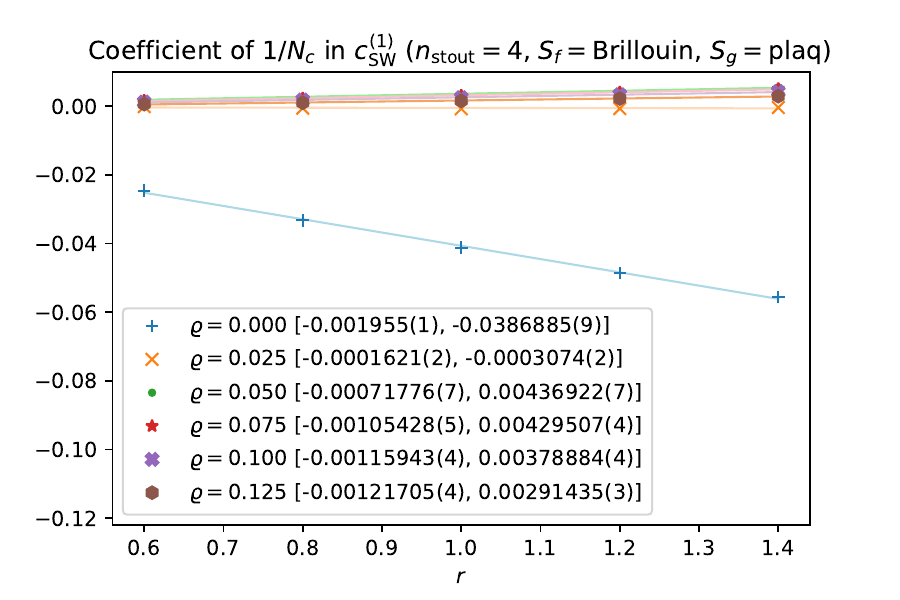}
\hspace*{7.35cm}\
\caption{Same as in Figure~\ref{fig:csw_of_r_bri_stout} but for the part of $\csw^{(1)}$ linear in $1/N_c$, or
the same as Figure~\ref{app_fig:csw_ncinv_of_r_wil_stout} but for Brillouin fermions.
\label{app_fig:csw_ncinv_of_r_bri_stout}}
\end{figure}

In this section we give some more data for $\csw^{(1)}$ with stout smearing.
Table~\ref{app_tab:csw_stout} gives the results for $\csw^{(1)}$ with $N_c=3$ in terms of the coefficients in front of various powers of the smearing parameter $\varrho$.
For different values of $N_c$ our Tables~\ref{app_tab:csw_nc_stout} and \ref{app_tab:csw_ncinv_stout} give the coefficients of $N_c$ and $1/N_c$, respectively.
Figures ~\ref{app_fig:csw_nc_of_r_wil_stout} and \ref{app_fig:csw_ncinv_of_r_wil_stout} display the $r$-dependence of the coefficients for Wilson fermions.
Figures ~\ref{app_fig:csw_nc_of_r_bri_stout} and \ref{app_fig:csw_ncinv_of_r_bri_stout} serve the same purpose for Brillouin fermions.

\begin{table}[!htb]
\centering
\begin{tabular}{|c|c|c|c|c|c|}
\hline
& Action &  $n=1$ & $n=2$ & $n=3$ & $n=4$ \\
\hline
\multirow{4}{*}{$\varrho^0$} & Wil./plaq & 0.2685882(5) & 0.26859(3) & 0.26859(3) & 0.26859(1)\\
& Wil./sym & 0.19624(3) & 0.19624(3) & 0.19624(3) & 0.19624(3)\\
& Bri./plaq & 0.1236257(8) & 0.12362(2) & 0.12362(2) & 0.12362(2)\\
& Bri./sym & 0.08859(3) & 0.08859(3) &  & \\
\hline
\multirow{4}{*}{$\varrho^1$} & Wil./plaq & -3.634574(8) & -7.2691(5) & -10.9037(2) & -14.5383(8)\\
& Wil./sym & -2.5330(3) & -5.0659(5) & -7.5989(3) & -10.132(1)\\
& Bri./plaq & -1.69335(1) & -3.3867(3) & -5.0801(3) & -6.7734(6)\\
& Bri./sym & -1.1693(2) & -2.3387(3) &  & \\
\hline
\multirow{4}{*}{$\varrho^2$} & Wil./plaq & 15.312641(2) & 91.8758(4) & 229.6894(8) & 428.7537(5)\\
& Wil./sym & 10.40969(4) & 62.4581(4) & 156.1454(9) & 291.471(2)\\
& Bri./plaq & 7.194783(3) & 43.1685(5) & 107.921(2) & 201.453(2)\\
& Bri./sym & 4.85384(6) & 29.1231(5) &  & \\
\hline
\multirow{4}{*}{$\varrho^3$} & Wil./plaq &  & -584.2124(2) & -2921.0621(7) & -8178.974(2)\\
& Wil./sym &  & -390.84777(8) & -1954.2388(5) & -5471.869(2)\\
& Bri./plaq &  & -279.5358(4) & -1397.679(2) & -3913.501(3)\\
& Bri./sym &  & -185.6528(3) &  & \\
\hline
\multirow{4}{*}{$\varrho^4$} & Wil./plaq &  & 1512.550(1) & 22688.25(2) & 105878.50(5)\\
& Wil./sym &  & 1000.402(2) & 15006.04(2) & 70028.17(7)\\
& Bri./plaq &  & 740.3087(9) & 11104.63(2) & 51821.62(7)\\
& Bri./sym &  & 486.2348(6) &  & \\
\hline
\multirow{4}{*}{$\varrho^5$} & Wil./plaq &  &  & -99737.99(7) & -930887.4(8)\\
& Wil./sym &  &  & -65387.62(4) & -610283.8(9)\\
& Bri./plaq &  &  & -49950.77(3) & -466206.5(9)\\
& Bri./sym &  &  &  & \\
\hline
\multirow{4}{*}{$\varrho^6$} & Wil./plaq &  &  & 191197.5(4) & 5353537(6)\\
& Wil./sym &  &  & 124452.9(3) & 3484681(8)\\
& Bri./plaq &  &  & 97831.9(2) & 2739293(4)\\
& Bri./sym &  &  &  & \\
\hline
\multirow{4}{*}{$\varrho^7$} & Wil./plaq &  &  &  & -18245358(47)\\
& Wil./sym &  &  &  & -11804589(26)\\
& Bri./plaq &  &  &  & -9518117(43)\\
& Bri./sym &  &  &  & \\
\hline
\multirow{4}{*}{$\varrho^8$} & Wil./plaq &  &  &  & 28031442(48)\\
& Wil./sym &  &  &  & 18041502(41)\\
& Bri./plaq &  &  &  & 14877196(7271)\\
& Bri./sym &  &  &  & \\
\hline
\end{tabular}
\caption{Coefficients of powers of $\varrho$ in $\csw^{(1)}$ for Wilson and Brillouin clover fermions with $N_c=3$.
\label{app_tab:csw_stout}}
\end{table}

\begin{table}[!htb]
\centering
\begin{tabular}{|c|c|c|c|c|c|}
\hline
& Action &  $n=1$ & $n=2$ & $n=3$ & $n=4$ \\
\hline
\multirow{4}{*}{$\varrho^0$} & Wil./plaq & 0.0988425(2) & 0.098843(9) & 0.098843(9) & 0.098842(3)\\
& Wil./sym & 0.071869(8) & 0.071869(8) & 0.071869(8) & 0.071869(8)\\
& Bri./plaq & 0.0457855(3) & 0.045782(6) & 0.045782(6) & 0.045782(6)\\
& Bri./sym & 0.032596(7) & 0.032596(7) &  & \\
\hline
\multirow{4}{*}{$\varrho^1$} & Wil./plaq & -1.385889(3) & -2.7718(2) & -4.15766(5) & -5.5436(3)\\
& Wil./sym & -0.96522(8) & -1.9304(2) & -2.89567(9) & -3.8609(4)\\
& Bri./plaq & -0.665239(4) & -1.3305(1) & -1.99572(8) & -2.6610(2)\\
& Bri./sym & -0.45953(5) & -0.91906(8) &  & \\
\hline
\multirow{4}{*}{$\varrho^2$} & Wil./plaq & 5.9430979(4) & 35.6586(1) & 89.1464(3) & 166.4067(2)\\
& Wil./sym & 4.04484(2) & 24.2690(2) & 60.6726(3) & 113.2555(6)\\
& Bri./plaq & 2.9145387(9) & 17.4872(2) & 43.7179(4) & 81.6068(7)\\
& Bri./sym & 1.97257(2) & 11.8355(2) &  & \\
\hline
\multirow{4}{*}{$\varrho^3$} & Wil./plaq &  & -228.81860(5) & -1144.0929(3) & -3203.4604(6)\\
& Wil./sym &  & -153.37163(2) & -766.8582(1) & -2147.2028(4)\\
& Bri./plaq &  & -114.9556(1) & -574.7778(5) & -1609.3777(7)\\
& Bri./sym &  & -76.68337(8) &  & \\
\hline
\multirow{4}{*}{$\varrho^4$} & Wil./plaq &  & 595.1488(4) & 8927.232(5) & 41660.43(2)\\
& Wil./sym &  & 394.4523(5) & 5916.784(7) & 27611.66(3)\\
& Bri./plaq &  & 306.4193(3) & 4596.289(5) & 21449.35(3)\\
& Bri./sym &  & 202.1925(2) &  & \\
\hline
\multirow{4}{*}{$\varrho^5$} & Wil./plaq &  &  & -39331.71(3) & -367095.8(3)\\
& Wil./sym &  &  & -25837.98(2) & -241154.3(3)\\
& Bri./plaq &  &  & -20718.029(9) & -193368.0(3)\\
& Bri./sym &  &  &  & \\
\hline
\multirow{4}{*}{$\varrho^6$} & Wil./plaq &  &  & 75472.9(2) & 2113244(2)\\
& Wil./sym &  &  & 49217.98(9) & 1378103(3)\\
& Bri./plaq &  &  & 40574.85(6) & 1136096(2)\\
& Bri./sym &  &  &  & \\
\hline
\multirow{4}{*}{$\varrho^7$} & Wil./plaq &  &  &  & -7204527(15)\\
& Wil./sym &  &  &  & -4669018(8)\\
& Bri./plaq &  &  &  & -3943331(12)\\
& Bri./sym &  &  &  & \\
\hline
\multirow{4}{*}{$\varrho^8$} & Wil./plaq &  &  &  & 11068717(16)\\
& Wil./sym &  &  &  & 7134378(12)\\
& Bri./plaq &  &  &  & 6154392(2424)\\
& Bri./sym &  &  &  & \\
\hline
\end{tabular}
\caption{Same as Table~\ref{app_tab:csw_stout} but for the part of $\csw^{(1)}$ linear in $N_c$.
\label{app_tab:csw_nc_stout}}
\end{table}

\begin{table}[!htb]
\centering
\begin{tabular}{|c|c|c|c|c|c|}
\hline
& Action &  $n=1$ & $n=2$ & $n=3$ & $n=4$ \\
\hline
\multirow{4}{*}{$\varrho^0$} & Wil./plaq & -0.0838175(2) & -0.08382(2) & -0.08382(2) & -0.08382(2)\\
& Wil./sym & -0.05809(2) & -0.05809(2) & -0.05809(2) & -0.05809(2)\\
& Bri./plaq & -0.04119223(6) & -0.04119(3) & -0.04119(3) & -0.04119(3)\\
& Bri./sym & -0.02760(3) & -0.02760(3) &  & \\
\hline
\multirow{4}{*}{$\varrho^1$} & Wil./plaq & 1.5692766(2) & 3.13856(7) & 4.7078(2) & 6.2771(2)\\
& Wil./sym & 1.08814(6) & 2.1763(2) & 3.2644(2) & 4.3526(3)\\
& Bri./plaq & 0.90710168(4) & 1.81420(3) & 2.72131(9) & 3.62840(6)\\
& Bri./sym & 0.62772(2) & 1.25545(3) &  & \\
\hline
\multirow{4}{*}{$\varrho^2$} & Wil./plaq & -7.549958(5) & -45.2998(2) & -113.2494(5) & -211.3989(3)\\
& Wil./sym & -5.17448(3) & -31.0469(2) & -77.6172(4) & -144.8854(6)\\
& Bri./plaq & -4.64650062(3) & -27.87899(2) & -69.69752(8) & -130.1020(2)\\
& Bri./sym & -3.191648(2) & -19.149881(7) &  & \\
\hline
\multirow{4}{*}{$\varrho^3$} & Wil./plaq &  & 306.73007(4) & 1533.6503(2) & 4294.2212(7)\\
& Wil./sym &  & 207.8014(2) & 1039.007(2) & 2909.219(3)\\
& Bri./plaq &  & 195.9927(3) & 979.963(1) & 2743.897(3)\\
& Bri./sym &  & 133.1918(2) &  & \\
\hline
\multirow{4}{*}{$\varrho^4$} & Wil./plaq &  & -818.6903(7) & -12280.35(1) & -57308.34(7)\\
& Wil./sym &  & -548.8632(5) & -8232.950(7) & -38420.428(8)\\
& Bri./plaq &  & -536.8471(2) & -8052.701(6) & -37579.30(2)\\
& Bri./sym &  & -361.0282(2) &  & \\
\hline
\multirow{4}{*}{$\varrho^5$} & Wil./plaq &  &  & 54771.40(3) & 511200(1)\\
& Wil./sym &  &  & 36379.01(3) & 339537.3(1)\\
& Bri./plaq &  &  & 36609.95(4) & 341692.9(4)\\
& Bri./sym &  &  &  & \\
\hline
\multirow{4}{*}{$\varrho^6$} & Wil./plaq &  &  & -105663.9(5) & -2958589(2)\\
& Wil./sym &  &  & -69603.1(2) & -1948888(4)\\
& Bri./plaq &  &  & -71678.0(2) & -2006983(5)\\
& Bri./sym &  &  &  & \\
\hline
\multirow{4}{*}{$\varrho^7$} & Wil./plaq &  &  &  & 10104664(41)\\
& Wil./sym &  &  &  & 6607391(43)\\
& Bri./plaq &  &  &  & 6935625(78)\\
& Bri./sym &  &  &  & \\
\hline
\multirow{4}{*}{$\varrho^8$} & Wil./plaq &  &  &  & -15524128(37)\\
& Wil./sym &  &  &  & -10084901(70)\\
& Bri./plaq &  &  &  & -10757942(118)\\
& Bri./sym &  &  &  & \\
\hline
\end{tabular}
\caption{Same as Table~\ref{app_tab:csw_stout} but for the part of $\csw^{(1)}$ linear in $1/N_c$.
\label{app_tab:csw_ncinv_stout}}
\end{table}

%%%%%%%%%%%%%%%%%%%%%%%%%%%%%
\section{Additional Data for $\mathbf{c_{SW}^{(1)}}$ with Wilson Flow\label{app:csw_wf}}
%%%%%%%%%%%%%%%%%%%%%%%%%%%%%

\begin{figure}[!htb]
\centering
\includegraphics[scale=0.48]{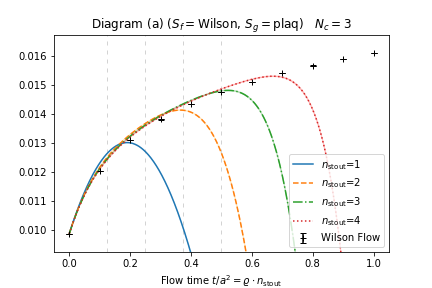}
\includegraphics[scale=0.48]{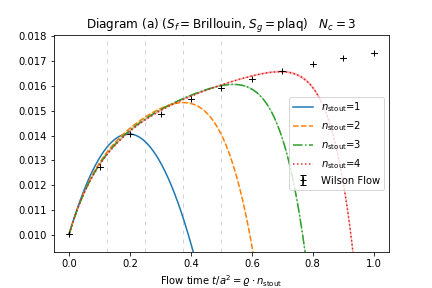}\\
\includegraphics[scale=0.48]{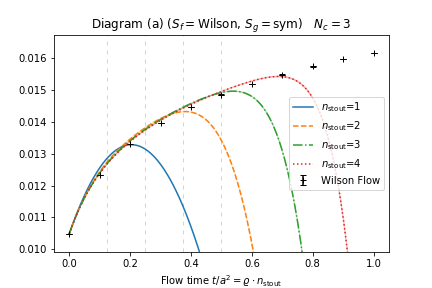}
\includegraphics[scale=0.48]{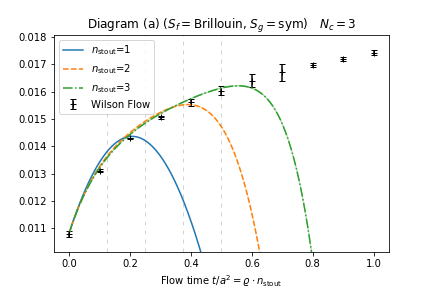}
\caption{The constant part of diagram (a) with both Wilson flow and up to four stout smearing steps.
Results are plotted against the flow time $t$ or $n_\mathrm{stout}\cdot\varrho$, respectively.
All four combinations of Wilson or Brillouin fermion action and plaquette or Symanzik glue are shown.
\label{app_fig:csw_a_flow}}
\end{figure}

\begin{figure}[!htb]
\centering
\includegraphics[scale=0.48]{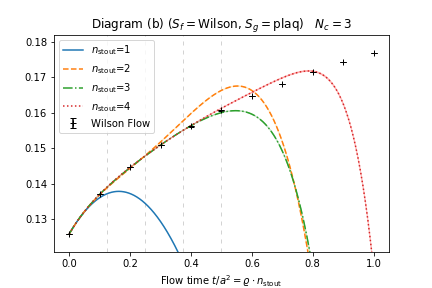}
\includegraphics[scale=0.48]{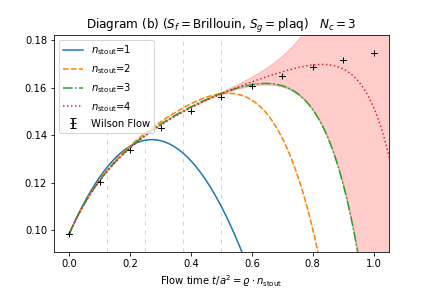}\\
\includegraphics[scale=0.48]{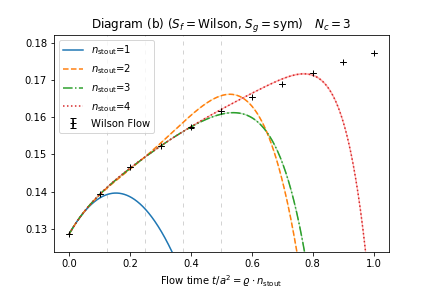}
\includegraphics[scale=0.48]{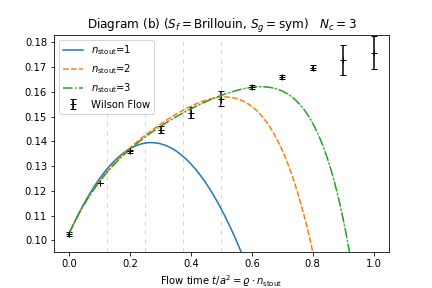}
\caption{The same as Figure~\ref{app_fig:csw_a_flow} but for diagram (b).
\label{app_fig:csw_b_flow}}
\end{figure}

\begin{figure}[!htb]
\centering
\includegraphics[scale=0.48]{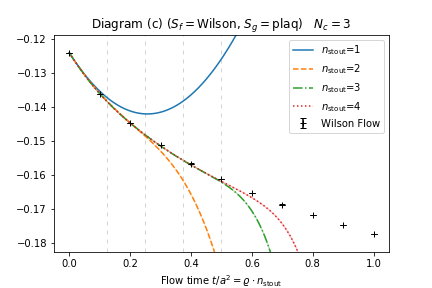}
\includegraphics[scale=0.48]{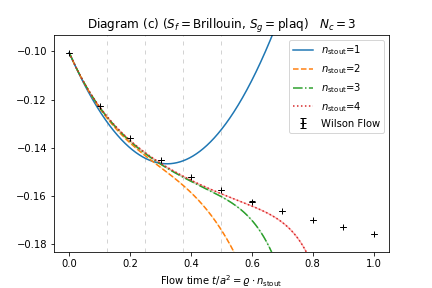}\\
\includegraphics[scale=0.48]{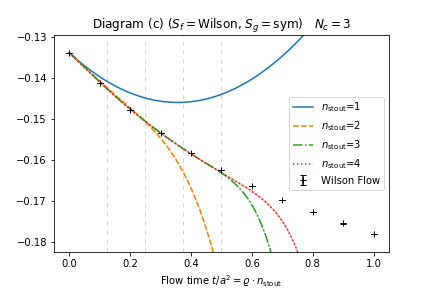}
\includegraphics[scale=0.48]{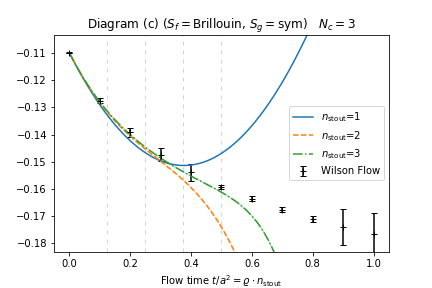}
\caption{The same as Figure~\ref{app_fig:csw_a_flow} but for diagram (c).
\label{app_fig:csw_c_flow}}
\end{figure}

\begin{figure}[!htb]
\centering
\includegraphics[scale=0.48]{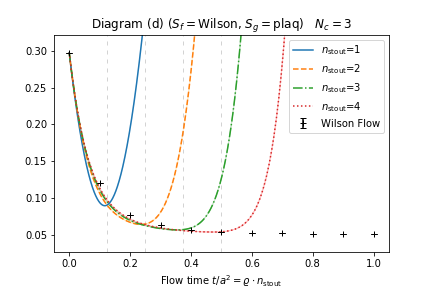}
\includegraphics[scale=0.48]{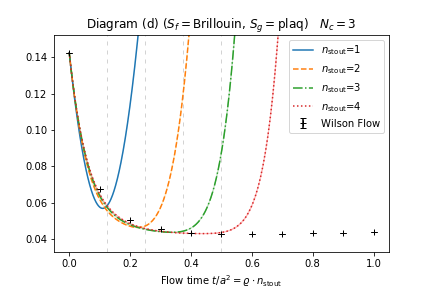}\\
\includegraphics[scale=0.48]{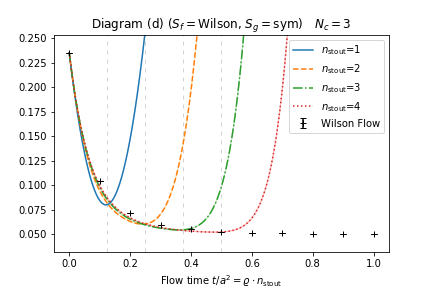}
\includegraphics[scale=0.48]{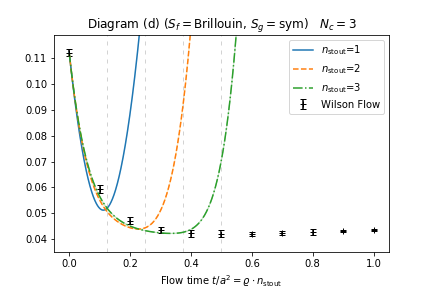}
\caption{The same as Figure~\ref{app_fig:csw_a_flow} but for diagram (d).
\label{app_fig:csw_d_flow}}
\end{figure}

\begin{figure}[!htb]
\centering
\includegraphics[scale=0.48]{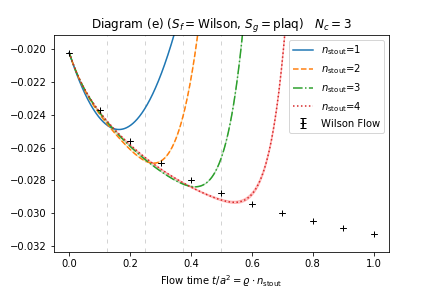}
\includegraphics[scale=0.48]{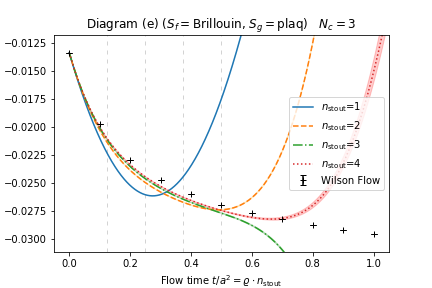}\\
\includegraphics[scale=0.48]{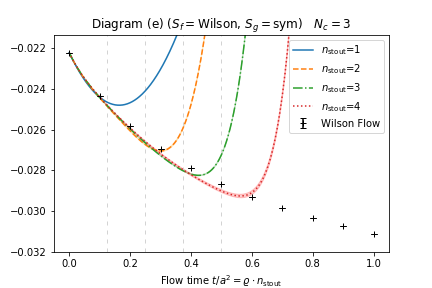}
\includegraphics[scale=0.48]{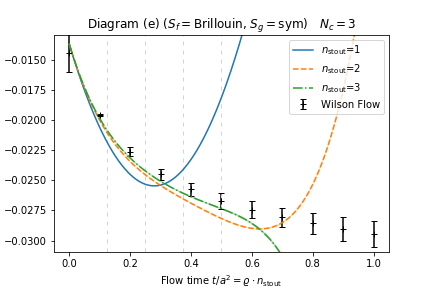}
\caption{The same as Figure~\ref{app_fig:csw_a_flow} but for diagram (e); diagram  (f) gives the same result.
\label{app_fig:csw_e_flow}}
\end{figure}

Here we show the individual results of the six diagrams that contribute to $\csw^{(1)}$ named (a) through (f), as depicted in Figure~\ref{fig:diagrams}.
Figures~\ref{app_fig:csw_a_flow}-\ref{app_fig:csw_e_flow} show the constant part of each diagram plotted against $t/a^2=n_\mathrm{stout}\cdot\varrho$ with stout smearing up to $n_\mathrm{stout}=4$ and flow time $t/a^2\leq1$.
The absolute values for the divergent diagrams [all except for diagram (d)] refer to our choice of regularization (see Ref.~\cite{Ammer:2023otl}).

\begin{figure}[!htb]
\centering
\includegraphics[scale=0.48]{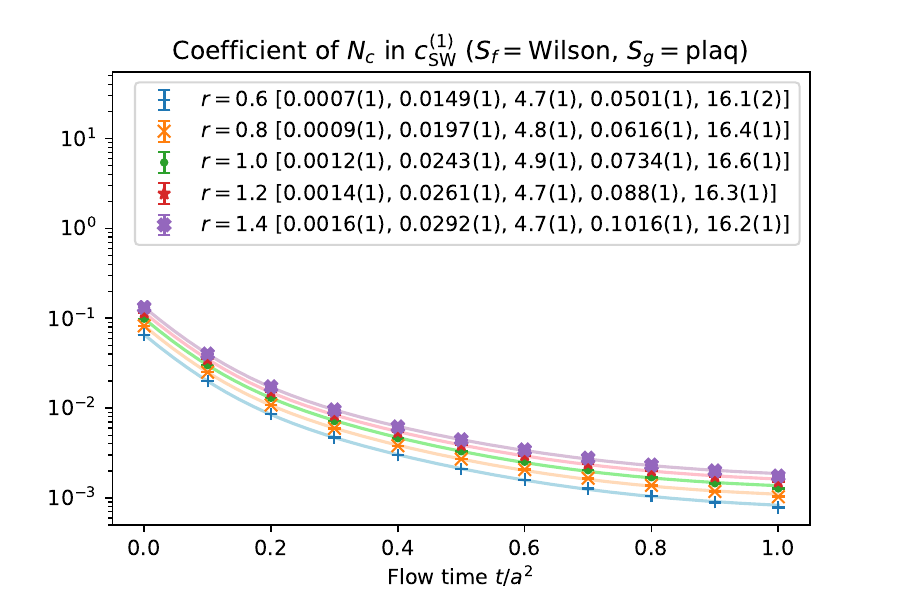}
\includegraphics[scale=0.48]{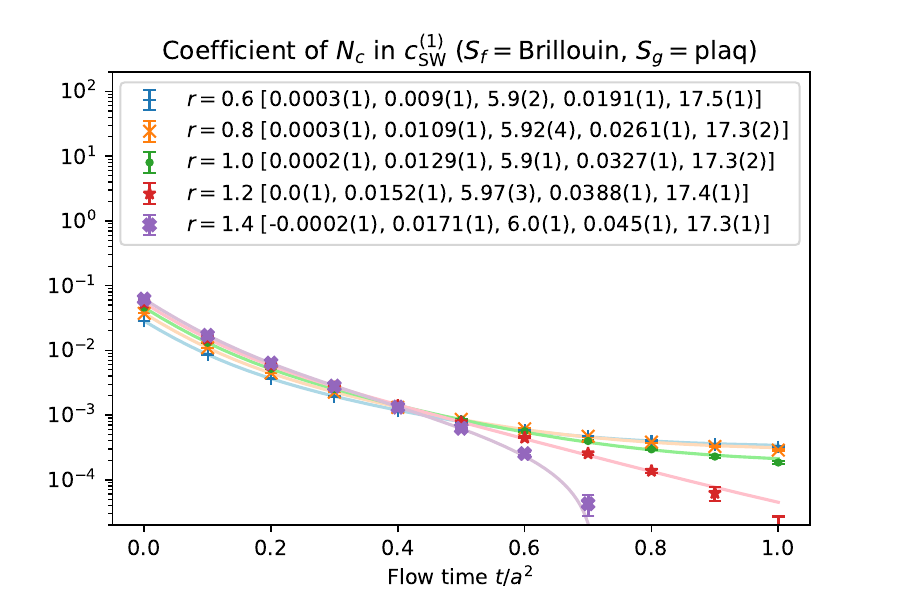}\\
\includegraphics[scale=0.48]{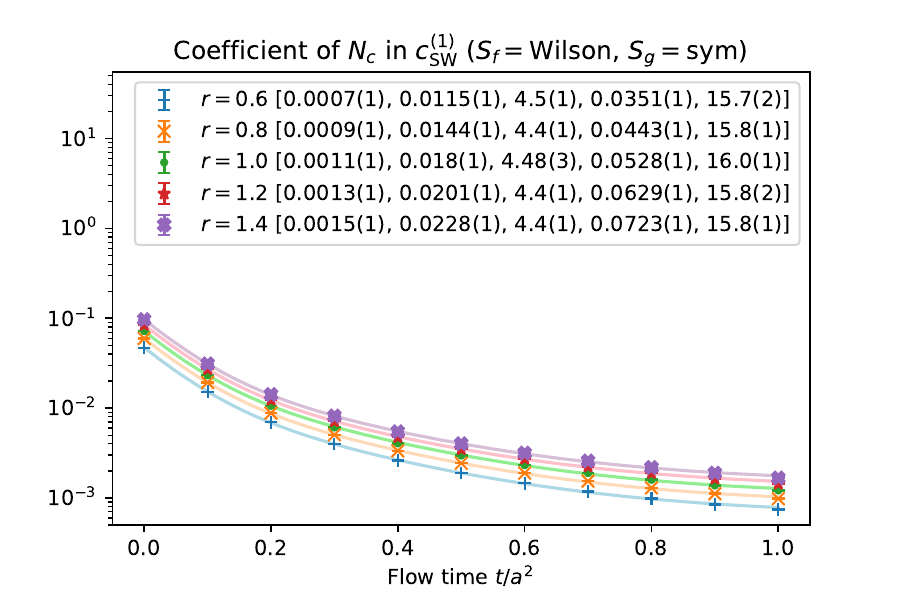}
\includegraphics[scale=0.48]{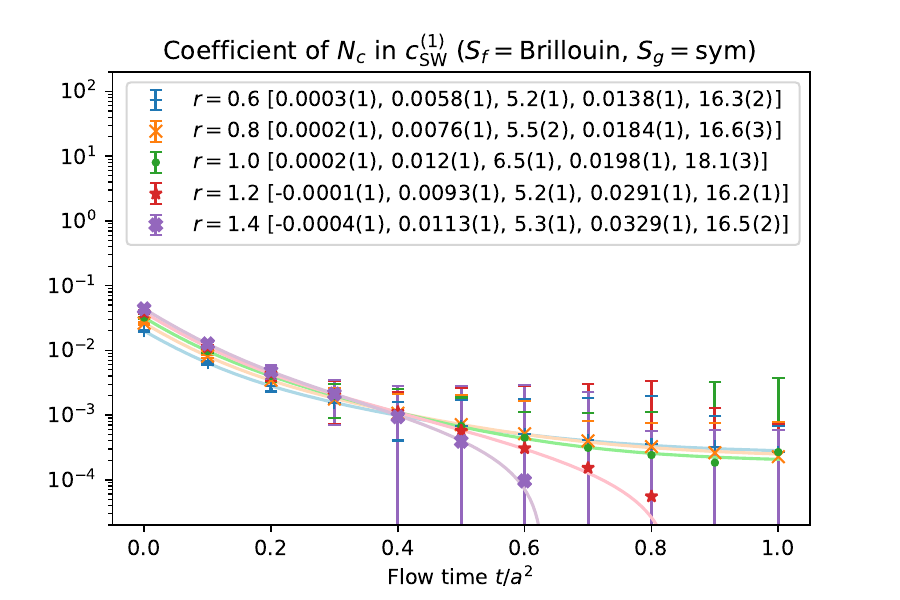}
\caption{Logarithmic plot of the coefficient of $N_c$ in $\csw^{(1)}$ with Wilson flow as a function of the flow time $t/a^2$ for five different values of $r$.
Two-exponential fits of the form $c_0+c_1e^{-c_2t}+c_3e^{-c_4t}$ are also shown and the coefficients $[c_0,c_1,c_2,c_3,c_4]$ are given in brackets.
\label{app_fig:csw_nc_flow_fits}}
\end{figure}

\begin{figure}[!htb]
\centering
\includegraphics[scale=0.48]{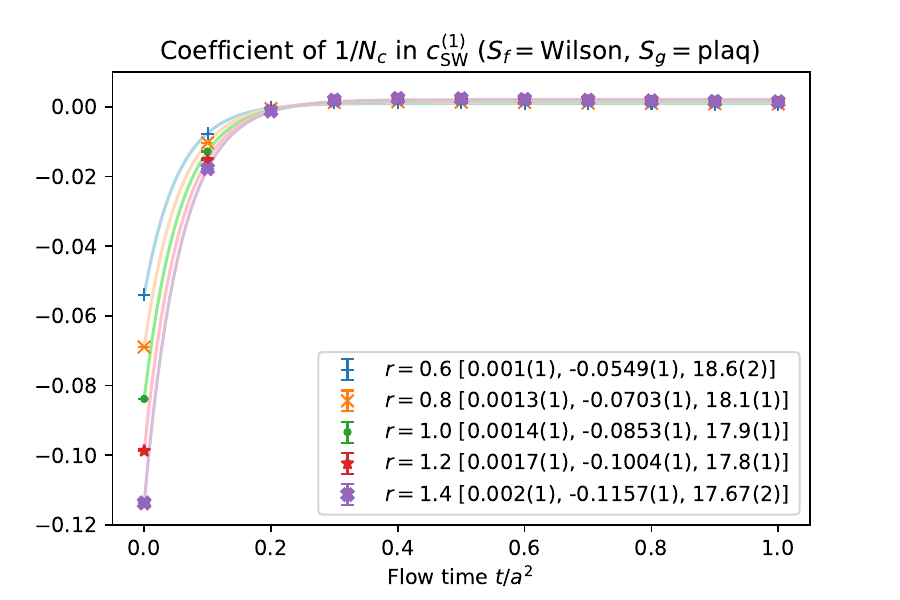}
\includegraphics[scale=0.48]{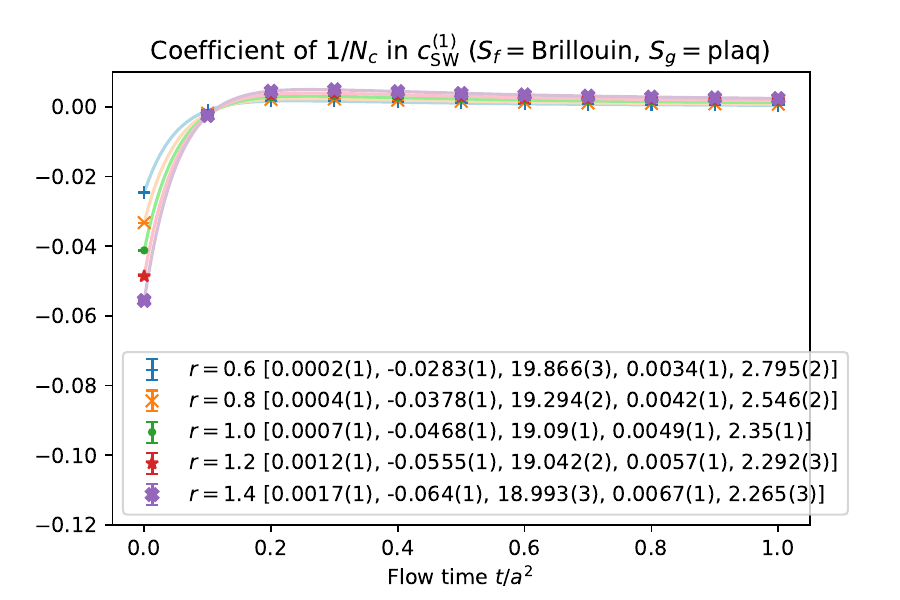}\\
\includegraphics[scale=0.48]{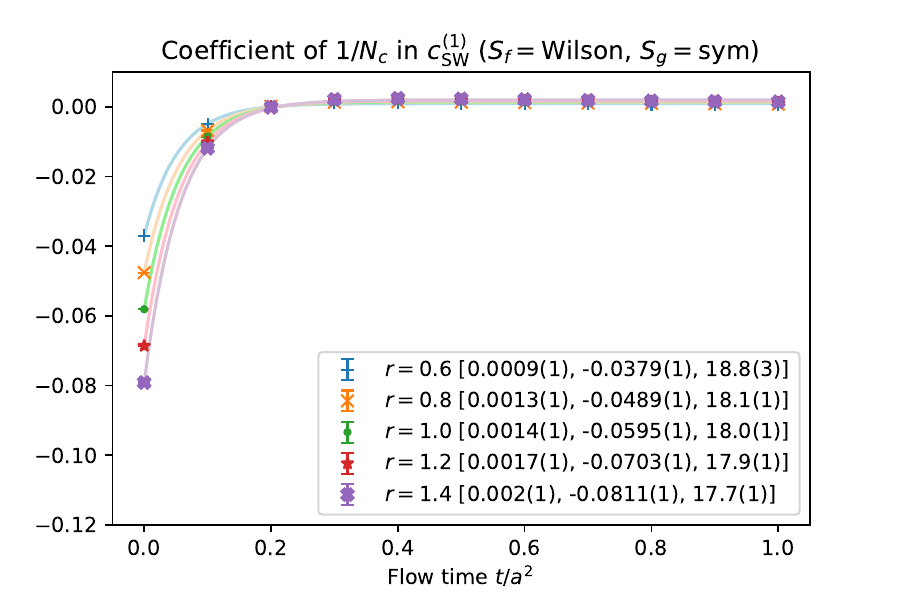}
\includegraphics[scale=0.48]{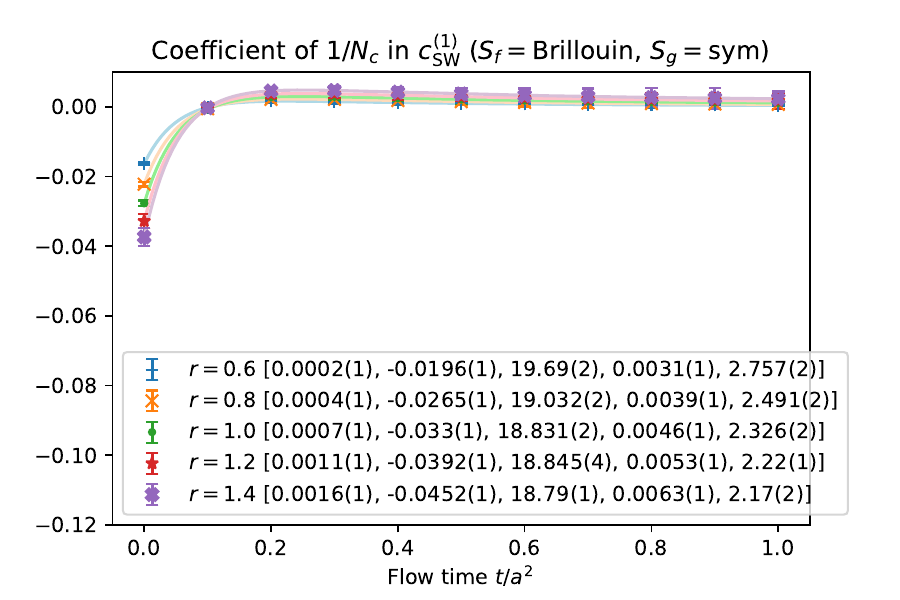}
\caption{The same as Figure~\ref{app_fig:csw_nc_flow_fits} but for the coefficient of $1/N_c$.
\label{app_fig:csw_ncinv_flow_fits}}
\end{figure}

\begin{figure}[!htb]
\centering
\includegraphics[scale=0.48]{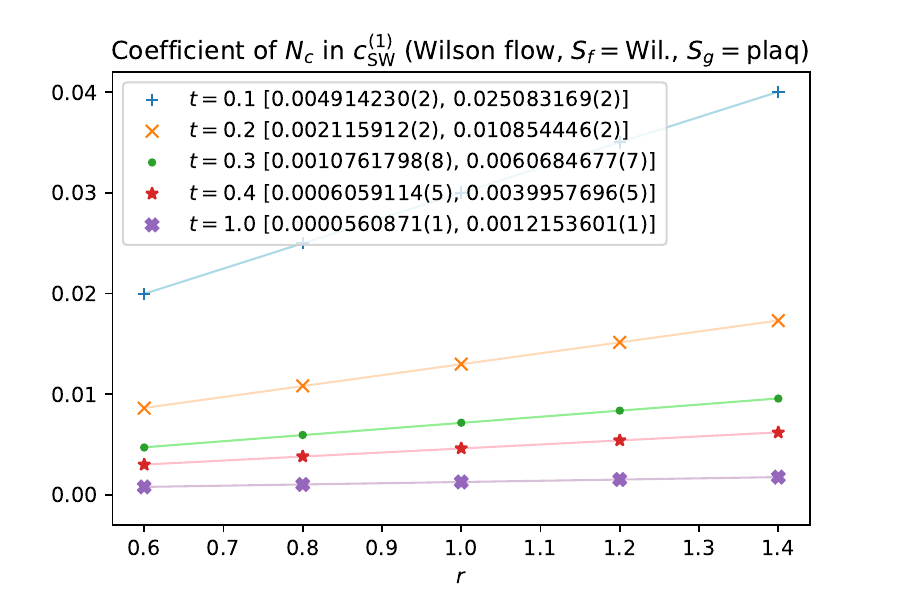}
\includegraphics[scale=0.48]{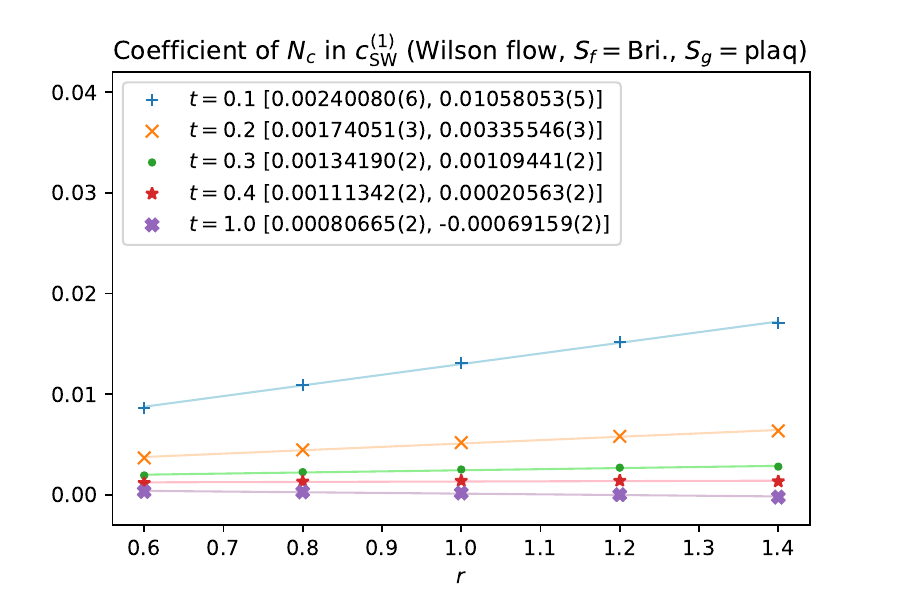}\\
\includegraphics[scale=0.48]{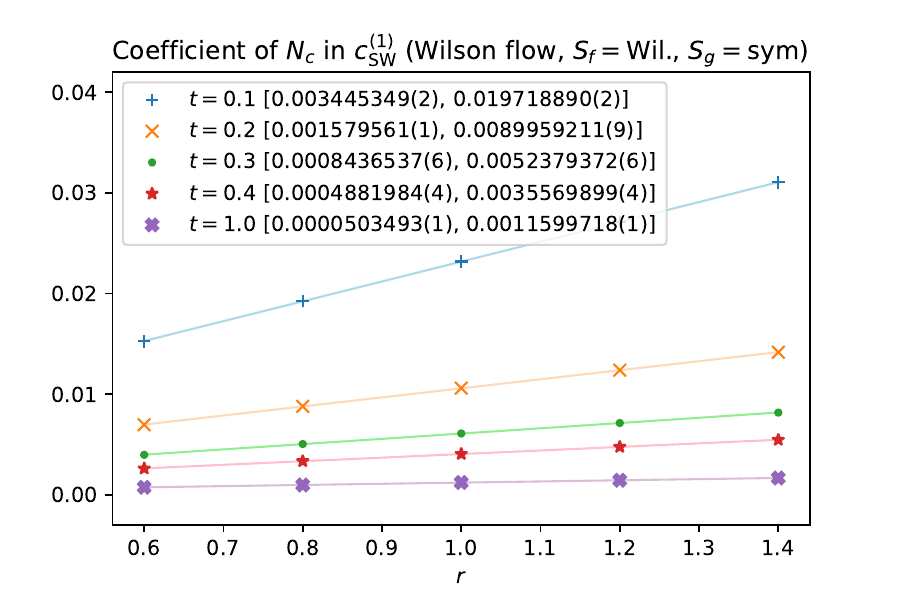}
\includegraphics[scale=0.48]{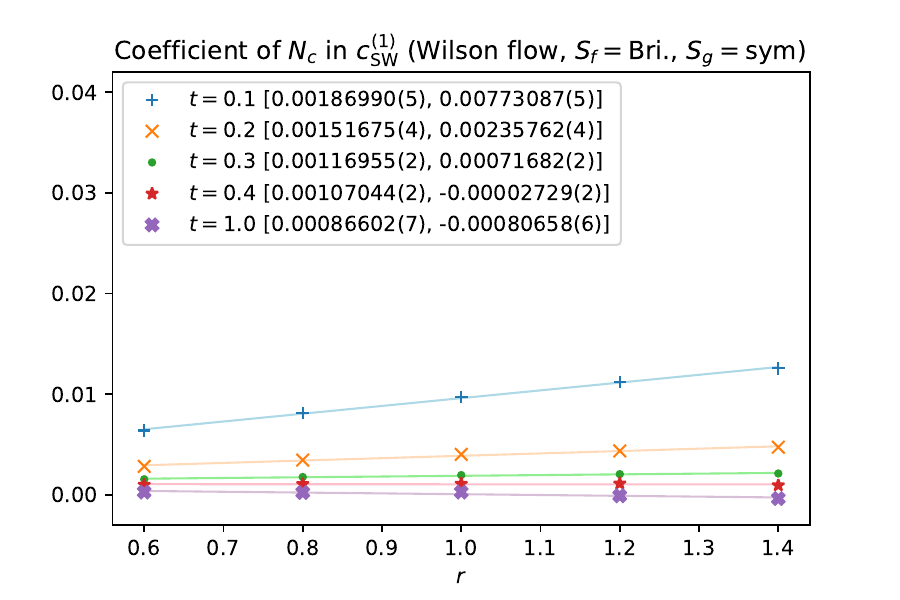}
\caption{The coefficient of $N_c$ in $\csw^{(1)}$ as a function of $r$ for a few values of the flow time $t/a^2$.
Linear least-squares fits of the form $c_0+c_1\cdot r$ are also shown and the coefficients $[c_0,c_1]$ are given in brackets.
\label{app_fig:csw_nc_of_r_flow}}
\end{figure}

\begin{figure}[!htb]
\centering
\includegraphics[scale=0.48]{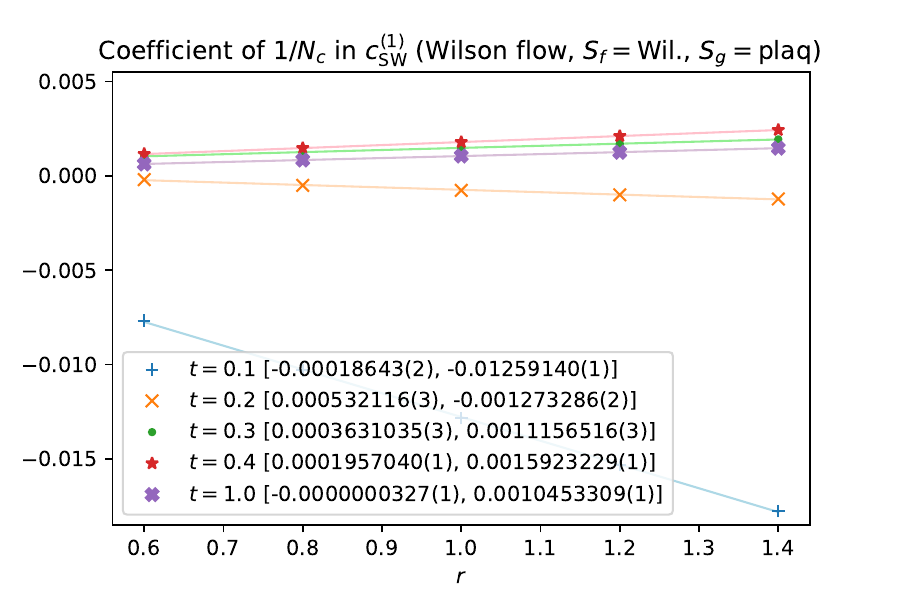}
\includegraphics[scale=0.48]{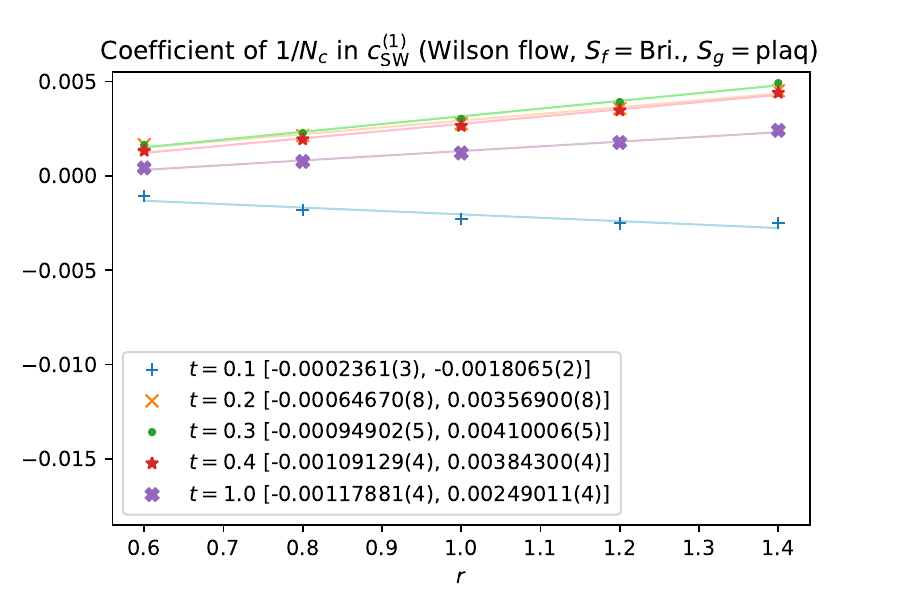}\\
\includegraphics[scale=0.48]{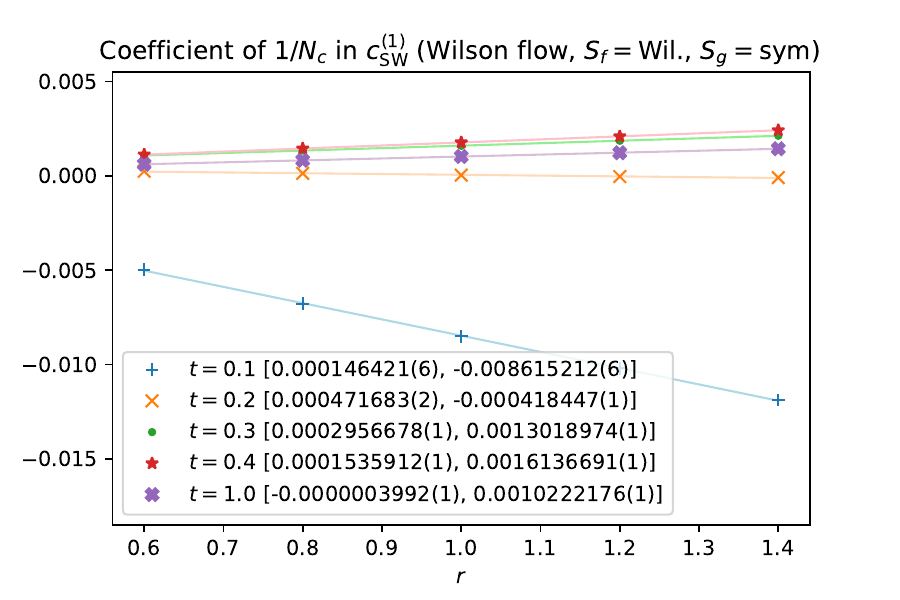}
\includegraphics[scale=0.48]{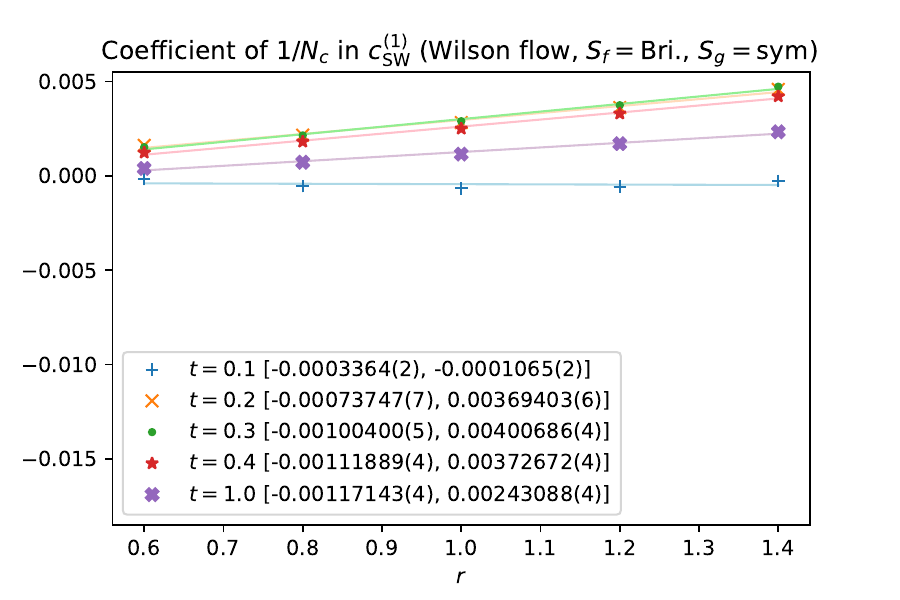}
\caption{The same as Figure~\ref{app_fig:csw_nc_of_r_flow} but for the coefficient of $1/N_c$.
\label{app_fig:csw_ncinv_of_r_flow}}
\end{figure}

Figures~\ref{app_fig:csw_nc_flow_fits} and \ref{app_fig:csw_ncinv_flow_fits} show the coefficients of $N_c$ and $1/N_c$, respectively, with Wilson flow for five different values of $r$ including double-exponential fit functions.
Figures~\ref{app_fig:csw_nc_of_r_flow} and \ref{app_fig:csw_ncinv_of_r_flow} show a part of the same data (for five flow times $t/a^2$) as a function of $r$ with linear fits.

\clearpage

%\printbibliography

%%%%%%%%%%%%%%%%%%%%%%%%%%%%%%%%%%%%%%%%%%%%%%%%%%%%%%%%%%%%%%%%%%%%%%%%%%%%%%%

%%%%%%%%%%%%%%%%%%%%%%%%%%%%%%%%%%%%%%%%%%%%%%%%%%%%%%%%%%%%%%%%%%%%%%%%%%%%%%%

\end{document}